\documentclass[twoside, 11pt]{article}
\usepackage{geometry}
\usepackage{amsmath, amsthm, amssymb, dsfont, mathrsfs, BOONDOX-cal, upgreek}
\usepackage{graphicx, caption, float, multirow}
\usepackage{chngcntr, etoolbox}
\usepackage[title]{appendix}
\usepackage{natbib}

\usepackage[hidelinks]{hyperref}
\usepackage[dvipsnames]{xcolor}
\hypersetup{colorlinks=true, linkcolor=RoyalBlue, citecolor=RoyalBlue}

\usepackage{txfonts}
\usepackage[T1]{fontenc}

\usepackage{fancyhdr}
\counterwithin*{equation}{section}
\counterwithin*{equation}{subsection}
\renewcommand\theequation{\ifnumgreater{\value{subsection}}{0}{\thesubsection.}{\thesection.}\arabic{equation}}

\allowdisplaybreaks

\newtheoremstyle{theorem}
  {10pt}
  {10pt}
  {\sl}
  {\parindent}
  {\bf}
  {. }
  { }
  {}
\theoremstyle{theorem}
\newtheorem{assumption}{Assumption}[section]
\newtheorem{definition}{Definition}[section]
\newtheorem{lemma}{Lemma}[section]
\newtheorem{theorem}{Theorem}[section]
\newtheorem{proposition}{Proposition}[section]

\newtheorem{example}{Example}[section]

\begin{document}
\title{\textsc{\textbf{Wealth or Stealth? The Camouflage Effect \\ in Insider Trading}}}
\author{Jin Ma\footnote{Department of Mathematics, University of Southern California}~~\thanks{Email: \underline{jinma@usc.edu}} \and Weixuan Xia\footnotemark[1]~~\thanks{Email: \underline{weixuanx@usc.edu}} \and Jianfeng Zhang\footnotemark[1]~~\thanks{Email: \underline{jianfenz@usc.edu}}}
\date{This Version: \today}

\maketitle

\begin{abstract}
  We consider a Kyle-type model where insider trading takes place among a potentially large population of liquidity traders and is subject to legal penalties. Insiders exploit the liquidity provided by the trading masses to ``camouflage'' their actions and balance expected wealth with the necessary stealth to avoid detection. Under a diverse spectrum of prosecution schemes, we establish the existence of equilibria for arbitrary population sizes and a unique limiting equilibrium. A convergence analysis determines the scale of insider trading by a stealth index $\gamma$, revealing that the equilibrium can be closely approximated by a simple limit due to diminished price informativeness. Empirical aspects are derived from two calibration experiments using non-overlapping data sets spanning from 1980 to 2018, which underline the indispensable role of a large population in insider trading models with legal risk, along with important implications for the incidence of stealth trading and the deterrent effect of legal enforcement. \medskip
\end{abstract}

\newcommand{\dd}{{\rm d}}
\newcommand{\pd}{\partial}
\newcommand{\PP}{\mathbb{P}}
\newcommand{\E}{\mathbb{E}}
\newcommand{\Var}{\mathrm{Var}}
\newcommand{\1}{\mathds{1}}
\newcommand{\erfc}{\mathrm{erfc}}
\newcommand{\sgn}{\mathrm{sgn}}
\renewcommand{\Phi}{\varPhi}
\renewcommand{\Psi}{\varPsi}
\renewcommand{\Lambda}{\varLambda}

\bigskip

\section{Introduction}\label{sec:1}

The main objective of the present paper is to conduct an in-depth analysis of the intrinsic link between an insider's trade size choices and the number of liquidity traders (or noise traders) during concurrent trading episodes, in uncovering and quantifying the level of stealth that insiders exercise in illicit trading activity. The central methodology is based on a newly-constructed game-theoretic framework (following Kyle (\citeyear{K85})) for insider trading that adapts to a wide variety of detection mechanisms and penalty functions while incorporating the population size of liquidity traders. In this general setting, the assumption of normally distributed liquidity trades is well supported by the large size of the liquidity trading crowd and the independence among individual trades, and we shall show that an equilibrium exists for any finite population of liquidity traders -- a result that is in itself highly nontrivial. As the population size tends to infinity, we also establish the uniqueness of the corresponding limiting equilibrium -- a critical outcome of considering the entire trading crowd. The convergence properties lead to the discovery of a stealth index $\gamma$ that sets up a rigorous yet simple link between insider trading and stealth trading. Such quantification sheds light upon the prevalence of an often overlooked moderate stealth level in illicit insider trading, revealing an incentive whose dual dependence on both regulatory scrutiny and potential legal penalties illuminates practical interconnections between market behavior and the legal enforcement landscape.

The canonical model of Kyle (\citeyear{K85}) provides a foundational framework for understanding informed trading in financial markets by exploring how an insider (informed trader) with private information about an asset's value trades strategically to maximize profit. It was demonstrated that the insider's trades are gradual and proportional to market liquidity, balancing profitability with informational concealment,\footnote{Developments in continuous time further highlight a gradual reduction in the tendency to conceal information towards the time when the asset's value becomes publicly known; see Back (\citeyear{B92}), Back and Baruch (\citeyear{BB04}), and Caldentey and Stacchetti (\citeyear{CS10}), among others.} and the equilibrium reached highlights a tradeoff between information efficiency and risks from market manipulation.

While the Kyle model does not explicitly address illicit insider trading -- focusing instead on informational concealment for profit maximization rather than avoidance of legal penalties -- many recent studies have advanced the framework to incorporate the modeling of legal risk associated with illicit trading for a comprehensive exploration of the price impact and strategic behavior influenced by the \textsl{dual} objectives of wealth expansion and penalty mitigation. For instance, Carr\'{e}, Collin-Dufresne, and Gabriel (\citeyear{CCDG22}) explicitly extended Kyle's one-period model to a setting subject to legal penalties and analyzed insider trading regulations striking a balance between market liquidity and price informativeness, and Kacperczyk and Pagnotta (\citeyear{KP24}), using data sourced from the \text{U.S.} Securities and Exchange Commission (SEC) case files across 1995 to 2018, argued that (illegal) insiders adjust trading strategies as they internalize legal risk from regulators, and that legal enforcement is indeed effective in deterring insider trading aggression and containing price informativeness.\footnote{Additionally, \c{C}etin (\citeyear{C25}), adopting a continuous-time approach, found that legal risks can impede the optimality for insiders to bring prices to the true valuation at the end of the trading episode.}

A notable feature in much of the literature along these lines, whether or not legal risks are considered, is the treatment of non-insider trades collectively as a single entity. In essence, these trades can be thought of as being placed by a single (representative) liquidity (or noise) trader.\footnote{For example, in the original Kyle model, liquidity (noise) trading quantities are assumed to be normally distributed; we also refer to Bagnoli, Viswanathan, and Holden (\citeyear{BVH01}) and Boulatov, Kyle, and Livdan (\citeyear{BKL13}) regarding the existence and uniqueness properties of a (linear) equilibrium.} Indeed, without legal risks, the number of liquidity traders is a relatively minor concern because the only potential harm from insider trading -- from an insider's viewpoint -- lies in reduced profits caused by their price impact; as a result, it is optimal to trade at the same scale as the representative liquidity trader, or equivalently, all liquidity traders combined. As legal risks arise, however, (illegal) insiders should suppress their trade sizes further, as trading at comparable levels to the entirety of liquidity traders is likely to expose them to legal repercussions. More precisely, it increases the likelihood of drawing regulatory attention and triggering red flags (see, e.g., Picardo (\citeyear{P22})). This raises the following question: \textsl{How do insiders exploit the presence of a large crowd of liquidity traders (representing normal trading activity) to moderate their trade sizes, thereby reducing detection risk while simultaneously mitigating price impact to maximize profits?} In particular, with a sizeable population of liquidity traders, it can be reasonably suspected that in equilibrium, while insiders still internalize legal risk when devising their trading strategies, they no longer account for the risks through the profit channel, as their consideration of price impact diminishes under the deterrent effect of legal consequences.

Although intuitive and seemingly natural, the understanding that the number of liquidity traders should far exceed that of insider traders -- on insider trading days -- can be justified by various legal and societal factors. First, laws against insider trading create a high barrier for engaging in it, and criminal and civil penalties, including fines, disgorgement, and prison sentences, deter most traders from engaging in insider trading (see, e.g., Patel and Put\c{n}in\v{s} (\citeyear{PP21}) \text{Sect.} 1 and Kacperczyk and Pagnotta (\citeyear{KP24}) \text{Tab.} I).\footnote{According to recent reports, the SEC prosecuted a total of 583 enforcement actions in 2024, obtaining orders for \$8.2 billion in financial remedies. See \underline{https://www.sec.gov/newsroom/press-releases/2024-186}.}
In addition, insider trading is contingent on access to material, non-public information, which is generally restricted to a small group (such as executives and directors) within firms, and only a minority of individuals within companies, or those heavily tied, have access to such sensitive information. Present legal risk, the large pool of liquidity traders can effectively provide insiders with the ``necessary'' cover to engage in illicit trading, while simultaneously reducing the likelihood of immediate detection, and contrarily, absence of a substantial trading crowd would readily render such illegal activities apparent and swiftly prosecuted, \text{e.g.} due to the relatively low costs associated with investigations, including legal and data analysis efforts (Picardo (\citeyear{P22})).

In psychological terms, this phenomenon is commonly known as the ``\textsl{camouflage effect},'' which describes situations where individuals blend into a crowd to avoid detection while engaging in wrongful or harmful acts; see, e.g., Griskevicius \textsl{et al.} (\citeyear{GGMCK06}). In the insider trading context, this ``simple'' act of hiding was originally hinted at in the model of Kyle (\citeyear{K85}), with the implication that when maximizing profit, (illicitly) informed traders tend to camouflage their information by splitting up their trades over time, while Admati and Pfeiderer (\citeyear{AP88}) argued that the same can also be achieved by purposefully engaging in trading amid high liquidity volume.

Noteworthily, the camouflage effect is closely associated with the well-studied concept of ``stealth trading,'' where informed traders strategically concentrate their trades in medium sizes. Intuitively, faced with legal risk, informed traders tend to execute volumes smaller than the total volume during periods of heavy trading activity to obscure their identities and reduce the likelihood of detection and prosecution. Meanwhile, driven by a clear profit motive, they are also inclined to trade in significantly larger volumes compared to the average liquidity trader. This phenomenon was early examined in Meulbroek (\citeyear{M92}) in studying the price impact of informed trades using data sourced from the SEC case files over the 1980s. In particular, a notable finding is that the (daily) median ratio of the insider trading volume to the target firms' total trading volume is about 11\%. A later study by Del Guercio, Odders-White, and Ready (\citeyear{DGOWR17}), utilizing SEC case files data from 2003 to 2011, has shown a significantly reduced price impact in contrast with the finding of Meulbroek (\citeyear{M92}).\footnote{This is ascribed to gradually increased prudence among insiders after 2000, likely in response to enhanced regulatory measures (such as increased regulatory budgets and the introduction of the SEC Whistleblower Program).}

This consideration also leads to what is known as the ``stealth trading hypothesis'' in the literature, originally proposed by Barclay and Warner (\citeyear{BW93}) in addressing the issue of informed traders' choices of trade sizes. The hypothesis predicts that medium-size (precisely defined as 500 to 9,900 shares) trades come with disproportionately large cumulative price changes. Apart from their own empirical evidence from a sample of NYSE firms between 1981 and 1984, there have been numerous studies over the years to confirm the predictions of this hypothesis. To name a few, using audit trail data from NYSE firms, Chakravarty (\citeyear{C01}) highlighted the disproportionately large role played by informed traders utilizing medium-sized trades in driving price movements. Anand and Chakravarty (\citeyear{AC07}) confirmed this preference for medium-sized trades in the options markets as well, particularly for high-leverage options. However, it is worth noting that this tendency toward moderate trade sizes is not consistently observed in non-\text{U.S.} markets, likely due to elevated levels of price manipulation associated with liquidity trading (see Cai, Cai, and Keasey (\citeyear{CCK06})). Another empirical study by Frino \textsl{et al.} (\citeyear{FSWZ13}) shows that insiders' trade sizes are largely affected by both the probability of detection and expected penalty, and that insider volumes surpass liquidity trades at the individual level for a specific expected return while falling below the aggregate level.

A natural implication from the last line is that if detection of insider trading is triggered with abnormal order flow imbalances exceeding certain watermarks (see, e.g., DeMarzo, Fishman, and Hagerty (\citeyear{DFH98}) and Kacperczyk and Pagnotta (\citeyear{KP24}) \text{Sect.} II), then it would be optimal for insiders to execute trades at medium levels where normal trades are conspicuous to reduce the likelihood of detection. Alexander and Peterson (\citeyear{AC07}) referred to such occurrence as trade-size clustering, documenting that NYSE and NASDAQ trades typically cluster around multiples of 500, 1,000, or 5,000 shares, associated with the greatest price impact, and the clustering strengthens with aggregate trading activity; see also Chen (\citeyear{C19}) for similar considerations for cryptocurrency trading. On the other hand, using futures trading data, Chang, Pinegar, and Schachter (\citeyear{CPS97}) showed that when large speculative trades are singled out, a significantly stronger price--volume relationship is observed; see also Blau (\citeyear{B17}). The concentration of medium-sized trades has the econometric implication of enhanced sparsity in trade size distributions despite significant overall trading volumes, as modeled and demonstrated by Fei and Xia (\citeyear{FX24}). From different angles, these studies consistently support the presence of stealth trading, confirming that informed traders camouflage their activities by favoring medium-sized trades during periods of heavy market activity.

With the above considerations, we aim to develop a Kyle-type model (Section \ref{sec:2}) with legal penalties (compare Carr\'{e}, Collin-Dufresne, and Gabriel (\citeyear{CCDG22}) \text{Sect.} 2 and \c{C}etin (\citeyear{C25}) \text{Sect.} 2) that incorporates the liquidity trading population for a formal study of the camouflage effect in insider trading. We quantify the camouflage effect by introducing a measure termed ``stealth index,'' gauging the level of stealth (or caution) that insiders exercise in illicit trading (Section \ref{sec:3}). Determination of this stealth index relies in large part on the convergence of the corresponding equilibria as the liquidity trading population grows, which deciphers how insiders' trade sizes vary with the number of liquidity traders in addition to the severity of legal penalties. We would like to highlight that the resulting limiting equilibrium can effectively reproduce a significantly reduced price impact which is adequate for practical purposes when justified by the presence of a large trading crowd.

It is worth noting that such a study would not be possible with only a scale parameter tracking the size of all liquidity trades (as adopted in Kyle (\citeyear{K85}) \text{Sect.} 2). The reason is that with abnormal order flow imbalance-based detection of insider trading (DeMarzo, Fishman, and Hagerty (\citeyear{DFH98})), market liquidity is well factored by insiders, whereas regulatory investigations are conducted on a case-by-case basis, i.e., on the individual trader level rather than the volume of trades.\footnote{What complicates the investigation process is not the total orders from liquidity traders, but their sheer number, thus giving insiders the opportunity to ``camouflage.''} Also, given that each liquidity trader places trades independently responding to idiosyncratic needs, the total liquidity order flow is in proportion to their average trade size but \textsl{not} their population size. This means that the particular size of liquidity trades does not directly contribute to the camouflage effect, which requires a sufficiently large population to manifest.\footnote{This perspective will be further clarified by the formal mathematical treatment in Section \ref{sec:2}.}

From a modeling perspective, the present paper also runs parallel to research on regulatory approaches to insider trading, including investigation schemes and the imposition of penalties. We consider a flexible detection mechanism (Section \ref{sec:2.1}) that is adaptive to the detection mechanism postulated by DeMarzo, Fishman, and Hagerty (\citeyear{DFH98}), who were among the first to consider the optimal design of insider trading regulation. Besides, our choice of the penalty function covers combinations of civil penalties and criminal penalties, with civil penalties determined based on insiders' illicit trading profit and criminal penalties depending primarily on their trading strategies. Aside from reflecting practical scenarios encompassing both civil and criminal cases (detailed discussion in Section \ref{sec:2.2}), this consideration is broad enough to include most forms of penalty functions discussed in the literature, such as linear penalties (as in Carr\'{e}, Collin-Dufresne, and Gabriel (\citeyear{CCDG22}) \text{Sect.} 3.4.2), quadratic penalties (as in Shin (\citeyear{S96}), Carr\'{e}, Collin-Dufresne, and Gabriel (\citeyear{CCDG22}) \text{Sect.} 3.4.1, and \c{C}etin (\citeyear{C25})), and multiples of illicit profit (as in Kacperczyk and Pagnotta (\citeyear{KP24}) \text{Sect.} II).\footnote{Notably, while criminal penalties are naturally nondecreasing in the (insider) trading strategy, in the case of civil penalties, the expected penalty function is generally not symmetric or nondecreasing (see Section \ref{sec:3.2}), and so it operates outside the framework of Carr\'{e}, Collin-Dufresne, and Gabriel (\citeyear{CCDG22}) and \c{C}etin (\citeyear{C25}).} In Section \ref{sec:3}, we shall also demonstrate that such a penalty combination is materially significant by both accounting for the nuances of realized illicit gains and assessing the pre-trade intent in determining the appropriate legal charges.

In a theoretical construct, this paper focuses on the following main contributions:
\begin{itemize}
  \item Construct a Kyle-type model with legal risk incorporating a flexible detection mechanism as well as both civil and criminal penalties, in which the number of liquidity traders can be large (Section \ref{sec:2}).
  \item Propose a stealth index to quantify the camouflage effect in insider trading, implying that in equilibrium, insiders prefer a trading intensity that lies strictly between that of an average liquidity trader and the combined activity of all liquidity traders (Section \ref{sec:3} and Section \ref{sec:5}).
  \item Conduct a thorough convergence analysis to show that using the limiting equilibrium (with diminished price impact) as an approximate equilibrium offers great ease in computation while not altering the equilibrium implications (Section \ref{sec:3} and Section \ref{sec:5}).
\end{itemize}

Our model is also amenable to calibration to market data on insider trading volumes, while also aligning with unresolved empirical findings from the literature. With the calibration experiments in Section \ref{sec:4} based on two data sets, we confirm insiders' internalization of legal risk from regulators as well as the effectiveness of legal enforcement in deterring insider trading, along with reduced price informativeness (Del Guercio, Odders-White, and Ready (\citeyear{DGOWR17}) and Kacperczyk and Pagnotta (\citeyear{KP24})). On the other hand, insiders adjust their trade sizes according to the size of liquidity trades (e.g., Barclay and Warner (\citeyear{BW93}) and Chakravarty (\citeyear{C01})) in attempting to disguise their trades for legal risk avoidance, to such extent that their price impact asymptotically vanishes in equilibrium. All main proofs are provided in Appendices \ref{A}, \ref{B}, and \ref{C}, while the proofs of additional results and other auxiliary details are presented in Supplemental Appendices \ref{SA} and \ref{SB}.

\medskip

\section{A Kyle model with legal risk}\label{sec:2}

We begin with a description of our model framework. Following the standard one-period model of Kyle (\citeyear{K85}), the market features three types of agents trading a single risky asset with fundamental value $V$, which is revealed, at time 1, to be 1 with probability $0<p<1$ and 0 otherwise.

There are a total of $N$ \textbf{liquidity traders} who will trade the asset non-strategically for exogenous needs. At time 0, the liquidity traders have no prior access to the asset value $V$ and submit independent orders with sizes being \text{i.i.d.} random variables with zero mean and variance $\sigma^{2}$. Thus, with the central limit theorem in place, we assume that the total (net) order flow from liquidity trading can be written as $\sqrt{N}W$, where $W\overset{\rm d.}{=}\text{Normal}(0,\sigma^{2})$ is independent of $V$. There is a single (risk-neutral) \textbf{insider trader} who observes the value $V$ at time 0 at no cost and in a strategic manner submits his order $Z(V)$ as a function of $V$, knowing that placing the order is likely to introduce a contemporary price impact on the risky asset. At time 0, a competitive \textbf{market maker} sees the total (net) order flow $\sqrt{N}W+Z(V)=Y$ and executes orders at some price $P(Y)$ as a function of $Y$.\footnote{While taking account of his own price impact, the insider cannot design his trading strategy based on the orders from the liquidity traders; in other words, the insider's order is submitted no later than the realization of $W$.}

One major distinction from the standard Kyle model (and its variations with legal risk) is in the \textbf{population size} of liquidity traders $N$,\footnote{If the one period (from time 0 to time 1) represents a generic insider trading episode, $N$ can also be regarded as the total number of liquidity trades, assuming that each trade is placed by a different (liquidity) trader.} which we expect to be ``very large.'' Beyond the practical reasons for a large number of liquidity traders discussed in Section \ref{sec:1}, a modeling perspective also supports choosing a normal distribution for $W$. Indeed, if the population size were small, any specific distributional assumption (whether binomial, uniform, normal, etc.) would risk being overly restrictive; on the other hand, with a sufficiently large population size, independence among the liquidity traders sets the ground for the normal approximation.\footnote{\label{CLT}We may use 30 as a practical threshold -- following conventional statistical guidelines -- to show that the population size is large enough for this approximation to take hold. In our convergence analysis later, however, a larger value of $N$ may be required to ensure effectiveness of the convergence.}

\smallskip

\subsection{Prosecution mechanism}\label{sec:2.1}

The insider trader is also aware of legal risk that he faces in the presence of a \textbf{regulator}, who initiates investigations into his trading behavior by various means, such as detections of abnormal trading activity (DeMarzo, Fishman, and Hagerty (\citeyear{DFH98})), market surveillance associated with corporate events (Patel and Put\c{n}in\v{s} (\citeyear{PP21})), or informant reports (such as the SEC Whistleblower Program). Once an investigation starts, prosecution of the insider will ensue depending on the actual trading behavior. A general description of the prosecution mechanism event could be based on an independent Bernoulli random variable $B_{N}$ with strategy-dependent parameter, $\mathcal{p}_{N}$; that is, given his order $Z(V)=z$ placed at time 0, the insider will be charged with illicit trading with probability $\mathcal{p}_{N}(z)$ at time 1, where the dependence on $z$ is implied by the regulator gaining access to the insider's trading account. More precisely, we consider this probability parameter, or the probability of successful prosecution, in the following form:
\begin{equation}\label{2.1.1}
  \mathcal{p}_{N}(z)=1-e^{-\lambda_{N}(z)},\quad z\in\mathbb{R},
\end{equation}
where $\lambda_{N}(z)\geq0$ is interpreted as the \textbf{hazard rate} and is a \textsl{strictly} increasing function in the magnitude of $z$; contrarily, $1-\mathcal{p}_{N}(z)=e^{-\lambda_{N}(z)}$ is the probability of no investigation or a dismissal from court -- or broadly, the corresponding ``survival'' probability. The structure (\ref{2.1.1}) allows us to focus on the prosecution intensity, directly related to the likelihood, and it immediately implies that successful prosecution is never guaranteed, which is also practical given the complex circumstances of each insider trading case. Meanwhile, it is important to impose that $\lambda_{N}(0)=0$, for, if the insider does not trade at all, then there is no room for prosecution, while the increasing feature of $\lambda_{N}$ signifies that aggressive (insider) trading strategies necessarily result in a heightened probability of prosecution.

It is also crucial that the hazard rate (and thus the prosecution probability) explicitly depends on the population size $N$, which is to incorporate the idea that the probability of identifying an insider trader should decrease as the total number of normal traders increases. Indeed, a large number of liquidity traders can create an elevated noise level, making it harder to distinguish abnormal trading patterns from normal trading activity. Since the insider's trades are usually detected as they represent outliers in terms of order flow imbalances (DeMarzo, Fishman, and Hagerty (\citeyear{DFH98})), with an increasing number of liquidity trades, the threshold for what counts as a statistical outlier also rises, in turn making it more challenging to detect his trades. The insider can then ``camouflage'' more easily, with his suspicious behavior becoming less conspicuous. To examine the scale effect of $N$ on the insider's strategies, we impose a general power-type structure,
\begin{equation*}
  \lambda_{N}(z)=\lambda(N^{-\beta}z),
\end{equation*}
where $\lambda\equiv\lambda_{1}$ is a \textsl{size-modulated} hazard rate and $\beta\geq0$ is a power coefficient governing the scale.

As an important example, the probability of prosecution may be conceptualized as arising from the sequential processes of abnormal order flow imbalance-triggered investigation and subsequent successful legal actions, i.e.,
\begin{equation}\label{2.1.2}
  \mathcal{p}_{N}(z)=D(N^{-\frac{1}{2}}z)\PP\{|W+N^{-\frac{1}{2}}z|\geq\bar{y}\},
\end{equation}
where $D:\mathbb{R}\mapsto[0,1]$ is a size-modulated function satisfying that $D(0)=0$, representing the probability of prosecution conditional on investigation, and $\bar{y}>0$ is a custom detection threshold. Then, we have the following functional form for the hazard rate:
\begin{align}\label{2.1.3}
  \lambda_{N}(z)&=-\log(1-\mathcal{p}_{N}(z)) \nonumber\\
  &=-\log\bigg(1-\frac{D(N^{-\frac{1}{2}}z)}{2}\bigg(\erfc\bigg(\frac{\bar{y}+N^{-\frac{1}{2}}z}{\sqrt{2\sigma^{2}}}\bigg) +\erfc\bigg(\frac{\bar{y}-N^{-\frac{1}{2}}z}{\sqrt{2\sigma^{2}}}\bigg)\bigg)\bigg),
\end{align}
where $\erfc(\cdot)$ is the complementary Gauss error function; in this case, $\beta=1/2$ exactly. Detailed derivations are found in Supplemental Appendix \ref{B}. In Kacperczyk and Pagnotta (\citeyear{KP24}), the prosecution probability is precisely $\mathcal{p}_{1}$ from (\ref{2.1.2}), except with $D\in[0,1]$ being a constant probability.

In general, the power coefficient $\beta$ allows us to consider a flexible dependence structure of the prosecution probability with respect to the aggregate noise level, which is crucial to examining its impact on the behavior of the insider's strategy $Z$ when the population size becomes large.

\smallskip

\subsection{Composition of legal penalties}\label{sec:2.2}

Once successfully prosecuted, the insider trader is subject to legal penalties depending on the jurisdiction and the specific circumstances of the case. First, a common component of insider trading penalties is \textsl{disgorgement} of illicit profits, which involves the return of any profits gained from the illicit trading activity, ensuring unprofitability. The amount of disgorgement is usually calculated based on the total profits earned by the insider, equaling $(V-P)Z(V)$ upon revelation of the asset value.

Another notable aspect is the adoption of so-called ``penalty multipliers,'' which are imposed in some cases as an enhancement of the deterrent effect of the penalties, based on factors such as the egregiousness of the violation or the level of harm caused to investors or the market. For example, in the U.S., the maximum civil penalty for insider trading is up to three times the profit gained or loss avoided, \text{a.k.a.} the ``treble damages'' provision under \textsc{the Insider Trading and Securities Fraud Enforcement Act of 1988 (ITSFEA)}. Similarly, as amended by the ITSFEA, depending on evidence of the insider's level of intent and trading behavior, significant penalties can be triggered in criminal cases to deter future misconduct; this includes prison sentences and steep fines -- e.g., up to 20 years and up to \$5 million for individuals in the U.S.

With these factors in mind, we believe that a comprehensive model for legal penalties should allow for explicit dependence on \textsl{both} the insider trader's illicit profit (as a civil penalty) and his trading strategy (as a criminal penalty). Thus, it is desirable that a penalty function should be a bivariate function of the insider's trading strategy ($z$) as well as his profit gained ($(v-P(y))z$), taking the general form $C(z,(v-P(y))z)$. In the present paper, we focus on the below composition:
\begin{equation}\label{2.2.1}
  C\big(z,(v-P(y))z\big)=C_{0}(z)+\chi\big((v-P(y))z\big)^{+},\quad v\in\{0,1\},
\end{equation}
where $(\cdot)^{+}$ denotes the positive part. On the right side of (\ref{2.2.1}), the first component with $C_{0}$ incorporates \textbf{criminal penalties}, including imprisonment and fines,\footnote{In the case of imprisonment, $C_{0}$ can be considered equivalent to a capital penalty due to its severe, lifelong impact on one's economic, social, and personal ``capital.'' Such an impact could manifest as opportunity costs, social and economic disconnection, lasting stigma, etc.} that directly depends on the insider's trading strategy, while the second stands for \textbf{civil penalties} (with regard to the profit) controlled by a \textbf{penalty multiplier} $\chi$. For a meaningful consideration, we require $C_{0}(z)$ to be increasing (not necessarily strictly) in the magnitude of $z$, and it makes sense to force $C_{0}(0)=0$ in the case of no trading. Additionally, we require that $\chi\geq1$, which implies disgorgement of the insider's illicit profit following successful prosecution.\footnote{This feature is also in keeping with the institutional parametric condition made in Kacperczyk and Pagnotta (\citeyear{KP24}) \text{Sect.} II, ensuring a penalty no less than the profit gained.}  Clearly, when $\chi=1$, only $C_{0}$ will remain on top of disgorgement, and penalization is largely strategy-based, while by taking $C_{0}\equiv0$, the second component dominates, leading to profit-based (civil) penalties.

\smallskip

\subsection{Equilibrium definition}\label{sec:2.3}

With the aforementioned legal risk specifications, we now formulate the insider trader's optimization problem and give a formal definition of the equilibrium, for any population size $N$ of liquidity traders.

Accessing the value $V$ at time 0, the insider trader takes the price function $P$ from the market maker as given and designs a trading strategy to maximize his expected net profit. His objective function is given by
\begin{equation}\label{2.3.1}
  J_N(P;z,v):=\E\Big[(v-P(\sqrt{N}W+z))z ~ - ~\mathcal{p}_{N}(z)C\big(z,(v-P(\sqrt{N}W+z))z\big)\Big],
\end{equation}
for $z\in\mathbb{R}$, $v\in\{0,1\}$, which he seeks to maximize over $z$ for each value of $v$, namely $z=Z(v)$. At the same time, given the insider's trading strategy $Z=(Z(0),Z(1))$, the market maker observes the total order flow $Y$ and sets a rational price function for break-even,\footnote{This can be justified by a Bertrand competition argument; see \text{e.g.} Kyle (\citeyear{K85}). An alternative interpretation is that the market maker chooses $P$ to minimize the squared error $\E[(V-P(\sqrt{N}W+Z(V)))^2]$.} namely setting $P(y)=P_N(Z;y)$, where
\begin{equation}\label{P*}
  P_{N}(Z;y):=\E\big[V|\sqrt{N}W+Z(V)=y\big],\quad y\in\mathbb{R}.
\end{equation}
Since $V$ only takes values $0$ and $1$, by applying Bayes' rule, we easily obtain
\begin{align}\label{2.3.2}
  P_{N}(Z;y)&=\PP\big\{V=1|\sqrt{N}W+Z(V)=y\big\} =\frac{p e^{-\frac{(y-Z(1))^{2}}{2N\sigma^{2}}}}{p e^{-\frac{(y-Z(1))^{2}}{2N\sigma^{2}}}+(1-p)e^{-\frac{(y-Z(0))^{2}}{2N\sigma^{2}}}} \nonumber\\
  &=\frac{1}{1+q e^{\frac{(2y-Z(0)-Z(1))(Z(0)-Z(1))}{2N\sigma^{2}}}},\quad\mbox{with }q:=\frac{1-p}{p}.
\end{align}

With the population size $N$ fixed, the (original) hazard rate $\lambda_{N}$ and the liquidity order flow $\sqrt{N}W\overset{\rm d.}{=}\text{Normal}(0,N\sigma^{2})$ are both well-defined and finite, and the notion of equilibrium can be stated as in the standard Kyle model setting.

\begin{definition}\label{def:1}
For any fixed $N\geq1$, an \underline{equilibrium} is a couple $(Z^{\ast}_{N},P^{\ast}_{N})$ of trading strategy and price function such that: \\
(i) Given $P^{\ast}_{N}$, the insider trader maximizes his expected net profit from trade (\ref{2.3.1}), i.e.,
\begin{equation}\label{Z*}
  J_N(P^{\ast}_{N};Z^{\ast}_{N}(v),v)=\sup_{z\in\mathbb{R}}J_N(P^{\ast}_{N};z,v),\quad v\in\{0,1\}.
\end{equation}
(ii) Given $Z^{\ast}_{N}$, the market maker sets a rational price function for break-even, namely
\begin{equation*}
   P^{\ast}_{N}(y) = P_{N}(Z^{\ast}_{N};y),  
   \quad y\in\mathbb{R}.
\end{equation*}
\end{definition}

According to condition (ii) above, for equilibrium analysis it is useful to rewrite the insider's objective function as
\begin{equation*}
  \acute J_N(Z;z,v):=J_N(P_N(Z;\cdot); z, v),
\end{equation*}
where $Z$ can be viewed as the market maker's target strategy. Since $V$ is supported on $\{0,1\}$, from (\ref{P*}) it is clear that $0<P_N<1$. Then, in light of (\ref{2.2.1}) and (\ref{2.3.1}), regarding the profit maximization \eqref{Z*}, the insider would rationally concentrate on sell strategies with $z\leq0$ when $v=0$ and buy strategies with $z\geq0$ when $v=1$; indeed, by abstaining from trading entirely ($z=0$), the investor guarantees a nonnegative net profit. However, a zero trading strategy is also trivial on account of legal risk and is demonstrably suboptimal for the insider.\footnote{Under the key Assumption \ref{as:1}, it can be shown that a zero strategy cannot be optimal in equilibrium; see Step 1 in the proof of Theorem \ref{thm:1} in Appendix \ref{B}.} With this consideration in mind, we may assume without loss of generality that the insider's strategy satisfies the constraint $Z=(Z(0), Z(1))\in(-\infty, 0)\times (0, \infty)$. As a result, the objective (\ref{Z*}) is equivalent to
\begin{equation}\label{Z*2}
  \acute J_N(Z^{\ast}_{N};Z^{\ast}_{N}(v),v)=\sup_{z\in\mathbb{R}_{v}}\acute J_N(Z^{\ast}_{N};z,v),\quad\text{with }
  \mathbb{R}_{v}:=
  \begin{cases}
    (-\infty,0),&\;\text{if }v=0, \\
    (0,\infty),&\;\text{if }v=1.
  \end{cases}
\end{equation}
Besides, with $v\in\{0,1\}$ and $0<P_N<1$, we have in (\ref{2.2.1}) that
\begin{equation}\label{2.3.6}
  ((v-P(y))z)^{+}=(v-P(y))z~~\mbox{for all}~~ z\in \mathbb{R}_{v},\quad\mbox{and } P=P_N(Z;\cdot).
\end{equation}
Then, recalling (\ref{2.3.1}) and (\ref{2.2.1}), we define the insider's expected price, expected (gross) profit, and expected additional penalties respectively as
\begin{align}\label{2.3.3}
  \Phi_{N}(Z;z)&:=\E\big[P_{N}(Z;\sqrt{N}W+z)\big], \quad Q_N(Z; z, v):=\big(v-\Phi_{N}(Z;z)\big)z, \nonumber\\
  \Psi_{N}(Z;z,v)&:= 
  C_{0}(z)+(\chi-1)Q_N(Z; z, v),
\end{align}
which allow us to recast the objective function (\ref{2.3.1}) into the equivalent form
\begin{equation}\label{2.3.4}
  \acute J_{N}(Z;z,v)=e^{-\lambda_N(z)}Q_N(Z; z, v)-(1-e^{-\lambda_N(z)})\Psi_{N}(Z;z,v),\quad z\in\mathbb{R}_{v}.
\end{equation}
In particular, \eqref{2.3.2} shows that the function $\Phi_{N}$ is precisely the expected value of a Gaussian random variable under a sigmoid-type transformation.\footnote{Despite no known closed form, it can be expressed in terms of an infinite series containing Gaussian moments and explicitly computable coefficients by adopting an expansion argument.}

\medskip

\section{Limiting equilibrium and convergence}\label{sec:3}

In exercising stealth trading, the insider trader has the tendency to trade in larger quantities compared to a single liquidity trader, while his trade size cannot be ``too large'' at the same time in view of the risk of detection and prosecution. We again refer to Frino \textsl{et al.} (\citeyear{FSWZ13}) \text{Sect.} V.B, Meulbroek (\citeyear{M92}) \text{Sect.} E and Chakravarty (\citeyear{C01}) for related empirical evidence. This suggests that an insider's optimal trading strategy should, ideally, scale with the total population size to adequately capture its asymptotic behavior.

From a mathematical viewpoint, in the limit as $N$ goes to infinity, the optimal trading strategy $Z^{\ast}_{N}$ from Definition \ref{def:1} can well be infinity. Since we consider power-law growth in the hazard rate, it is reasonable to introduce the scaled strategy 
$\tilde Z^{\ast}_N =N^{-\gamma}Z^{\ast}_{N}$ for some (yet-to-be-determined) coefficient $\gamma\geq0$, subject to the requirement that the limiting scaled strategy $\tilde{Z}^{\ast}_{\gamma}:=\lim_{N\to\infty}\tilde Z^{\ast}_N$ exist with nonzero values in $\mathbb{R}_{0}\times\mathbb{R}_{1}$. In other words, we suspect the insider's (original) strategy to have the property that $Z^{\ast}_{N}\sim N^{\gamma}\tilde{Z}^{\ast}_{\gamma}$, i.e., $\lim_{N\to\infty}Z^{\ast}_{N}(v)/(N^{\gamma}\tilde{Z}^{\ast}_{\gamma}(v))=1$ for $v\in\{0,1\}$, as $N\to\infty$. The coefficient $\gamma$ bears direct connections to the camouflage effect as it directly reflects the insider's desire to balance his wealth expected from illicit trading and his level of stealth to avoid detection and prosecution. A smaller value is associated with a smaller chance of detection, or equivalently, a higher stealth level. In particular, we formally name it the \textbf{stealth index} and, considering the aforementioned moderateness of insider trade sizes, a desirable -- but yet-to-be-verified -- value range for this index will be $(0,1/2)$.\footnote{Clearly, $\gamma=0$ would imply that the insider's trade size is comparable to that of a single liquidity trader, while if $\gamma=1/2$, he would trade at the same scale as all the liquidity traders combined.} Nevertheless, in what follows we shall still consider the general value range $[0,\infty)\ni\gamma$.

To formalize the notion of a limiting equilibrium based on the scaled trading strategy, $\tilde{Z}=N^{-\gamma}Z$, we need to consider the limit of the price function (\ref{2.3.2}) as $N\to\infty$. Since for $z$ at the scale of $N^{\gamma}$, $y=\sqrt{N}W+z$ is at the scale of $N^{\max\{\gamma,1/2\}}$, we can introduce the scaled order flow $\tilde y = N^{- \max\{\gamma,1/2\}} y$ and re-parameterize the equilibrium price function in (\ref{P*}) and (\ref{2.3.2}) as
\begin{equation}\label{PNgamma}
  \tilde {P}^\gamma_{N}(\tilde{Z}; \tilde y)=P_{N}(N^\gamma \tilde Z;N^{\max\{\gamma,1/2\}}\tilde y),\quad \tilde Z=(\tilde{Z}(0),\tilde{Z}(1))\in \mathbb{R}_{0}\times\mathbb{R}_{1},\quad \tilde y\in \mathbb{R}.
\end{equation}
Throughout the paper, we use the tilde notation ($\tilde{\;}$) to indicate the scaled quantities. The next proposition shows that the corresponding limit generally depends on $\gamma$.

\begin{proposition}\label{pro:1}
For $\tilde {P}^\gamma_{N}$ as in (\ref{PNgamma}), $\tilde Z\in \mathbb{R}_{0}\times\mathbb{R}_{1}$,
and $\tilde y\in \mathbb{R}$, it holds that
\begin{equation}\label{3.1}
  \lim_{N\to\infty}\tilde {P}^\gamma_{N}(\tilde{Z}; \tilde y)=  \tilde P^\gamma_\infty(\tilde{Z}; \tilde y) :=
  \begin{cases}
    p,&\;\text{if }\gamma\in\big[0,\frac{1}{2}\big), \\
    P_{1}(\tilde{Z};\tilde y),&\;\text{if }\gamma=\frac{1}{2}, \\
    \1_{\{2\tilde y > \tilde Z(0) + \tilde Z(1)\}} + p\1_{\{2\tilde y = \tilde Z(0) + \tilde Z(1)\}},&\;\text{if }\gamma>\frac{1}{2}.
  \end{cases}
\end{equation}
\end{proposition}

Proposition \ref{pro:1} demonstrates that for the interesting range $\gamma\in(0,1/2)$ of the stealth index, the limiting price function no longer depends on the insider's trading strategy and is, in particular, constant, equal to the expected value of $V$. In the case $\gamma=1/2$, as the insider trades at the scale of all liquidity trades, the limit is, unsurprisingly, invariant to the population size $N$, which can be normalized to 1, as considered in the literature. The case $\gamma>1/2$ points to an audacious insider trading strategy, with volumes far exceeding typical aggregate levels, and the asset price is set in the usual way (equal to $p$) when the market maker perceives no insider involvement, and is otherwise adjusted to the extremal values (0 or 1), making detection and prosecution nearly inevitable (no stealth). This observation suggests $[0,1/2]\ni\gamma$ as a plausible range and is to be verified through equilibrium analysis.

The price limits in Proposition \ref{pro:1} allow to examine the corresponding limiting behaviors of the expected price, profit, and penalties, $\Phi_{N}$, $Q_{N}$, and $\Psi_{N}$ in equilibrium. However, simply analyzing these limits appears to offer little insight into the correct choice of the stealth index $\gamma$ from the insider's perspective. Instead, we shall re-scale the insider's objective function (or expected net profit): Given a price function $\tilde P$ on the scaled order flow $\tilde y$, we consider the following limiting scaled objective function:
\begin{equation}\label{3.2}
  \tilde J^\gamma_{\infty}(\tilde{P};\tilde{z},v):=\lim_{N\to\infty}N^{-\gamma}J_{N}( P^\gamma_N;N^{\gamma}\tilde{z},v), \quad\tilde{z}\in\mathbb{R}_{v},\;v\in\{0,1\},
\end{equation}
where $J_{N}$ is the original objective function in (\ref{2.3.4}), and $P^\gamma_N(y):=\tilde P(N^{-\max\{\gamma,1/2\}} y)$ is the recovered price function. On the right side of (\ref{3.2}), $J_{N}$ is scaled exactly by $N^{-\gamma}$ because based on (\ref{2.3.4}), when $P^\gamma_N(y)= P_N(N^\gamma \tilde Z; y)$, the expected price $\Phi_{N}$ is bounded and the expected additional penalty $\Psi_{N}$ must be controlled not to blow up either. Hence, the limiting equilibrium paired with Definition \ref{def:1} can be defined with $\gamma$ chosen such that the limiting objective function (\ref{3.2}) is well-defined, and to ensure that the effect of $\gamma$ is meaningful, it is important to restrict attention to nonzero limiting strategies as well as nonzero limiting objective function values.

\begin{definition}\label{def:2}
A \underline{$\gamma$-limiting equilibrium} is a couple $(\tilde{Z}^{\ast}_{\gamma},\tilde{P}^{\ast}_{\gamma})$ of scaled trading strategy and price function such that: \\
(i) The insider trader maximizes his limiting scaled expected net profit (\ref{3.2}), i.e.,
\begin{equation*}
  \tilde J^\gamma_{\infty}(\tilde{P}^{\ast}_{\gamma};\tilde{Z}^{\ast}_{\gamma}(v),v)= \sup_{\tilde{z}\in\mathbb{R}_{v}}\tilde{J}^\gamma_{\infty}(\tilde{P}^{\ast}_{\gamma};\tilde{z},v)\neq0,\quad\tilde Z^*_\gamma(v)\neq 0,\quad v\in\{0,1\}.
\end{equation*}
(ii) The market maker sets a rational price function according to (\ref{3.1}), $\tilde{P}^{\ast}_{\gamma} =  \tilde P^\gamma_\infty(\tilde{Z}^\ast_\gamma; \cdot)$.
\end{definition}

On paper, from the objective function forms (\ref{2.3.4}) and (\ref{3.2}), a sophisticated choice of the stealth index $\gamma$ should exhibit reliance on the growth rate of the size-modulated hazard rate $\lambda$, controlled by $\beta\geq0$, as well as that of the criminal penalty component $C_{0}$, with respect to the insider's trading strategy.\footnote{On the other hand, note that from (\ref{2.2.1}), the civil penalty component is always of linear growth as $|z|\to\infty$ due to the boundedness of the price function $P_{N}$.} Intuitively, with light-to-moderate criminal penalties, the expected price $\Phi_{N}$ tends to overshadow potential legal risk, prompting the insider to trade at larger scales and worry less about being detected, thereby using a larger $\gamma$, while severe criminal penalties are likely to place a deterrent effect on the trading behavior and yields a relatively small $\gamma$. Condition (ii) in Definition \ref{def:2} also reveals that if the insider trader exercises stealth trading with $\gamma<1/2$, then in the limit of the population size $N$, the equilibrium price function becomes constant and \textsl{decoupled} from the trading strategy due to diminished price informativeness (see again Chakravarty (\citeyear{C01}) and Kacperczyk and Pagnotta (\citeyear{KP24})). This decoupling feature significantly simplifies subsequent equilibrium analysis.

The above suggests that the exact structure of the penalty function $C$ plays an important role in determining the correct stealth index with an increasing population size of liquidity traders. To provide a thorough understanding of convergence towards the limiting equilibria, instead of providing a balanced treatment of all penalty types, the following analysis concentrates on the predominant scenario of civil penalties, which represents the most common enforcement outcome in insider trading cases and matches the empirical nature of the data sets to be presented in Section \ref{sec:4}. The examination of scenarios involving criminal and mixed penalties is reserved for further discussions in Section \ref{sec:5}, also helping align with various cases considered in the literature (Carr\'{e}, Collin-Dufresne, and Gabriel (\citeyear{CCDG22}), Kacperczyk and Pagnotta (\citeyear{KP24}), and \c{C}etin (\citeyear{C25})), apart from highlighting the technical differences. 

\medskip

\section{Predominant scenario: Civil penalties}\label{sec:3.2}

In the predominant scenario, upon successful prosecution, the insider trader faces civil penalties only (as considered in Kacperczyk and Pagnotta (\citeyear{KP24}) \text{Sect.} 2). The penalties are imposed on his illicit profit ($(V-P_{N})Z(V)$) through the penalty multiplier $\chi\geq1$. In the absence of criminal charges, we shall take $C_{0}=0$ in (\ref{2.2.1}).

Given a population size $N\geq1$, the equilibrium objective function (\ref{2.3.4}) reduces to
\begin{align}\label{3.2.1}
  \acute J_{N}(Z;z,v) &=e^{-\lambda_N(z)}Q_{N}(Z;z,v)-(1-e^{-\lambda_N(z)})\Psi_{N}(Z;z,v) \nonumber\\
  & =e^{-\lambda_N(z)}Q_{N}(Z;z,v) -\chi_{0}(1-e^{-\lambda_N(z)})Q_{N}(Z;z,v),\quad z\in\mathbb{R}_{v},
\end{align}
where $\chi_{0}:=\chi-1\geq0$ is the \textsl{disgorgement-adjusted} penalty multiplier. Based on the expression of $Q_N$ in (\ref{2.3.3}), a notable feature of the civil penalties is their explicit dependence on the asset value $v\in\{0,1\}$, and the insider designs his trading strategies knowing that the total order flow ($Y=\sqrt{N}W+Z(V)$) observed by the market maker will automatically absorb uncertainty embedded in such strategies.

On a closer look, the expected (additional) penalty $\Psi_{N}$ in (\ref{3.2.1}) can increase at most linearly with the insider's trade size ($|z|$), since the expected price $\Phi_{N}$ is bounded, and more importantly, it is not a monotone function of $z$,\footnote{This non-monotonicity property is what makes the structure of civil penalties inherently different from that of criminal penalties -- hence not regardable as a special instance of the latter; see Section \ref{sec:5}.} which arises from a reduction in price informativeness that a significantly larger number of trades placed by the insider inevitably shifts the total order flow towards revelation of the fundamental asset value ($V$), thereby pushing the market maker to adjust the price closer to this true value. This adjustment can then substantially reduce or even wipe out the insider's profit, leading to a smaller penalty (in addition to disgorgement). From another viewpoint, considering that civil penalties are typically less severe than criminal penalties (reserved for serious violations), if only civil penalties are in force, the insider is likely to trade large amounts, knowing that the penalty might decrease as the profits diminish.\footnote{From the proof of Theorem \ref{thm:1} we shall see that in the present scenario with only civil penalties, for every $N\geq1$ and both $v\in\{0,1\}$, $\Psi_{N}(Z;z,v)\to0$ as $|z|\to\infty$, meaning that the insider can technically reduce the penalty to none by increasing his trade size indefinitely.} For this reason, sole reliance on profit-based civil penalties may seem ``overly idealistic'' in that it fails to adequately address and penalize extremely violent insider trading actions, as the absence of significant illicit profit should not absolve the insider of his malicious behavior from the outset. This key observation further motivates a more general consideration combining civil and criminal penalties, to be addressed in Section \ref{sec:5}.


To precisely determine the stealth index $\gamma$, we make the following technical assumption on the (size-modulated) hazard rate, which is also essential for establishing existence and uniqueness.

\begin{assumption}\label{as:1}
For $v\in\{0,1\}$, $\lambda$ is continuously differentiable and convex (not necessarily strictly) on $\mathbb{R}_{v}$, with $\lambda'<0$ on $(-\infty, 0)$ and $\lambda'>0$ on $(0, \infty)$.
\end{assumption}




The requirement that $\lambda$ be convex (apart from its strict increase) is nonrestrictive, which signifies a growing inclination to detect and prosecute the insider trader as his trade size increases. This is, for example, the case for the order flow imbalance-based detection mechanism (DeMarzo, Fishman, and Hagerty (\citeyear{DFH98})): For (\ref{2.1.3}) one can show that whenever $D(z)=K_{D}|z|^{\theta_{D}}$ for some $K_{D}>0$ and $\theta_{D}\geq1$, then $\lambda'$ is guaranteed to be increasing, and $|\lambda'(z)|\geq K'_{\theta}|z|^{\theta_{D}}$ for some $K'_{\theta}>0$. Details are given in Supplemental Appendix \ref{SB}.


The following theorem, serving as the main result of this section, details the existence and uniqueness of the finite-$N$ equilibrium and the limiting equilibrium, along with the desired convergence.

\begin{theorem}\label{thm:1}
Consider the setting of (\ref{3.2.1}), and let Assumption \ref{as:1} hold true. Then, we have the following three assertions. \medskip\\
(i) For every $N\geq1$, there exists an equilibrium $(Z^{\ast}_{N},P^{\ast}_{N})$ in the sense of Definition \ref{def:1}. \\
(ii) There exists a $\gamma$-limiting equilibrium  in the sense of Definition \ref{def:2} if and only if
\begin{equation}\label{3.2.2}
  \gamma=\min\Big\{~\! \beta,~ \frac{1}{2}~\!\Big\}.
\end{equation}
Moreover, when $\gamma<1/2$, the $\gamma$-limiting equilibrium $(\tilde{Z}^{\ast}_{\gamma},\tilde{P}^{\ast}_{\gamma})$ is unique with $\tilde{P}^{\ast}_{\gamma}=p$. \\
(iii) Let $\gamma<1/2$ and $(\tilde{Z}^{\ast}_{\gamma},\tilde{P}^{\ast}_{\gamma}\equiv p)$ be the unique $\gamma$-limiting equilibrium from assertion (ii). Then, there exists a constant $K>0$, depending only on the model parameters but not on $N$, such that, for any equilibrium $(Z^{\ast}_{N},P^{\ast}_{N})$ from assertion (i),
\begin{equation}\label{3.2.3}
  |N^{-\gamma}Z^{\ast}_{N}(v)-\tilde{Z}^{\ast}_{\gamma}(v)| \le K N^{2\gamma-1},\quad v\in\{0,1\},
\end{equation}
and, with $\tilde{P}^{\ast}_{\gamma}=p$,
\begin{align}\label{3.2.4}
  |P^{\ast}_{N}(y)-p|&\le K\big(|y| N^{\gamma-1} + N^{2\gamma-1}\big),\quad y\in \mathbb{R}, \nonumber\\
  |P^{\ast}_{N}(\sqrt{N}W + Z^*_N(V))-p|&\le K\big(|W| N^{\gamma-1\slash 2} + N^{2\gamma-1}\big)\le K(|W|+1) N^{\gamma-\frac{1}{2}},\quad\PP\text{-a.s.}
\end{align}
\end{theorem}

An immediate implication from the first two assertions is that Assumption \ref{as:1} provides sufficient conditions to ensure the existence of a finite-$N$ equilibrium and a limiting equilibrium, the latter also being unique when $\gamma<1/2$ -- as is the interesting case with stealth trading. Technically, analyzing the finite-$N$ equilibrium is significantly more involved than the limiting equilibrium and relies on a key lemma (Lemma \ref{lem:1} in Appendix \ref{B}), which elucidates a fundamental property of the function governing the insider's expected (gross) profit.\footnote{Although this property inherently follows from the normal distribution of total liquidity order flow, it is in fact shared by a broad class of log-concave distributions (including many infinitely divisible distributions); see Saumard and Wellner (\citeyear{SW14}) and Yamazato (\citeyear{Y78}).} Besides, according to (\ref{3.2.2}), the stealth index $\gamma$ for the trading strategy is capped at $1/2$, meaning that the insider deliberately refrains from overly aggressive trading -- at quantities exceeding the total orders -- inevitably exposing himself.


Assertion (iii) from Theorem \ref{thm:1} expounds the convergence of any finite-$N$ equilibrium towards the (unique) $\gamma$-limiting equilibrium, in terms of both the insider's optimal strategy and the market maker's price function, on the assumption that $\gamma<1/2$. 
For the price function, the rate of convergence, identified as the inverse power of $N$, is $\gamma-1/2<0$, proportionate to the stealth index $\gamma$. The intuition behind this is clear: As $\gamma\searrow0$, the insider's trade size, scaled to match that of a single liquidity trader, becomes negligible within the total order flow, leaving the population size $N$ as the dominant factor; conversely, as $\gamma\nearrow1/2$, the trade size effectively matches the entire population of liquidity traders, nullifying the size impact, and convergence is out of question.

The convergence rate ($2\gamma-1$) for the insider's optimal strategies is proportionally related to how the population size obscures investigation and prosecution, as measured by the coefficient $\beta$, and stems directly from the convergence of the price function. An important regulatory insight from this result is that the intensity of investigation (linked to $\beta$) can significantly impact how closely insiders' trade sizes approach their limit associated with diminished price impact. More specifically, when investigations are highly effective ($\beta=0$) -- nearly unaffected by the trading population -- informed traders' strategies tend to converge rapidly for any given stealth level and are driven by the price function's behavior.

Pertaining to Definition \ref{def:2}, Theorem \ref{thm:1} further demonstrates that if $\gamma$ happens to be less than $1/2$, the limiting equilibrium price function becomes constant, decoupled from insider trading information, hence making itself much easier to analyze than that with a finite population, which is even explicitly solvable in specific settings. We give the following illustrative example, which considers a quadratic hazard rate while adhering to the ``treble damages'' provision. The proof is given in Supplemental Appendix \ref{SB}.



\begin{example}\label{ex:1}
Let $p\in(0,1)$ be arbitrary, and let $\lambda(z)=z^{2}$ (satisfying Assumption \ref{as:1}) and $\chi=3$. Suppose that $\gamma=\beta\in[0,1/2)$. Then, the $\gamma$-limiting equilibrium is uniquely determined as
\begin{equation}\label{4.5}
  \tilde{Z}^{\ast}_{\gamma} =\Big(-\sqrt{\tfrac{1}{2}-\mathrm{W_0}\big(\tfrac{\sqrt{e}}{3}\big)},\sqrt{\tfrac{1}{2}-\mathrm{W_0}\big(\tfrac{\sqrt{e}}{3}\big)}~\Big) \approx(-0.350753,0.350753),\quad \tilde{P}^{\ast}_{\gamma}=p,
\end{equation}
where $\mathrm{W}_{0}(\cdot)$ denotes the Lambert $\mathrm{W}$ function (\text{a.k.a.} the product logarithm).
\end{example}

An interesting observation in Example \ref{ex:1} is that regardless of the probability $p$ of revealing the asset value $V=1$, at equilibrium, the insider sticks to trading the same amount in both states ($|\tilde{Z}^{\ast}_{\gamma}(0)|=\tilde{Z}^{\ast}_{\gamma}(1)$). An explanation is that as the civil penalties are tied to the insider's illicit profit, they can effectively remove his incentive to compare and exploit the price differences in the two states, even though the contingent expected profits may still be different.


Following Theorem \ref{thm:1}, the approximation power of the limiting equilibrium towards any finite-population equilibrium can be further illuminated with the concept of $\epsilon$-equilibria, connected to Definition \ref{def:2}, which we show in Definition \ref{def:3}.\footnote{Generally speaking, the convergence of non-unique equilibria in stronger senses (e.g., with respect to the Hausdorff distance) cannot be established without imposing additional technical conditions. A thorough exploration of this convergence issue in the insider trading context is left for further research.} 

\begin{definition}\label{def:3}
For any fixed $N\geq1$ and $\epsilon\geq0$, an \underline{$\epsilon$-equilibrium} is a couple $(Z^{\ast,\epsilon}_{N},P^{\ast,\epsilon}_{N})$ of trading strategy and price function such that, based on (\ref{2.3.2}), (\ref{PNgamma}), and (\ref{3.2}),
\begin{equation}\label{3.1.7}
  N^{-\gamma} J_{N}(P^{\ast,\epsilon}_{N};Z^{\ast,\epsilon}_{N}(v),v) \ge \sup_{z\in\mathbb{R}_{v}} N^{-\gamma}J_{N}(P^{\ast,\epsilon}_{N};z,v) - \epsilon,\quad v\in\{0,1\},
\end{equation}
and
\begin{equation}\label{3.1.8}
  |P^{\ast,\epsilon}_{N}(N^{\max\{\gamma, \frac{1}{2}\}}\tilde y)-P_{N}(Z^{\ast,\epsilon}_{N}; N^{\max\{\gamma, \frac{1}{2}\}}\tilde y)|\leq\epsilon (1+  |\tilde y|),\quad \tilde y\in\mathbb{R}.
\end{equation}
\end{definition}



We have the next important result about the approximation of any finite-population equilibrium via an arbitrary limiting equilibrium, with no uniqueness required.

\begin{theorem}\label{thm:2}
Consider the setting of Theorem \ref{thm:1} assertion (iii), with $\gamma<1/2$ and $(\tilde{Z}^{\ast}_{\gamma}, p)$ being the unique $\gamma$-limiting equilibrium. Then, for every $N\geq1$, $(N^{\gamma}\tilde{Z}^{\ast}_{\gamma},p)$ is an $\epsilon_{N}$-equilibrium in the sense of Definition \ref{def:3}, where for some constant $K$ depending only on the model parameters,
\begin{equation}\label{3.2.5}
  \epsilon_{N} = K N^{\gamma-\frac{1}{2}}.
\end{equation}
\end{theorem}


Theorem \ref{thm:2} asserts that any limiting equilibrium defined under Definition \ref{def:2} automatically qualifies as an $\epsilon_{N}$-equilibrium within the actual, finitely populated market, for which the magnitude of $\epsilon_{N}$, as a function of the population size $N$, carries the same economic implications in terms of the stealth index $\gamma$ as the equilibrium strategies. The statement highlights the central idea that the limiting equilibrium is a practical and robust approximation across a wide range of detection schemes, whenever the population of liquidity traders is justifiably large enough.

\medskip

\section{Empirical perspectives}\label{sec:4}

In this section, we address the issue of estimating the population size of liquidity traders, $N$, and the stealth index $\gamma$, using available data on insider trading cases. The main idea of this empirical analysis, as mentioned in Section \ref{sec:1}, is to verify that the population of liquidity traders is, as expected, considerably large for any generic risky asset and -- more importantly -- that insiders favor trading with medium intensity (compared to all liquidity traders present in the same trading episodes) in the presence of legal risk by adopting a moderate stealth level, therefore confirming the deterrent effect of legal risk on insiders' trade size choices, associated with diminished price informativeness (as implied from the convergence properties in Section \ref{sec:3}); see, again, Barclay and Warner (\citeyear{BW93}), Chakravarty (\citeyear{C01}), and Kacperczyk and Pagnotta (\citeyear{KP24}). We conduct the analysis with two calibration experiments, which rely on different calibration conditions and yet confirm the same phenomena of interest.

\smallskip

\subsection{Calibration experiment I}\label{sec:4.1}

Our first calibration experiment makes use of insider trading volume data collected and analyzed in one of the earliest financial studies of its kind, conducted by Meulbroek (\citeyear{M92}). Specifically, the data set consists of a list of 320 defendants formally charged with insider trading by the SEC in civil or administrative cases from 1980 to 1989, with the (daily) trading volumes for these defendants sourced from (both public and non-public) SEC documents, along with the target firms' total trading volumes on the days of insider trading (sourced from Iterative Data Services' Investment Statistical Listing Tapes and from S\&P's Daily Stock Price Record).\footnote{A more detailed breakdown of the data set, including a yearly analysis of reported insider trading volumes, can be found in Meulbroek (\citeyear{M92}) \text{Tab.} I.}

Table \ref{tab:1} summarizes useful statistics for this data set, extracted from Meulbreuk (\citeyear{M92}) \text{Sect.} III.E. It is worth mentioning that the data set only covers detected insider trading violations (as reported to the SEC), and the volumes are precisely those traded by the defendants (who have been detected and prosecuted), while much insider trading remains undetected over the observation period. On average, the (detected) insider trading volume constitutes about 8\% of the total volume, both in terms of shares traded and dollar value, and so we shall utilize the share volume statistics exclusively.

\begin{table}[H]\small
  \centering
  \captionsetup{justification=centering}
  \caption{Statistics for (daily) insider trading volume data (1980--1989) \\ {\small (Source: Meulbroek (\citeyear{M92}))}}
  \label{tab:1}
  \begin{tabular}{c|c}
    \hline
    Average insider share volume & 9,819 \\ \hline
    Average insider dollar volume & \$300,023 \\ \hline
    Average total share volume & 113,909 \\ \hline
    Standard error for average total share volume & 10,246 \\ \hline
    Average total dollar volume & \$4,121,533 \\ \hline
    Standard error for average total dollar volume & \$594,327 \\ \hline
    Median insider-to-total volume ratio & 11.3\% \\
    \hline
  \end{tabular}
\end{table}

Under the equilibrium model framework, we adopt a \textsl{method-of-moment}-type estimation approach, which starts by deriving the theoretical counterparts of the statistics in Table \ref{tab:1} as functions of $N$ and $\gamma$, and then setting them to be equal to the sample statistics, respectively, on the assumption that the underlying economy is in equilibrium. It is hence understood that the (single) insider trader in our model acts as a \textsl{representative agent} of insider trading during any insider trading episode, with $N$ tracking the relative size of liquidity (non-inside or normal) traders compared to insiders. Also, as only defendants formally charged with insider trading are included in the data set, the statistics must be interpreted as conditional on prosecution -- the reported average insider volumes exclude undetected or unprosecuted cases and unequivocally underestimate the actual insider trading volumes.

The theoretical counterparts of the above statistics are tractable to compute within the model by appealing to the approximate normality in the effect of large $N$, as discussed below; detailed derivations are in Supplemental Appendix \ref{SB}. Recall that $B_{N}$ is the (independent) Bernoulli random variable representing the prosecution mechanism (see (\ref{2.1.1})). Then, given the insider's strategy $Z$, the estimated prosecution probability is
\begin{equation}\label{4.2.1}
  \PP\big\{B_N(Z(V))=1\big\}= \sum^{1}_{v=0} p(v)(1-e^{-\lambda_{N}(Z(v))}),\quad\mbox{with}~   p(v) := \begin{cases}1-p, &~\text{if }v=0,\\  p, &~\text{if }v=1.\end{cases}
\end{equation}
The average insider trading volume (in shares) corresponds to the conditional expectation of the (representative) insider's 
trade size upon successful prosecution, namely
\begin{equation}\label{4.1.1}
  \E\Big[|Z(V)|\big|B_{N}(Z(V))=1\Big]=\frac{\sum^{1}_{v=0} p(v)|Z(v)|\big(1-e^{-\lambda_{N}(Z(v))}\big)} {\sum^{1}_{v=0} p(v)\big(1-e^{-\lambda_{N}(Z(v))}\big)}.
\end{equation}

To construct a meaningful match for the average total volume (in shares), let us recall that according to the model framework (Section \ref{sec:2}), all the liquidity traders place independent orders, each with zero mean and variance $\sigma^{2}$. Assuming that such orders are successfully filled, a reasonable proxy for the total liquidity volume is then the absolute sum of these independent random variables, and by the central limit theorem again, this sum can be approximated by a normal random variable $X_{N}\overset{\rm d.}{=}\text{Normal}(N\mu,N(\sigma^{2}-\mu^{2}))$ for some parameter $\mu>0$ (again, under the large-$N$ assumption).\footnote{As each liquidity trader places an order $\xi_{i}$, $i=1,\dots,N$, \text{i.i.d.} with zero mean and variance $\sigma^{2}$, the total liquidity order flow is $\sum^{N}_{i=1}\xi_{i}\approx\sqrt{N}W$ (assuming that $N$ is large), while the total liquidity volume is approximated as $\sum^{N}_{i=1}|\xi_{i}|=X_{N}$. Also, with $\E[|\xi_{1}|]=\mu$, we have $\Var(|\xi_{1}|)=\E[\xi^{2}_{1}]-\E[|\xi_{1}|]^{2}=\sigma^{2}-\mu^{2}$.} This coefficient, which depends on the exact trade distribution of an average liquidity trader, is yet to be determined, while for now we take it as given. Hence, the average total volume (in shares) should be matched to
\begin{equation}\label{4.1.2}
  \E\Big[X_{N}+|Z(V)|\big|B_N(Z(V))=1\Big]=N\mu+\frac{\sum^{1}_{v=0} p(v)|Z(v)|\big(1-e^{-\lambda_{N}(Z(v))}\big)} {\sum^{1}_{v=0} p(v)\big(1-e^{-\lambda_{N}(Z(v))}\big)}.
\end{equation}


For the insider-to-total volume ratio, it is more convenient to consider its reciprocal (``total-to-insider volume ratio''), whose conditional tail distribution function on prosecution is given by
\begin{equation}\label{4.1.3}
  \PP\bigg\{\frac{X_N+|Z(V)|}{|Z(V)|}>x\Big|B_N(Z(V))=1\bigg\}=\frac{\sum^{1}_{v=0}p(v)\big(1-e^{-\lambda_{N}(Z(v))}\big) \erfc\frac{|Z(v)|(x-1)-N\mu}{\sqrt{2N\sigma^{2}}}}{2\sum^{1}_{v=0} p(v)\big(1-e^{-\lambda_{N}(Z(v))}\big)},\;\; x\geq1.
\end{equation}

Since the data set focuses on civil cases with pecuniary charges only, we follow the setting of Example \ref{ex:1} to design our calibration experiment (with $\gamma=\beta$). Assuming market efficiency, we fix the probability of revealing the upstate ($V=1$) at $p=1/2$. The consideration of a quadratic (size-modulated) hazard rate $\lambda(z)=Kz^{2}$, $z\in\mathbb{R}$, for some $K>0$, agrees with the tail behaviors of (\ref{2.1.3}) for abnormal order flow imbalance-based detection and can be seen as a reasonable approximation of the latter, and we have further $K=1/(2\sigma^{2})$; see Supplemental Appendix \ref{SB} for details. For civil penalties, we emphasize the ``treble damages'' provision again by testing three values of the penalty multiplier: $\chi\in\{1,2,3\}$ (or $\chi_{0}\in\{0,1,2\}$).

If $N$ happens to be large, then according to assertion (iii) in Theorem \ref{thm:1} and Theorem \ref{thm:2}, we can take the re-scaled limiting equilibrium $(N^{\gamma}\tilde{Z}^{\ast}_{\gamma},\tilde{P}^{\ast}_{\gamma})$ as a reasonable approximation for the actual (finite-$N$) equilibrium $(Z^{\ast}_{N},P^{\ast}_{N})$. This implicit assumption can be verified posteriorly after the calibration. In the limiting case, with the above specified parameters, by following Example \ref{ex:1} we obtain the following insider trading strategy at equilibrium:
\begin{equation}\label{4.1.4}
  (\tilde{Z}^{\ast}_{\gamma}(0),\tilde{Z}^{\ast}_{\gamma}(1))=(-\sigma\times\mathfrak{a},\sigma\times\mathfrak{a}),\quad \text{with }\mathfrak{a}=\sqrt{1-2\mathrm{W}_0\big(\tfrac{\sqrt{e}\chi_{0}}{2\chi}\big)},
\end{equation}
which is proportional to $\sigma$, i.e., the trade size of an average liquidity trader, with a constant limiting price function $\tilde{P}^{\ast}_{\gamma}=1/2$.

Next, we make the substitution $Z=N^{\gamma}\tilde{Z}^{\ast}_{\gamma}$ from (\ref{4.1.4}) in (\ref{4.1.1}), (\ref{4.1.2}), and (\ref{4.1.3}), and then establish three calibration conditions by equating (\ref{4.1.1}) and (\ref{4.1.2}) to the numbers 9819 and 113909, respectively, and setting (\ref{4.1.3}) to be equal to $1/2$ for 
$x=1/11.3\%$. However, it turns out that under (\ref{4.1.4}), these conditions are not able to determine the values of $N$ and $\sigma$ simultaneously, because the latter is inherently a scaling factor for the total order flow. Instead, since $N$ and $\sigma$ clearly exhibit an inverse relationship, it is always possible to obtain a lower bound estimate for $N$ by imposing an upper bound on $\sigma$. In accordance with the standard 100 round lot in the \text{U.S.} as well as the phenomenon of trade-size clustering (Alexander and Peterson (\citeyear{AP07})), we set $\sigma=1000$, which is a conservative enough estimate in this context given the (daily) average price around \$36.18 traded during the observation period -- amounting to \$36,180 worth of trades placed per trader each day. Using smaller values of $\sigma$ will necessarily lead to $N$ increasing, to which the value of $\gamma$ (as a power coefficient) is also resistant.

In addition, while the additional parameter $\mu>0$ may be determined in several ways, we prefer to take on a direct approach leveraging the standard deviation of the total volume in the same data set, estimated to be 248,452; again, see Supplemental Appendix \ref{SB} for details. By equating the standard deviation of $X_{N}$, or $\sqrt{N(\sigma^{2}-\mu^{2})}$, to this value and using the calibration condition from (\ref{4.1.2}), we arrive at an estimate $\hat\mu\approx1.68625$.\footnote{The estimate is numerically stable -- using the more complex conditions (\ref{4.1.1}) or (\ref{4.1.3}) instead of (\ref{4.1.2}) produces similar values in the range $[1,2]$, and so for succinctness we stick with this estimate.} This estimate is in keeping with the significant asymmetric and leptokurtic feature of trading volumes, which is a well-documented phenomenon in market microstructure; see, e.g., Mike and Farmer (\citeyear{MF08}) and \c{C}etin and Waelbroeck (\citeyear{CW24}).\footnote{This phenomenon indicates that while many trading episodes carry small or no trades, a few episodes contain very large trades, which also closely reflects the aforementioned trade size clustering (e.g., Chang, Pinegar, and Schachter (\citeyear{CPS97}), Alexander and Peterson (\citeyear{AC07}), and Fei and Xia (\citeyear{FX24})). Notably, the heavy-tailed behavior does not conflict with the approximate-normality treatment for the present analysis, as the stated statistics are not tail-dependent and thus robust; indeed, replacing normality with a heavy-tailed distribution for $X_{N}$ yields virtually no change to the results to follow.} In particular, at the individual level, the average trading volume is exceeded substantially by its standard deviation ($\sigma$) in value, the relative smallness persisting with smaller values of $\sigma$. Then, by setting $\mu=1.68625$ and $\sigma=1000$, we continue to solve exactly two calibration conditions at a time and compare the parameter estimates $(\hat{N},\hat{\gamma})$. Results are reported in Table \ref{tab:2}.

\begin{table}[H]\small
  \centering
  \captionsetup{justification=centering}
  \caption{Results on calibration experiment I}
  \label{tab:2}
  \begin{tabular}{c|c|c|c|c}
    \hline
    \multicolumn{2}{c|}{Equations used} & (\ref{4.1.1}) and (\ref{4.1.2}) & (\ref{4.1.1}) and (\ref{4.1.3}) & (\ref{4.1.2}) and (\ref{4.1.3}) \\ \hline
    \multirow{3}{*}{$(\hat{N},\hat{\gamma})$}
    & $\chi=1$ & $(61729, 0.207091)$ & $(45708, 0.21289)$ & $(59918, 0.23226)$ \\
    & $\chi=2$ & $(61729, 0.249565)$ & $(45708, 0.256553)$ & $(59918, 0.274849)$ \\
    & $\chi=3$ & $(61729, 0.270651)$ & $(45708, 0.27823)$ & $(59918, 0.295992)$ \\
    \hline
  \end{tabular}
\end{table}

Table \ref{tab:2} shows that estimates for the stealth index $\gamma$ are quite robust when using different calibration conditions, standing around 0.25, strictly lying between 0 and 0.5 as expected. These values represent a moderate stealth level and suggests that over the observation period, an insider tends to trade at significantly smaller scales than the entire population of liquidity traders while transcending an average liquidity trader. In the case $\chi=2$, for instance, by noting that $15<\hat{N}^{\hat{\gamma}}<21$, while $213<\sqrt{\hat{N}}<249$, the insider trade size is roughly 17 times larger than a liquidity trade size, albeit 13 times smaller than the size of all trades combined. This observation is in line with a key conclusion of Meulbroek (\citeyear{M92}) that in spite of abnormal volumes on insider trading days, the insider trading volume only makes up a small portion of the total. The (lower bound) estimates for the population size $N$ exhibit slightly larger variations across conditions but stand at a scale of $10^{4}\gg30$ (see Footnote \ref{CLT}),  and the equilibrium model estimates that, on average, one insider trade is present for (at least) approximately every 50,000 non-insider trades over the same trading episodes.

One seemingly counterintuitive observation is that the estimate for $\gamma$ exhibits an increasing trend with respect to the penalty multiplier $\chi$. While one might well expect heightened penalties to deter insider activity and thus reduce $\gamma$, this result is actually consistent with the calibration approach. As the calibration works by fixing the sample statistics and deriving implied model parameters, achieving the same insider trade size under stricter penalties necessitates more aggressive insider trading behavior. As such, Table \ref{tab:2} should not be interpreted as a guide for designing optimal penalty multipliers.

\begin{figure}[H]
  \centering
  \includegraphics[width=3in]{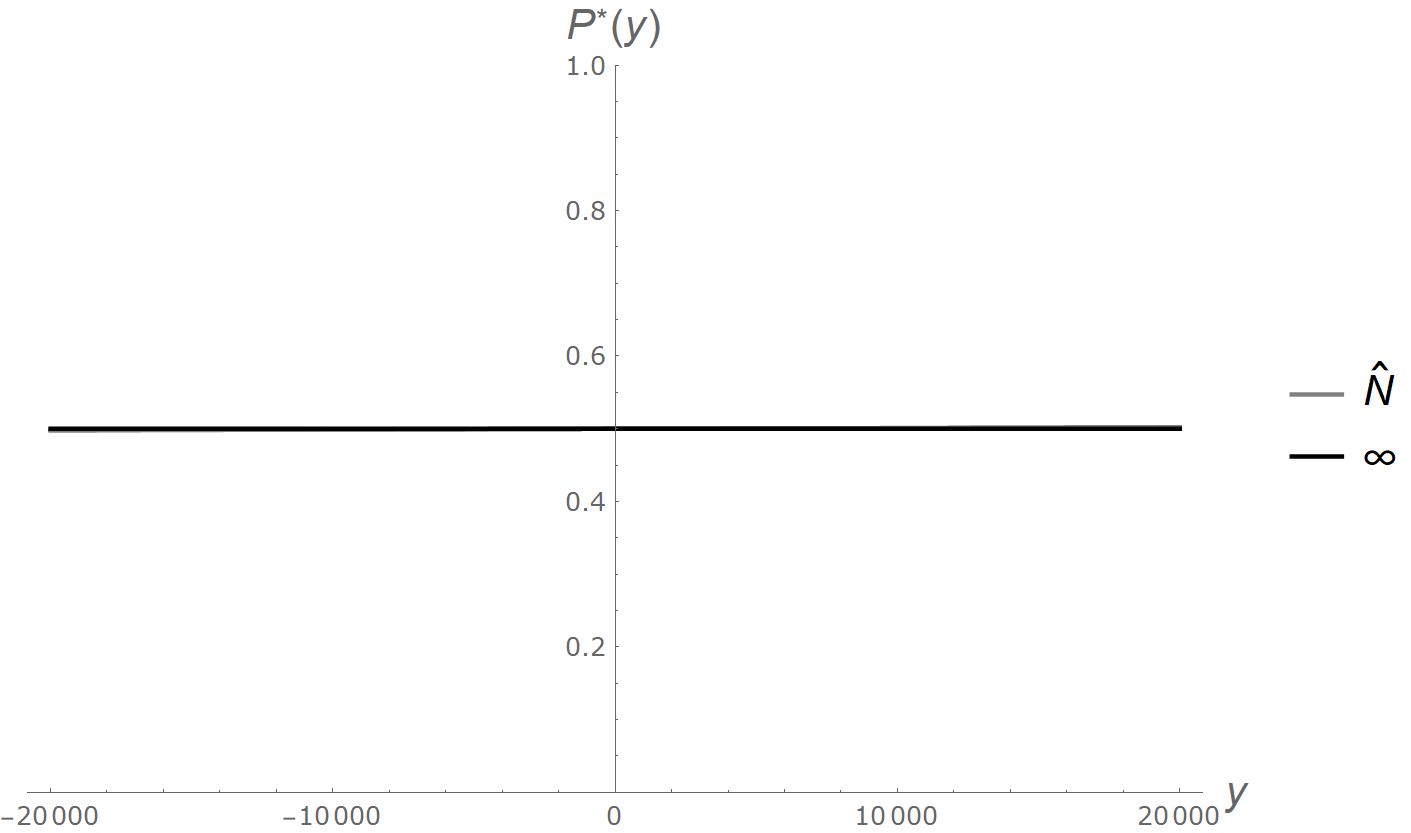}\\
  \caption{Comparison of equilibrium price function in calibration experiment I ($\chi=3$)}
  \label{fig:1}
\end{figure}

We proceed to solve the finite-population equilibrium (existent) with the values of $\hat{N}$ in Table \ref{tab:2} to check its closeness to the limiting equilibrium employed for the calibration, for which purpose we employ a \textsl{fixed-point algorithm} developed according to the equilibrium conditions in (\ref{B.10}) and (\ref{A.34}) in Appendix \ref{B}. For a concise presentation, we concentrate on the case $\chi=3$, with the resulting equilibrium objects presented in Table \ref{tab:3} and Figure \ref{fig:1} below -- results in the cases $\chi=1$ and $\chi=2$ are substantially no different. Clearly, the finite-$\hat{N}$ equilibrium objects are all very close to their limiting counterparts, verifying the validity of the approximation $(Z^{\ast}_{N},P^{\ast}_{N})\approx(N^{\gamma}\tilde{Z}^{\ast}_{\gamma},\tilde{P}^{\ast}_{\gamma})$ in the outset. From Theorem \ref{thm:1}, the rate of equilibrium convergence is proportionate to the stealth index $\gamma$.

\begin{table}[H]\small
  \centering
  \captionsetup{justification=centering}
  \caption{Comparison of equilibrium strategies in calibration experiment I ($\chi=3$)}
  \label{tab:3}
  \begin{tabular}{c|c|c|c}
    \hline
    Equations used & (\ref{4.1.1}) and (\ref{4.1.2}) & (\ref{4.1.1}) and (\ref{4.1.3}) & (\ref{4.1.2}) and (\ref{4.1.3}) \\ \hline
    $(Z^{\ast}_{\hat{N}}(0),Z^{\ast}_{\hat{N}}(1))$ & $(-9813,9813)$ & $(-9811,9811)$ & $(-12862,12862)$ \\
    $(\hat{N}^{\hat{\gamma}}\tilde{Z}^{\ast}_{\hat{\gamma}}(0),\hat{N}^{\hat{\gamma}}\tilde{Z}^{\ast}_{\hat{\gamma}}(1))$ & $(-9819,9819)$ & $(-9819,9819)$ & $(-12872,12872)$ \\
    \hline
  \end{tabular}
\end{table}

\smallskip

\subsection{Calibration experiment II}\label{sec:4.2}

In the second calibration experiment, we leverage available data from two recent studies, Kacperczyk and Pagnotta (\citeyear{KP24}) and Patel and Put\c{n}in\v{s} (\citeyear{PP21}), on legal risk in insider trading. First, using a data set of 530 insider trading cases prosecuted by the SEC between 1995 and 2018, Kacperczyk and Pagnotta (\citeyear{KP24}) constructed two key empirical proxies: the total dollar volume traded by the insider over a trading episode (in units of days) (denoted as $\widehat{\text{B}}\text{et}$ therein) and the corresponding (daily) insider-to-total volume ratio ($\widehat{\text{B}}\text{et}\text{Norm}$), the latter serving to provide a clearer indication of the market impact of insider trades. For these calculations, the total dollar volume represents the average daily dollar volume traded for the corresponding assets over the prior calendar year. Second, Patel and Put\c{n}in\v{s} (\citeyear{PP21}) constructed a data set spanning the period from 1996 to 2016 and utilized a two-stage detection-controlled estimation method to derive contemporaneous estimates for the probability of detection and prosecution of insider trading across different corporate events. Thus, these two data sets cover observation periods that are considerably close in range.\footnote{For more details regarding the two data sets we refer to Kacperczyk and Pagnotta (\citeyear{KP24}) \text{Tab.} I and Patel and Put\c{n}in\v{s} (\citeyear{PP21}) \text{Sect.} 2, respectively.}

We summarize some useful statistics for the calibration in Table \ref{tab:4}.\footnote{Although Kacperczyk and Pagnotta (\citeyear{KP24}) primarily focuses on dollar volume, for adequate comparison (with calibration experiment I) we still use share volume statistics.} The two statistics are gleaned from Kacperczyk and Pagnotta (\citeyear{KP24}) \text{Tab.} III \& \text{Fig.} IA.5.

The idea of the calibration is in principle identical to that of experiment I. In the setting of Example \ref{ex:1}, we set $p=1/2$ and consider three penalty multiplier values $\chi\in\{1,2,3\}$ for experimentation, along with a quadratic hazard rate $\lambda(z)=z^{2}/(2\sigma^{2})$.\footnote{The equilibrium simulation in Kacperczyk and Pagnotta (\citeyear{KP24}) is conducted in a very similar setting, with a binary asset value $V$ with range equal to 1 and $p=1/2$. The (size-modulated) hazard rate $\lambda_{1}$ from (\ref{2.1.3}) can be reasonably approximated by the quadratic hazard rate.} In this case, the limiting equilibrium strategy $\tilde{Z}^{\ast}_{\gamma}$ is as given in (\ref{4.1.4}), with $\tilde{P}^{\ast}_{\gamma}=1/2$.

\begin{table}[H]\small
  \centering
  \captionsetup{justification=centering}
  \caption{Statistics for insider trading volume data (1995--2018) \\ {\small (Source: Kacperczyk and Pagnotta (\citeyear{KP24}))}}
  \label{tab:4}
  \begin{tabular}{c|c}
    \hline
    Median insider volume & 4,900 \\ \hline
    Median insider-to-total volume ratio & 2.6\% \\
    \hline
  \end{tabular}
\end{table}

On the assumption that $N\gg1$, according to Table \ref{tab:4}, we can match (\ref{4.1.1}) to the number 4900 and equate (\ref{4.1.3}) to $1/2$ for $x=1/2.6\%$ in order to form two calibration conditions. Indeed, in Table \ref{tab:4}, we use the median insider volume instead of the average because the insider volume distribution over these observation periods is severely right-skewed and fat-tailed; the theoretical counterpart can still be represented by the conditional expectation (\ref{4.1.1}), as the sample median converges to this value asymptotically given that $V\overset{\rm d.}{=}\text{Bernoulli}(1/2)$. For the trade size of an average liquidity trader, we stick to the previous values, namely $\mu=1.68625$ and $\sigma=1000$, in order to obtain a comparable conservative lower bound estimate for $N$. As before, the insider volume data over the observation periods should be interpreted as being conditional on prosecution. The theoretical counterpart of the prosecution probability is given by \eqref{4.2.1}.
However, upon applying $Z=N^{\gamma}\tilde{Z}^{\ast}_{\gamma}$ (with $\gamma=\beta$), 
the expression in (\ref{4.2.1}) is not a function of the variables $N$ and $\gamma$ but can be used for later verifying the functional form of the hazard rate. Table \ref{tab:5} summarizes the calibration results in this experiment.

\begin{table}[H]\small
  \centering
  \captionsetup{justification=centering}
  \caption{Results on calibration experiment II}
  \label{tab:5}
  \begin{tabular}{c|c|c}
    \hline
    \multirow{3}{*}{$(\hat{N},\hat{\gamma})$}
    & $\chi=1$ & $(108858,0.137029)$ \\
    & $\chi=2$ & $(108858,0.177425)$ \\
    & $\chi=3$ & $(108858,0.19748)$ \\
    \hline
  \end{tabular}
\end{table}

\clearpage 

\begin{table}[H]\small
  \centering
  \captionsetup{justification=centering}
  \caption{Comparison of equilibrium strategies in calibration experiment II ($\chi=3$)}
  \label{tab:6}
  \begin{tabular}{c|c|c}
    \hline
    & Strategy & implied prosecution probability \\ \hline
    $(Z^{\ast}_{\hat{N}}(0),Z^{\ast}_{\hat{N}}(1))$ & $(-4900, 4900)$ & 11.572\% \\
    $(\hat{N}^{\hat{\gamma}}\tilde{Z}^{\ast}_{\hat{\gamma}}(0),\hat{N}^{\hat{\gamma}}\tilde{Z}^{\ast}_{\hat{\gamma}}(1))$ & $(-4900,4900)$ & 11.576\% \\
    \hline
  \end{tabular}
\end{table}

\begin{figure}[H]
  \centering
  \includegraphics[width=3in]{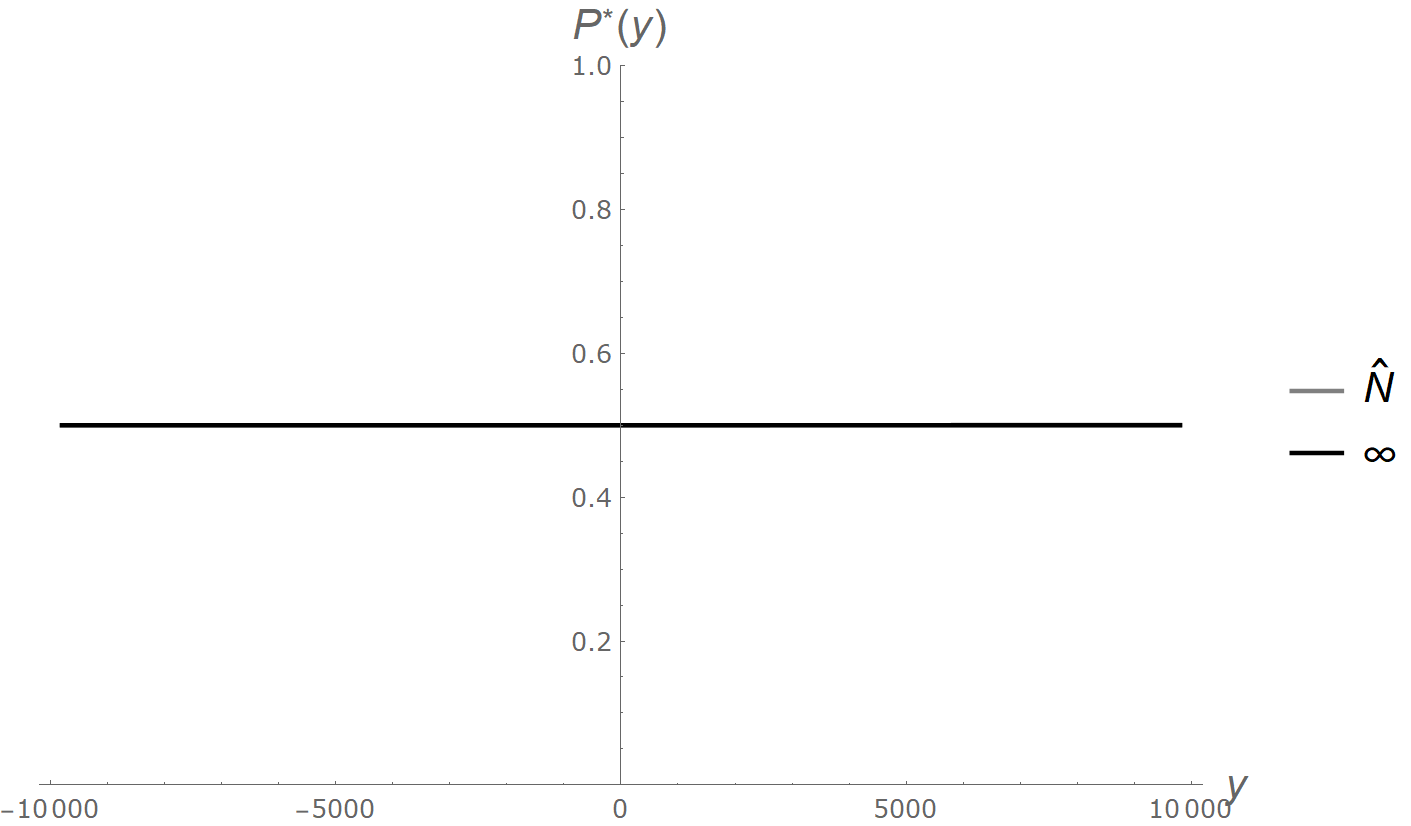}\\
  \caption{Comparison of equilibrium price function in calibration experiment II ($\chi=3$)}
  \label{fig:2}
\end{figure}

We see that based on statistics from this non-overlapping new observation period (1995--2018), the stealth index $\gamma$ is again estimated to strictly lie within the interval $(0,0.5)$, as expected, though the estimates suggest a somewhat higher stealth level compared to the estimates from the first experiment covering a pre-1990 period. This could indicate gradually increased prudence among insiders after 2000, likely in response to enhanced regulatory measures (such as increased regulatory budgets and the introduction of the SEC Whistleblower Program), which also agrees with the increasing trends of the prosecution probability as discussed in Patel and Put\c{n}in\v{s} (\citeyear{PP21}) \text{Sect.} 4.3. In addition, the implied lower bound of the population size, $\hat{N}$, turns out to be larger than those from the first experiment (Table \ref{tab:2}), standing at the $10^{5}$ scale.

As before, we compute the finite-population equilibrium (existent) using the estimates $(\hat{N},\hat{\gamma})$ from Table \ref{tab:5} to justify the validity of employing the limiting equilibrium; again, we only illustrate the case $\chi=3$. Table \ref{tab:6} and Figure \ref{fig:2} verify the closeness between the equilibrium objects, respectively. In the last column, the model-implied prosecution probabilities are computed by plugging the equilibrium strategy values into (\ref{4.2.1}), and notably, are at a comparable level to the average detection rate of 15\% reported by Patel and Put\c{n}in\v{s} (\citeyear{PP21}) \text{Sect.} 2, a key conclusion of their study (with data range 1996--2016).\footnote{The implied values both fall within the 95\% confidence interval for the average detection rate for insider trading ahead of earnings announcements; see Patel and Put\c{n}in\v{s} (\citeyear{PP21}) \text{Sect.} 4.2.} This observation also empirically validates the quadratic form of the hazard rate.

\medskip

\section{Further discussions: Criminal and mixed penalties}\label{sec:5}

In this section, we discuss the general scenario where criminal charges are within consideration. Suppose that on successful prosecution, besides losing all of his illicit profit, the insider trader faces a mixture of both civil and criminal penalties. While civil penalties are calculated with penalty multipliers applied to the insider's illicit profit as before, criminal penalties are determined based on his realized trading strategy ($Z$). Therefore, we are looking at the general form (\ref{2.2.1}) of the penalty function.

For any population size $N\geq1$, the insider's objective function is given by (\ref{2.3.4}) in equilibrium, which is detailed as, for $v\in\{0,1\}$ and $z\in\mathbb{R}_{v}$,
\begin{equation}\label{3.3.1}
  \acute J_{N}(Z;z,v)=e^{-\lambda_N(z)} Q_N(Z; z, v)-(1-e^{-\lambda_N(z)})\big(C_0(z)+\chi_{0} Q_N(Z; z, v)\big).
\end{equation}
Recall that $\chi_{0}=\chi-1$ is the adjusted penalty multiplier, and $Q_N$ denotes the insider's expected profit in (\ref{2.3.3}). The last term in (\ref{3.3.1}) shows the total penalty besides disgorgement, reflecting a balanced mechanism that accounts for both the insider's malicious intent in engaging in illicit trading and his realized profit, without either directly offsetting the other. We emphasize that under the general form (\ref{2.2.1}), any leniency or discretion exercised by the regulator can be incorporated through the functional form of the criminal penalty function ($C_{0}$) or the penalty multiplier ($\chi_{0}$) governing civil penalties, and so the approach by adding both types of penalties is sufficiently general from a practical standpoint as well. 

By setting the penalty multiplier $\chi=1$ in the structure (\ref{2.2.1}), we come to a situation where the insider faces criminal penalties exclusively on top of disgorgement. 
Then, with $\chi_0=0$, (\ref{3.3.1}) becomes
\begin{equation}\label{3.1.1}
  \acute J_{N}(Z;z,v)=e^{-\lambda_N(z)}Q_N(Z; z, v)-(1-e^{-\lambda_N(z)})C_0(z), \quad z\in\mathbb{R}_{v},\;v\in\{0,1\}.
\end{equation}
A comparison between (\ref{3.1.1}) and (\ref{3.2.1}) shows that, different from the predominant scenario involving civil penalties only, the penalty function $C_{0}$ has no explicit dependence on the asset value $v\in\{0,1\}$, focusing instead on the insider's trading behavior for a specific asset value -- and not that in the alternative situation, e.g., through the price function $P_{N}$ (or the general form of $\Psi_{N}$). This is reasonable, as in criminal cases, insider trading behavior is assessed based on the realized asset value (to which the insider has prior access), rather than evaluated in hypothetical situations (Picardo (\citeyear{P22})). The insider is also aware that penalties necessarily increase with his actual trading activity, governed by the monotonicity of the penalty function $C_{0}$.

As noted earlier, the asymptotic growth of $C_{0}$ should be decisive of the asymptotic behaviors of the trading strategies. This leads to the following conditions in addition to Assumption \ref{as:1}.

\begin{assumption}\label{as:1e}
The following three conditions hold. \medskip\\
(i) For $v\in\{0,1\}$, both $\lambda$ and $C_{0}$ are continuously differentiable and convex (not necessarily strictly) on $\mathbb{R}_{v}$, with $\lambda'<0$ and $C'_0\le0$ on $(-\infty, 0)$, while $\lambda'>0$ and $C'_0\ge0$ on $(0, \infty)$. \\
(ii) There exist constants $\theta\geq1$, $\theta'>0$, and $K_{\theta}>0$,  such that\footnote{Given the convexity of $\lambda$, an application of the monotone density theorem (see, e.g., Bingham, Goldie, and Teugels (\citeyear{BGT89})) ensures that $\lambda'(z) = K_\theta \theta z|z|^{\theta-2}$ if $\lambda(z)\sim K_{\theta}|z|^{\theta}$, both as $|z|\searrow0$; however, there are no direct implications for the asymptotic behavior of the remainder ($\lambda'(z)-K_\theta \theta z|z|^{\theta-2}$), hence the imposition of (\ref{lambda}).}
\begin{equation}\label{lambda}
  \lambda(z)=K_{\theta}|z|^{\theta}(1+o(1)),\quad \lambda'(z) = K_\theta \theta z|z|^{\theta-2} + O( |z|^{\theta(1+\theta')-1}),\quad\mbox{as}~ |z|\searrow0.
\end{equation}
(iii) There exist constants $\alpha\geq1$ and $K_{\alpha}>0$ and, if $\alpha>1$, an additional constant $\alpha'>0$, such that
\begin{equation}\label{C0}
  C_{0}(z)=K_{\alpha}|z|^{\alpha}(1+o(1)),\quad C_0'(z) = K_\alpha \alpha z|z|^{\alpha-2} +O( |z|^{\alpha-1-\alpha'}),\quad \mbox{as}~ |z|\to\infty.
\end{equation}
\end{assumption}

Since the choice of the penalty function is entirely up to the regulator, it is customizable and can be adjusted flexibly on a case-by-case basis.
By contrast, the small argument behavior assumption regarding the hazard rate is natural and covers abnormal order flow imbalance-based detection (DeMarzo, Fishman, and Hagerty (\citeyear{DFH98})). For example, in (\ref{2.1.3}), since the sum of the two complementary error functions is asymptotically equivalent to $z^{2}$ as $|z|\searrow0$, we have that $\theta=\min\{\theta_{D},2\}$ if $D(z)\sim K_{D}|z|^{\theta_{D}}$ as $|z|\searrow0$ for $K_{D}>0$ and $\theta_{D}\geq1$; details are in Supplemental Appendix \ref{SB}. The value of $\alpha$ is likely inversely related to $\gamma$, and in fact, the assumption on $\lambda$ is only dominant when $\alpha$ becomes ``sufficiently large,'' as the insider will judiciously reduce the prosecution probability so as to offset the fast-growing (additional) penalty term in (\ref{3.1.1}). Thus, Assumption \ref{as:1e} includes small arguments because the scaled quantity $N^{-\beta}z$ can still be small (as $N\to\infty$) even if $z$ is large.



Under Assumption \ref{as:1e}, we are able to extrapolate the main results in the predominant scenario (Section \ref{sec:3.2} Theorem \ref{thm:1}) to the present general scenario concerning both civil and criminal penalties. In particular, a finite-population equilibrium $(Z^{\ast}_{N},P^{\ast}_{N})$ in the sense of Definition \ref{def:1} continues to exist;
see Supplemental Appendix \ref{SA} for details. Most striking is the existence of a unique $\gamma$-limiting equilibrium $(\tilde{Z}^{\ast}_{\gamma},\tilde{P}^{\ast}_{\gamma})$ in the sense of Definition \ref{def:2}, which is also unique when $\gamma<1/2$; in particular, (\ref{3.2.2}) is to be replaced by the more elaborate condition
\begin{equation}\label{3.1.2}
  \gamma=\min\Big\{\frac{\beta\theta}{\theta+\alpha-1},\frac{1}{2}\Big\}.
\end{equation}
Based on (\ref{3.1.2}), the stealth index still has the upper bound $1/2$, while it further exhibits a clear inverse relationship with $\alpha$. If $\alpha=1$, the insider trader views the criminal penalties as relatively minor and simply chooses $\gamma=\min\{\beta,1/2\}$, based entirely on the scale effect from $\lambda$ (equivalently, the prosecution probability), as in (\ref{3.2.2}); in this case, the small argument behavior of $\lambda$ (namely $\theta$) plays no role. Conversely, as $\alpha$ exceeds 1, indicating severe criminal penalties, the insider begins to place greater weight on the consequences of prosecution and, accordingly, trades less aggressively with a smaller value $\gamma<\beta$.
%
{These results are formally stated in Theorem \ref{thm:3} in Supplemental Appendix \ref{SA}, along with the corresponding convergence rates that are jointly determined by all the coefficients $\beta$, $\theta$, $\alpha$, $\theta'$, and $\alpha'$.}


We give the following illustrative example to highlight the analytical simplicity of the limiting equilibrium.

\begin{example}\label{ex:2}
Let $p=1/3$, $\lambda(z)=C_{0}(z)=|z|$, and $\chi=1$, which satisfy Assumption \ref{as:1e}, with $\theta=\alpha=1$. Suppose that $\beta\in[0,1/2)$, so that $\gamma=\beta<1/2$. Then, with the Lambert $\mathrm{W}$ function $\mathrm{W}_{0}$, the $\gamma$-limiting equilibrium is uniquely determined as
\begin{equation*}
  \tilde{Z}^{\ast}_{\gamma}\equiv(\tilde{Z}^{\ast}_{\gamma}(0),\tilde{Z}^{\ast}_{\gamma}(1)) =\Big(\mathrm{W}_0\big(\tfrac{3e}{4}\big)-1,1-\mathrm{W}_0\big(\tfrac{3e}{5}\big)\Big)\approx(-0.138547,0.23844),\quad \tilde{P}^{\ast}_{\gamma}=\tfrac{1}{3}.
\end{equation*}
\end{example}

In Example \ref{ex:2}, the (size-modulated) hazard rate and the criminal penalty function are both perfectly proportional to the insider's trading quantities, and the stealth index $\gamma$ does not alter the form of the limiting equilibrium (Definition \ref{def:2}), while only affecting the convergence rate of the finite-population equilibrium (Theorem \ref{thm:3} assertion (iii)) -- through the scaled equilibrium strategy $\tilde{Z}^{\ast}_{\gamma}$. With the asset value $V$ more likely revealed to be 0 than 1 ($p<1/2$), at equilibrium the insider also tends to trade more if $V=1$ (with $\tilde{Z}^{\ast}_{\gamma}(1)>|\tilde{Z}^{\ast}_{\gamma}(0)|$) to exploit the price difference, which behavior differs fundamentally from what we have seen in Example \ref{ex:1} involving only civil penalties. From the regulator's viewpoint, this fundamental difference speaks to a crucial benefit of retaining criminal penalties: to prosecute based on the intent of illicit trading -- a factor that can exhibit significant asymmetry across different realizations (see, e.g., \"{O}berg (\citeyear{O14})).

%

The next example shows that if the convexity conditions in Assumption \ref{as:1e} are violated, then the uniqueness of the limiting equilibrium is not guaranteed.

\begin{example}\label{ex:3}
Let $p=1/3$, $\beta=0$, and $\chi=1$. Consider  $\lambda(z)=\log(|z|+1)$, which is concave (violating Assumption \ref{as:1e}),
and the piecewise penalty function
\begin{equation*}
  C_{0}(z)=
  \begin{cases}
    \frac{1}{4z}+\frac{1}{12},&\quad\text{if }z\leq-6, \\
    -\frac{z}{144},&\quad\text{if }-6<z\leq0, \\
    \frac{25z}{144},&\quad\text{if }0<z\leq\frac{6}{5}, \\
    \frac{5}{12}-\frac{1}{4z},&\quad\text{if }z>\frac{6}{5},
  \end{cases}
\end{equation*}
which is strictly increasing and continuously differentiable but again concave in $|z|$ on $\mathbb{R}_{v}$, $v\in\{0,1\}$. 
Then, there is a continuum of $0$-limiting equilibria:
\begin{equation}
\label{SB.5}
  \tilde{Z}^{\ast}_{0}\in(-\infty,-6]\times\big[\tfrac{6}{5},\infty\big),\quad \tilde{P}^{\ast}_{0}=\tfrac{1}{3}.
\end{equation}
\end{example}

Example \ref{ex:3} illustrates an extreme case where the prosecution mechanism is considerably weak and ineffective, with the maximum penalty capped, leading to little-to-zero deterrent effect on insider trading. Consequently, the insider's optimal strategies can involve unboundedly large trade sizes as the population of liquidity traders grows.

\medskip

\section{Conclusions}\label{sec:6}

This paper has introduced a Kyle-type model for insider trading with legal risk featuring a flexible scheme of legal detection and penalty and the presence of a large liquidity trading crowd. 
Besides justifying the normal distribution of liquidity trade sizes, the population size $N$ of liquidity traders has spawned a series of limiting equilibria that provide a profound understanding of the camouflage effect in insider trading, particularly uncovering insiders' choices of trading scales when taking advantage of the surrounding trading masses in avoidance of detection. These trading scales are quantified by the stealth index $\gamma$, which is heavily associated with the population size over concurrent trading episodes.

The equilibrium analysis covering various types of penalty functions has shown that the stealth index implied by insiders' trading behavior is largely dependent on the mechanisms of detection as well as the imposition of penalties from regulators' side (Picardo (\citeyear{P22})). In particular, if only civil penalties are imposed -- based on illicit profit from trade -- then in equilibrium, insiders choose the stealth index solely based on the scale effect from the perceived probability of prosecution (see (\ref{3.2.2})), as long as the resulting trade size does not reach the level of the combined activity of all liquidity traders. On the other hand, when severe criminal penalties are devised (with $\alpha>1$), insiders are prone to adopting an elevated stealth level (see (\ref{3.1.2})) compared to what the underlying detection mechanism implies. This striking disparity highlights the importance of reserving criminal sanctions in insider trading cases, due to their effectiveness in deterring illicit trading driven by a highly strategic intent such as market abuse (\"{O}berg (\citeyear{O14}) and Dalko and Wang (\citeyear{DW16})).

On a technical level, the proven equilibrium convergence properties (Theorem \ref{thm:2} and assertions (iii) in Theorems \ref{thm:1} and \ref{thm:3}) convey a key message that employing a limiting equilibrium significantly enhances the ease of analysis without altering the equilibrium's fundamental implications. The empirical perspectives in Section \ref{sec:4} further demonstrate the ease of equilibrium calibration to insider trading volume data and confirm the prevalence of stealth trading amid sizeable normal trading activity, revealing a moderate stealth level in insider trading (Barclay and Warner (\citeyear{BW93}) and Chakravarty (\citeyear{C01})) accompanied by a significant reduction in price informativeness (Kacperczyk and Pagnotta (\citeyear{KP24})) -- hence conforming to the camouflage effect.

\medskip

\section*{Acknowledgements}

The authors are grateful to Marcin Kacperczyk for kindly sharing supplementary statistics for their data set in Kacperczyk and Pagnotta (\citeyear{KP24}). Jin Ma and Jianfeng Zhang are supported in part by \text{U.S.} NSF grant \#DMS-2510403.

\bigskip

\begin{appendices}

\renewcommand{\theequation}{A.\arabic{equation}}

\section{Proof of Proposition \ref{pro:1}}\label{A}

Based on (\ref{2.3.2}), we can write
\begin{align}\label{A.1}
  & \tilde{P}^\gamma_{N}(\tilde{Z}; \tilde y)=P_{N}(N^\gamma \tilde Z;N^{\max\{\gamma,\frac{1}{2}\}}\tilde y)=\frac{1}{1+q e^{\tilde\kappa_N(\tilde y)}}, \nonumber\\
  &\mbox{with}\quad \tilde\kappa_N(\tilde y):=\frac{(2N^{\max\{\frac{1}{2}-\gamma,0\}}\tilde y - \tilde z_0-\tilde z_1) (\tilde z_0-\tilde z_1)}{2N^{1-2\gamma}\sigma^{2}},\quad \tilde z_v := \tilde Z(v).
\end{align}

First, if $\gamma<1/2$, then the exponent $\tilde\kappa_N(\tilde y)$ above is
\begin{equation}\label{kNorder}
  \tilde\kappa_N(\tilde y)=\frac{(2N^{\frac{1}{2}-\gamma}\tilde y - \tilde z_0-\tilde z_1) (\tilde z_0-\tilde z_1)}{2N^{1-2\gamma}\sigma^{2}} = \frac{\tilde z_0-\tilde z_1}{ 2\sigma^2}\bigg(\frac{2\tilde y}{N^{\frac{1}{2}-\gamma}} -  \frac{ \tilde z_0+\tilde z_1}{N^{1-2\gamma}}\bigg),
\end{equation}
which tends to $0$ as $N\to\infty$. Thus, (\ref{A.1}) becomes $\lim_{N\to\infty} \tilde {P}^\gamma_{N}(\tilde{Z}; \tilde y) = 1/(1+q) = p$.

Second, if $\gamma=1/2$, the terms involving $N$ cancel out, and we have for (\ref{A.1}) that $\tilde {P}^\gamma_{N}(\tilde{Z}; \tilde y)=1/(1+q e^{(2\tilde y - \tilde z_0-\tilde z_1) (\tilde z_0-\tilde z_1)/(2\sigma^{2})})= P_1(\tilde{Z}; \tilde{y})$.

Third, let $\gamma>1/2$. Then $\tilde\kappa_N(\tilde y)=(2\tilde y - \tilde z_0-\tilde z_1)(\tilde z_0-\tilde z_1)/(2N^{1-2\gamma}\sigma^{2})$. If $\tilde z_0=\tilde z_1=0$ or $2\tilde y - \tilde z_0-\tilde z_1=0$,  it is clear that $\tilde\kappa_N(\tilde y)=0$, and thus $\tilde {P}^\gamma_{N}(\tilde{Z}; \tilde y)=p$. Otherwise, we have $\tilde z_0-\tilde z_1<0$, and so if $2\tilde{y}-\tilde z_0-\tilde z_1 >0$, then with the exponent $\tilde\kappa_N(\tilde y)$ tending to $0$, we have $ \tilde {P}^\gamma_{N}(\tilde{Z}; \tilde y)\to 1$, as $N\to\infty$, while if $2\tilde{y}-\tilde z_0-\tilde z_1<0$, the exponent $\tilde\kappa_N(\tilde y)$ goes to $\infty$, yielding that $ \tilde {P}^\gamma_{N}(\tilde{Z}; \tilde y)\to0$.

Summarizing the three cases, we have verified the limit expressions in (\ref{3.1}).
\qed

\renewcommand{\theequation}{B.\arabic{equation}}

\section{Proof of Theorem \ref{thm:1}}\label{B}

We start with the following lemma that governs the log-concavity property of the insider's expected profit function.

\begin{lemma}\label{lem:1}
For any given $N\geq1$,  $Z=(Z(0),Z(1))\in\mathbb{R}_{0}\times\mathbb{R}_{1}$, and $v\in\{0,1\}$, the function $ Q_{N}(Z;z,v)$ in \eqref{2.3.3}
is strictly \textsl{log-concave} in $z\in \mathbb{R}_{v}$, i.e., $\log Q_{N}(Z;\cdot,v)$ is a strictly concave function on $\mathbb{R}_{v}$.
\end{lemma}

\noindent\textbf{Proof.}\quad
Let $N\geq1$ and $Z=(Z(0),Z(1))\in\mathbb{R}_{0}\times\mathbb{R}_{1}$ be fixed. We first consider the case $v=0$, with $z<0$. By (\ref{2.3.2}), we write succinctly
\begin{align}\label{PNsimple}
  &\quad\quad\quad\quad\quad\quad\quad\quad\quad P_N(Z; y) = \frac{1}{1+e^{ay+b}},&\\
  &\mbox{where}\quad a:=\tfrac{Z(0)-Z(1)}{N\sigma^2}<0,\quad b:= \log q+\tfrac{Z(1)^2-Z(0)^2}{2N\sigma^{2}}. \nonumber
\end{align}
Then, $\log P_N(Z; y) = - \log\big(1+e^{ay+b}\big)$, and so
\begin{equation*}
  \partial_y \log P_N(Z; y) =- \frac{a e^{ay+b}}{1+ e^{ay+b}} = -a +\frac{a }{ 1+e^{ay+b}} ,\quad \partial_{yy} \log P_N(Z; y) = - \frac{a^2 e^{ay+b}}{(1+e^{ay+b})^2} <0,
\end{equation*}
which show that $P_N(Z; y)$ is strictly log-concave in $y$. Also, note that
\begin{equation}\label{A.e1}
  \Phi_N(Z;z)=\int_{\mathbb{R}}P_N(Z;\sqrt{N} x+z)\frac{e^{-\frac{x^2}{2\sigma^2}}}{\sqrt{2\pi\sigma^{2}}}\dd x. 
\end{equation}
Since it is well-known that the Gaussian distribution is log-concave, in the sense that the density function $e^{-x^2/(2\sigma^2)}\big/\sqrt{2\pi\sigma^{2}}$ is log-concave in $x\in\mathbb{R}$, an application of the Pr\'{e}kopa--Leindler theorem (see, e.g., Brascamp and Lieb (\citeyear{BL76})) yields that $\Phi_N(Z;z)$ in (\ref{A.e1}) is log-concave in $z$. Note further that, for $v=0$ and $z<0$, we have
\begin{equation*}
  \log Q_N(Z;z,0)= \log(-z)+ \log \Phi_{N}(Z;z).
\end{equation*}
Then, by the (strict) concavity of the logarithm, we see that $Q_N(Z;z,0)$ is (strictly) log-concave in $z<0$.

Next, consider the case $v=1$ and $z>0$, with
\begin{equation}\label{logQN}
  \log Q_N(Z;z,1)= \log z+ \log \big(1-\Phi_{N}(Z;z)\big),
\end{equation}
and $1- P_N(Z; y) = e^{ay+b}/(1+ e^{ay+b}) = e^{ay+b} P_N(Z;y)$, and so $\log(1-P_{N}(Z;y))=ay + b + \log P_N(Z;y)$. The log-concavity of $1-P_{N}(Z;y)$ in $y$, along with the Pr\'{e}kopa--Leindler theorem again, implies that $1-\Phi_{N}(Z;z)$ is log-concave in $z$. It then follows from (\ref{logQN}) that $Q_N(Z;z,1)$ is (strictly) log-concave in $z>0$ as well.
\qed

\medskip

\noindent \textbf{Proof of Theorem \ref{thm:1} assertion (i).}\quad
Consider the finite-$N$ equilibrium in Definition \ref{def:1}. We proceed in two steps.

\medskip

\noindent \underline{Step 1}\quad
This step is concerned with the insider's optimization problem -- that is, given $N\geq1$, $Z=(Z(0),Z(1))\in\mathbb{R}_{0}\times\mathbb{R}_{1}$, and $v\in\{0,1\}$, we shall apply Lemma \ref{lem:1} to show that the objective function $\acute J_N(Z; z,v)$ has a unique maximum point $z^*_v\in \mathbb{R}_{v}$. We assume for simplicity that $\lambda(z)\equiv\lambda_{1}(z)$ (and hence $\lambda_N(z)$) is twice-differentiable for $z\neq0$.\footnote{The twice-differentiability of $\lambda$ is mainly to ease verification of the log-concavity of $\Lambda_N$ in \eqref{lambda"}. The same property can be verified by slightly more involved arguments if $\lambda$ is only convex.} By (\ref{3.2.1}) we have
\begin{equation}\label{A.30}
 \acute J_{N}(Z;z,v)=(\chi e^{-\lambda_{N}(z)} -\chi_{0} )Q_{N}(Z;z,v),\quad z\in \mathbb{R}_{v}.
\end{equation}
We shall only prove the case $v=0$. The case $v=1$ can be proved analogously.

We now restrict to $z<0$, and denote $\Lambda_N(z):= \chi e^{-\lambda_{N}(z)}-\chi_{0}$. Recall that $Q_N>0$. By Assumption \ref{as:1}, since $\lambda'_{N}(z)<0$, the (continuous) function $\Lambda_N$ is strictly increasing, with $\lim_{z\to-\infty}\Lambda_N(z)=-\chi_0\leq0$ and $\lim_{z\nearrow0}\Lambda_N(z)=1$. Thus, if $\chi_{0}>0$, then $\Lambda_N$ has a unique zero $z^{\circ}_{0}<0$; if $\chi_{0}=0$, $\Lambda_N>0$ on $(-\infty,0)$, and we can simply set $z^{\circ}_{0}=-\infty$. In either case, on $(z^{\circ}_{0},0)$, we have $\Lambda_N>0$, and so $\acute J_{N}(Z;z,0)>0$. Note that
$\lim_{z\to z^{\circ}_{0}}\acute J_{N}(Z;z,0)=\lim_{z\nearrow0}\acute J_{N}(Z;z,0)=0$, and then $\acute J_{N}(Z;z,0)$ admits a local maximum point $z^*_0\in (z^{\circ}_{0},0)$.

To show that $z^*_0$ is the unique maximum point of $\acute J_{N}(Z;\cdot,0)$ on $\mathbb{R}_{0}=(-\infty,0)$, let us note further that $\acute J_{N}(Z;\cdot,0)\leq0$ on $(-\infty,z^{\circ}_{0}]$ if $\chi_{0}<0$, and so it suffices to consider $z\in(z^\circ_0,0)$. By Lemma \ref{lem:1}, the function $Q_{N}(Z;\cdot,0)>0$ is strictly log-concave. Since $\Lambda_N>0$ and $\lambda_N''\geq0$, then
\begin{equation}\label{lambda"}
  \frac{\dd^2}{\dd z^2} \log \Lambda_N(z) =- \frac{\chi \chi_{0} (\lambda_{N}'(z))^2 e^{-\lambda_{N}(z)}}{ (\Lambda_N(z))^2}  -\frac{\chi \lambda_{N}''(z) e^{-\lambda_{N}(z)}}{ \Lambda_N(z)} \leq0;
\end{equation}
that is, $\Lambda_N$ is log-concave on $(z^{\circ}_{0},0)$, and by \eqref{A.30} and Lemma \ref{lem:1}, $\acute J_{N}(Z;\cdot,0)$ is strictly log-concave. This implies that $\acute J_{N}(Z;\cdot,0)$ admits at most one maximum point on $(z^{\circ}_{0},0)$, which is $z^*_0$.

\medskip

\noindent \underline{Step 2}\quad
In this step, we show the existence of a finite-$N$ equilibrium (Definition \ref{def:1}). By continuous differentiability from Assumption \ref{as:1}, we take the first-order conditions for (\ref{A.30}). Then, following the analysis in Step 1, any equilibrium strategy $Z^{\ast}_{N}=(Z^{\ast}_{N}(0),Z^{\ast}_{N}(1))\in\mathbb{R}_{0}\times\mathbb{R}_{1}$ must satisfy the equilibrium conditions
\begin{align}\label{1storder2}
  &\breve J_{N}(Z^{\ast}_{N};Z^{\ast}_{N}(v),v):=\big[e^{\lambda_N(z)}\partial_z \acute J_N(Z^*_N;z,v)\big]\big|_{z=Z^*_N(v)} =0, \quad v\in\{0,1\}.
\end{align}
By \eqref{2.3.3}, we have
\begin{equation}\label{B.6}
  \breve J_{N}(Z; z,v) = -\chi \lambda_N'(z) (v-\Phi_N(Z;z))z - \big(\chi-\chi_0e^{\lambda_N(z)}\big)\big(\pd_z\Phi_N(Z;z) z - v+\Phi_N(Z;z)\big).
\end{equation}


\noindent \underline{Step 2.1}\quad Consider the first-order conditions in (\ref{1storder2}) for $v=0$ first. Denote $\zeta^*_N := Z^*_N(1)-Z^*_N(0)>0$. Note that, for fixed $\zeta>0$ and $z<0$, by \eqref{2.3.2} and \eqref{2.3.3} we have
\begin{align}\label{ST}
  &\Phi_N((z, \zeta+z); z) =  \bar \Phi_N(\zeta), \quad (\pd_z \Phi_N)((z, \zeta+z); z) =\bar \Phi'_N(\zeta),\quad \mbox{where} \nonumber\\
  & \bar \Phi_N(\zeta):=\E\bigg[\frac{1}{ 1+ q e^{\bar \eta_N(\zeta)}}\bigg]>0, \quad \bar \Phi'_N(\zeta):=\frac{\zeta}{ N \sigma^2}\E\bigg[\frac{q e^{\bar \eta_N(\zeta)}}{ \big(1+ q e^{\bar \eta_N(\zeta)}\big)^2}\bigg] >0, \nonumber\\
  & \bar \eta_N(\zeta) := \frac{\zeta(\zeta-2\sqrt{N}W)}{2N\sigma^{2}}.
\end{align}
Based on \eqref{B.6}, we define for any fixed $\zeta>0$
\begin{align}\label{A.33}
  & F^N_{0}(\zeta;z):=\breve J_{N}((z,\zeta+z);z,0) = g_{1,N}(z) \bar\Phi_{N}(\zeta) + g_{2,N}(z) \bar\Phi'_{N}(\zeta),\quad z<0, \nonumber\\
  &\mbox{where}\quad g_{1,N}(z):=\chi z\lambda'_{N}(z)-\chi+\chi_{0}e^{\lambda_{N}(z)},\quad g_{2,N}(z):=(\chi_{0}e^{\lambda_{N}(z)}-\chi)z.
\end{align}
Since $Z^*_N(0) \le - \zeta^*_N$, the effective domain of $F^N_0$ in (\ref{A.33}) should be $(-\infty, -\zeta]$; however, for technical reasons, we extend $F^N_0$ to the whole $(-\infty,0)$.

By Assumption \ref{as:1}, since $g'_{1,N}(z)=(\chi_{0}e^{\lambda_{N}(z)}+\chi)\lambda'_{N}(z)+\chi z\lambda''_{N}(z)<0$, the function $g_{1,N}$ is strictly decreasing on $(-\infty,0)$. Further, noting that $\lim_{z\to-\infty}g_{1,N}(z)=\infty$ and $\lim_{z\nearrow0}g_{1,N}(z)=-1$, then $g_{1,N}$ has a unique zero $z^\diamond_0<0$, and thus $g_{1,N}(z) <0$ for $z\in (z^\diamond_0, 0)$. We now focus on the interval $(z^\diamond_0, 0)$. Note that $g'_{2,N}(z)=\chi_{0}e^{\lambda_{N}(z)}(z\lambda'_{N}(z)+1)-\chi$ and $\lambda_N(z)> 0>\lambda_N'(z)$ for $z<0$, then for $z\in (z^\diamond_0, 0)$,
\begin{equation}\label{B.9}
  g_{1,N}(z)-g'_{2,N}(z) = z\lambda'_{N}(z) \big(\chi - \chi_{0}e^{\lambda_{N}(z)}\big) = z\lambda'_{N}(z) \big(z\lambda'_{N}(z) - g_{1,N}(z)\big) >0; 
\end{equation}
thus $g'_{2,N}(z) < g_{1,N}(z)<0$. Since $\bar\Phi_N,\bar \Phi'_N>0$ on $(0,\infty)$, this implies that $F^N_0(\zeta;
\cdot)$, along with $g_{2, N}$, is strictly decreasing on $(z^\diamond_0, 0)$. Moreover, since $\lim_{z\nearrow0}g_{2,N}(z)=0$, then $g_{2,N}(z^\diamond_0)>0$, and $F^N_0(\zeta; z^\diamond) = g_{2,N}(z^\diamond_0)  \bar \Phi'_N(\zeta)>0$, with $\lim_{z\nearrow 0} F^N_0(\zeta; z) = - \bar \Phi_N(\zeta) <0$. Thus, $F^N_{0}(\zeta;\cdot)$ has a unique zero in $(z^\diamond_0, 0)$, denoted as $\mathcal{z}_{0}(\zeta)$.

Denote $\mathcal{z}_{1}(\zeta):= \zeta + \mathcal{z}_{0}(\zeta)$ and  $\mathcal{Z}^\zeta:=(\mathcal{z}_{0}(\zeta),\mathcal{z}_{1}(\zeta))$. Recall the function $\Lambda_N$ and the point $z^\circ_0$ from Step 1. From \eqref{B.9} we also see that $\Lambda_N(z) = e^{-\lambda_N(z)}( \chi - \chi_{0}e^{\lambda_{N}(z)} )>0$ for $z\in (z^\diamond_0, 0)$, which implies that $(z^\diamond_0, 0) \subset (z^\circ_0, 0)$, and thus, from Step 1, $\acute J_{N}(\mathcal{Z}^\zeta; \cdot,0)$ is strictly log-concave on $(z^\diamond_0, 0)$. It then follows from the first-order condition \eqref{1storder2} and Step 1 that $\mathcal{z}_{0}(\zeta)$ is the (unique) maximum point of $\acute J_{N}(\mathcal{Z}^\zeta; \cdot,0)$ in $\mathbb{R}_{0}$. We remark that the above analysis does not require $\mathcal{z}_{1}(\zeta)> 0$, which will be verified later for the equilibrium however.

For the purpose of the next sub-step, let us also note that
\begin{equation}
\label{B.10}
 F^N_{0}(\zeta; \mathcal{z}_{0}(\zeta)) = g_{1,N}(\mathcal{z}_{0}(\zeta)) \bar\Phi_{N}(\zeta) + g_{2,N}(\mathcal{z}_{0}(\zeta)) \bar\Phi'_{N}(\zeta)=0.
 \end{equation}
Since $g_{1, N}, g_{2,N}$ are continuous and strictly decreasing on $(z^\diamond_0, 0)$, and $\bar \Phi_N, \bar \Phi'_N$ are obviously continuous in $\zeta$, it is clear that $\mathcal{z}_{0}(\zeta)$ is continuous in $\zeta$. Moreover, from \eqref{ST} one can easily see that $\bar \Phi_N(0)=1$,  $\bar \Phi'_N(0)=0$.
Hence, by \eqref{B.10}, we have that $\lim_{\zeta\searrow 0} g_{1,N}(\mathcal{z}_{0}(\zeta)) =0$, and so
\begin{equation}
\label{B.11}
 \lim_{\zeta\searrow 0} \mathcal{z}_{0}(\zeta) = z^\diamond_0. 
 \end{equation}

%


\noindent\underline{Step 2.2}\quad Now we turn to (\ref{1storder2}) for $v=1$, restricting to $\mathcal{Z}^\zeta:=(\mathcal{z}_{0}(\zeta),\mathcal{z}_{1}(\zeta))$ based on Step 2.1.
We stress again that at this point, we do not require $ \mathcal{z}_{1}(\zeta) > 0$. 
Note that
\begin{align}\label{hatPhiN}
  &\Phi_N((z-\zeta, z); z) =  \hat \Phi_N(\zeta), \quad (\pd_z \Phi_N)((z-\zeta, z); z) =\hat \Phi'_N(\zeta),\quad \mbox{where} \nonumber\\
  & \hat \Phi_N(\zeta):=\E\bigg[\frac{1}{ 1+ q e^{\hat \eta_N(\zeta)}}\bigg]>0, \quad \hat \Phi'_N(\zeta):=\frac{\zeta}{ N \sigma^2}\E\bigg[\frac{q e^{\hat \eta_N(\zeta)}}{ \big(1+ q e^{\hat \eta_N(\zeta)}\big)^2}\bigg] >0, \nonumber\\
  & \hat \eta_N(\zeta) := -\frac{\zeta(\zeta+2\sqrt{N}W)}{2N\sigma^{2}}.
\end{align}
Recalling \eqref{B.6} and \eqref{A.33}, we define
\begin{align}\label{A.34}
  G^{N}_1(\zeta):=  \breve J_{N}(\mathcal{Z}^\zeta;\mathcal{z}_{1}(\zeta),1) =  g_{1,N}(\mathcal{z}_{1}(\zeta))\big( \hat\Phi_{N}(\zeta) -1\big) + g_{2,N}(\mathcal{z}_{1}(\zeta)) \hat\Phi'_{N}(\zeta).
\end{align}
Note that $\mathcal{z}_{1}(\zeta)= \zeta+\mathcal{z}_{0}(\zeta)$ is continuous for $\zeta> 0$ and $\mathcal{z}_{0}(\zeta) > z^\diamond_0$. By \eqref{B.11} we have $\lim_{\zeta\searrow 0} \mathcal{z}_{1}(\zeta)= z^\diamond_0<0$ and $\lim_{\zeta\to\infty} \mathcal{z}_{1}(\zeta)=\infty$. It is implied that there exists $\breve{\zeta}>0$ such that $\mathcal{z}_{1}(\breve{\zeta})=0$, and we choose the largest such $\breve\zeta$, i.e., such that $\mathcal{z}_{1}(\zeta)>0$ for all $\zeta>\breve\zeta$. Since $\hat \Phi_N(\breve\zeta) <1$, then by \eqref{A.33} and \eqref{A.34} we have
\begin{equation}
\label{B.13}
\lim_{\zeta\searrow\breve{\zeta}} G^{N}_1(\zeta)=  \big(\chi - \chi_0e^{\lambda_N(0)} \big) \big(1- \hat \Phi_N(\breve \zeta)\big) = 1- \hat \Phi_N(\breve \zeta) >0.
\end{equation}

On the other hand, since $g_{1,N}(0) = -\chi+\chi_0=-1<0$, and clearly $\lim_{z\to\infty}g_{1,N}(z)=\infty$, then there exists $\hat z\in (0, \infty)$ such that $g_{1,N}(\hat z)=0$. In particular, this implies that $\chi_{0}e^{\lambda_{N}(\hat z)}-\chi=-\chi \hat z\lambda'_{N}(\hat z)<0$, and so $g_{2,N}(\hat z) =(\chi_{0}e^{\lambda_{N}(\hat z)}-\chi) \hat z<0$. Moreover, recalling again that $\mathcal{z}_{1}(\zeta)$ is continuous with $ \mathcal{z}_{1}(\breve\zeta)= 0$ and $\lim_{\zeta\to\infty} \mathcal{z}_{1}(\zeta)=\infty$, then there exists $\hat\zeta > \breve \zeta$ such that $\mathcal{z}_{1}(\hat\zeta) = \hat z$. Therefore,
\begin{equation*}
G^{N}_1(\hat\zeta)=  g_{1,N}(\hat z)\big( \hat\Phi_{N}(\hat\zeta) -1\big) + g_{2,N}(\hat z) \hat\Phi'_{N}(\hat\zeta) = g_{2,N}(\hat z) \hat\Phi'_{N}(\hat\zeta)<0.
\end{equation*}
This, together with \eqref{B.13}, implies that there exists $\zeta^*\in (\breve \zeta, \hat \zeta)$ such that $G^{N}_1(\zeta^*)=0$.

Finally, since $\breve \zeta$ is the largest zero of $\mathcal{z}_{1}$ and $\zeta^*>\breve\zeta$,  we have $\mathcal{z}_{1}(\zeta^*)>0$.  Then, by the reasoning in Step 2.1 and Step 1 again, it follows that $\mathcal{z}_{1}(\zeta^*)$ is also the unique maximum point of $\acute J_{N}(\mathcal{Z}^{\zeta^*};\cdot,1)$ on $\mathbb{R}_1$. This, together with Step 2.1, implies that $(\mathcal{Z}^{\zeta^*}, P_N(\mathcal{Z}^{\zeta^*};\cdot))$ is an equilibrium in the sense of Definition \ref{def:1}, as required.
\qed

\medskip

\noindent \textbf{Proof of Theorem \ref{thm:1} Assertion (ii).}\quad We shall prove the ``if'' direction and the ``only if'' direction separately. Let us recall Definition \ref{def:2} condition (ii), \eqref{2.3.1}, \eqref{2.2.1} with $C_0=0$, and \eqref{2.3.6}. We see that, for any $\gamma$-limiting equilibrium $(\tilde Z^*_\gamma,\tilde P^*_\gamma)$ ($\tilde{Z}^{\ast}_{\gamma}\equiv(\tilde{Z}^{\ast}_{\gamma}(0),\tilde{Z}^{\ast}_{\gamma}(1))$),
\begin{align}\label{B.15}
  \tilde J^\gamma_{\infty}(\tilde P^*_\gamma;\tilde{z},v) &= \lim_{N\to\infty} N^{-\gamma} J_N\big(\tilde P^\gamma_\infty(\tilde Z^*_\gamma; N^{-\max\{\gamma, \frac{1}{2}\}} \cdot); N^\gamma \tilde z, v\big)\nonumber\\
  &= \lim_{N\to\infty} N^{-\gamma}  ( N^\gamma \tilde z) ~\Big(e^{-\lambda_N(N^\gamma \tilde z)}  -\chi_0(1- e^{-\lambda_N(N^\gamma \tilde z)})\Big) \Big(v -\tilde \Phi^\gamma_N(\tilde Z^*_\gamma; \tilde z)\Big) \nonumber\\
  &= \lim_{N\to\infty} \tilde z~\big(\chi e^{-\lambda(N^{\gamma-\beta}\tilde z)}  -\chi_0\big)\big( v -\tilde \Phi^\gamma_N(\tilde Z^*_\gamma; \tilde z)\big), \quad \tilde z \in \mathbb{R}_{v},\;v\in\{0,1\}, \nonumber\\
  \mbox{where}\quad& \tilde \Phi^\gamma_N(\tilde Z^*_\gamma; \tilde z):= \E\big[\tilde P^\gamma_\infty(\tilde Z^*_\gamma; N^{-\max\{\gamma, \frac{1}{2}\}}(\sqrt{N}W + N^\gamma \tilde z))\big],
\end{align}
with $\lambda\equiv\lambda_{1}$. Under Assumption \ref{as:1e}, we can easily verify the following limits: For $\tilde z\neq 0$, with $z^*_\gamma:=(\tilde Z^*_\gamma(0) + \tilde Z^*_\gamma(1))/2$,
\begin{equation}\label{limit1}
  \lim_{N\to\infty}e^{-\lambda(N^{\gamma-\beta}\tilde z)} =\begin{cases}1,\quad&~\text{if }\gamma < \beta,\\ \lambda(\tilde z),&~\text{if }\gamma = \beta,\\ 0,\quad&~\text{if }\gamma > \beta;\end{cases} \quad \lim_{N\to\infty}\tilde \Phi^\gamma_N(\tilde Z^*_\gamma; \tilde z) = \begin{cases}p,&~\text{if }\gamma < \frac{1}{2},\\ \Phi_1(\tilde Z^*_\gamma; \tilde z) ,&~\text{if }\gamma = \frac{1}{2},\\ \1_{\{\tilde z > z^*_\gamma\}} +\frac{1}{ 2}\1_{\{\tilde z = z^*_\gamma\}},&~\text{if }\gamma >\frac{1}{2};\end{cases}
\end{equation}
in particular, the case $\gamma>1/2$ follows by using \eqref{3.1} and \eqref{B.15} so that
\begin{align*}
  \tilde \Phi^\gamma_N(\tilde Z^*_\gamma; \tilde z)= \E\big[\1_{\{N^{{1/2} -\gamma}W +\tilde z> z^*_\gamma\}} + p \1_{\{N^{{1/2} -\gamma}W +\tilde z =z^*_\gamma\}}\big] 
  \to \1_{\{\tilde z > z^*_\gamma\}} +\tfrac{1}{2}\1_{\{\tilde z = z^*_\gamma\}}.
\end{align*}

\medskip

\noindent\underline{Step 1}\quad In this step we prove the ``if'' direction. Assume $\gamma = \min\{\beta, 1/2\}$. We naturally examine (\ref{B.15}) in two cases.

\smallskip

\noindent\underline{Case 1}\quad $\gamma=\beta <1/2$.  In this case, by Definition \ref{def:2} (ii) and \eqref{3.2} from Proposition \ref{pro:1}, we must set $\tilde P^*_\gamma = p$, and it is unique. Further, (\ref{B.15}) gives that
\begin{align}\label{B.16}
  \tilde J^\gamma_{\infty}(\tilde P^*_\gamma;\tilde{z},v) =(v-p)  \tilde z~\big(\chi e^{-\lambda(\tilde z)}  -\chi_0\big), \quad \tilde z \in \mathbb{R}_{v}.
\end{align}
Similar to \eqref{1storder2}, we define for $\tilde z\in \mathbb{R}_{v}$
\begin{align}\label{breveJ}
  \breve J^{\gamma}_{\infty}(\tilde{z},v)&:= e^{\lambda(\tilde z)}\pd_{\tilde{z}}\tilde{J}^{\gamma}_{\infty}(\tilde{P}^*_{\gamma};\tilde{z},v) = (v-p)\big(\chi - \chi \tilde z \lambda'(\tilde z) - \chi_0 e^{\lambda(\tilde z)}\big); \nonumber\\
  \pd_{\tilde z}\breve J^{\gamma}_{\infty}(\tilde{z},v) &= -(v-p)\big(\chi \lambda'(\tilde z)+ \chi \tilde z \lambda''(\tilde z) + \chi_0 e^{\lambda(\tilde z)}\lambda'(\tilde z)\big).
\end{align}
When $v=0$, by Assumption \ref{as:1}, one can easily see that
\begin{align}\label{B.17}
  &  \lim_{\tilde{z}\nearrow0}\tilde{J}^{\gamma}_{\infty}(\tilde{P}^*_{\gamma};\tilde{z},0)=0,\quad \lim_{\tilde{z}\to-\infty}\tilde{J}^{\gamma}_{\infty}(\tilde{P}^*_{\gamma};\tilde{z},0)=-\infty; \nonumber\\
  & \lim_{\tilde{z}\nearrow0}\breve J^{\gamma}_{\infty}(\tilde{z},0) =-p<0,\quad \lim_{\tilde{z}\to-\infty}\breve J^{\gamma}_{\infty}(\tilde{z},0)=\infty;\quad
  \pd_{\tilde z}\breve J^{\gamma}_{\infty}(\tilde{z},0)<0, \quad\tilde z<0.
\end{align}
Based on the first and third limits in (\ref{B.17}), we have that $\tilde{J}^{\gamma}_{\infty}(\tilde{P}^*_{\gamma};\tilde{z},0)>0$ when $\tilde z<0$ is close to $0$. Along with the second limit, this implies that the function $\tilde{J}^{\gamma}_{\infty}(\tilde{P}^*_{\gamma};\tilde{z},0)$ has a maximum point $\tilde Z^*_\gamma(0)\in (-\infty, 0)$, with $\tilde{J}^{\gamma}_{\infty}(\tilde{P}^*_{\gamma};\tilde Z^*_\gamma(0),0)$>0. Clearly, $\pd_{\tilde{z}}\tilde{J}^{\gamma}_{\infty}(\tilde{P}^*_{\gamma};\tilde Z^*_\gamma(0),0)=0$. Also, from the second line of \eqref{B.17}, we see that $\breve J^{\gamma}_{\infty}(\tilde{z},0)$, and hence $\pd_{\tilde{z}}\tilde{J}^{\gamma}_{\infty}(\tilde{P}^*_{\gamma};\tilde{z},0)$, has a unique zero. Thus, $\tilde Z^*_\gamma(0)$ is the unique maximum point of $\tilde{J}^{\gamma}_{\infty}(\tilde{P}^*_{\gamma};\cdot,0)$ in $(-\infty,0)$.

Similarly, when $v=1$, we have
\begin{align*}
  &  \lim_{\tilde{z}\searrow0}\tilde{J}^{\gamma}_{\infty}(\tilde{P}^*_{\gamma};\tilde{z},1)=0,\quad \lim_{\tilde{z}\to\infty}\tilde{J}^{\gamma}_{\infty}(\tilde{P}^*_{\gamma};\tilde{z},1)=-\infty; \\
  & \lim_{\tilde{z}\searrow0}\breve J^{\gamma}_{\infty}(\tilde{z},1)=1-p>0,\quad \lim_{\tilde{z}\to\infty}\breve J^{\gamma}_{\infty}(\tilde{z},1)=-\infty; \quad
  \pd_{\tilde z}\breve J^{\gamma}_{\infty}(\tilde{z},1) <0, \quad \tilde z>0.
\end{align*}
By similar observations it follows that $\tilde{J}^{\gamma}_{\infty}(\tilde{P}^*_{\gamma};\tilde{z},1)$ has a unique maximum point $\tilde Z^*_\gamma(1)\in (0,\infty)$, with $\tilde{J}^{\gamma}_{\infty}(\tilde{P}^*_{\gamma};\tilde Z^*_\gamma(1),1)$>0. Therefore, $\tilde Z^*_\gamma = (\tilde Z^*_\gamma(0), \tilde Z^*_\gamma(1))$ and $\tilde P^*_\gamma\equiv p$ together constitute the unique $\gamma$-limiting equilibrium.

\smallskip

\noindent\underline{Case 2}\quad $\gamma=1/2 \le \beta$. In this case, by \eqref{3.1}, \eqref{B.15}, and \eqref{2.3.3} (with $N=1$), we have
\begin{align}\label{B.18}
  \tilde J^\gamma_{\infty}(\tilde P^*_\gamma;\tilde{z},v) =  \lim_{N\to\infty} \big(\chi e^{-\lambda(N^{\gamma-\beta}\tilde z)}  -\chi_0\big)Q_1(\tilde Z^*_\gamma; \tilde z, v), \quad \tilde z \in \mathbb{R}_{v}.
\end{align}
When $\beta =1/2$, this is the same as (\ref{A.30}) with $N=1$, and by assertion (i) of Theorem \ref{thm:1} we know that there exists a $\gamma$-limiting equilibrium.

Now consider the case $\beta>1/2=\gamma$. Then $ \tilde J^\gamma_{\infty}(\tilde P^*_\gamma;\tilde{z},v) = Q_1(\tilde Z^*_\gamma; \tilde z, v)$,  $ \tilde z \in \mathbb{R}_{v}$,
which is in the form of (\ref{A.30}) with $N=1$ and $\lambda=0$ (hence violating Assumption \ref{as:1}). In particular, \eqref{A.33} becomes (using $z$ instead of $\tilde z$ and fixed $\zeta>0$) $F^1_{0}(\zeta;z) = - \bar\Phi_{1}(\zeta) -z \bar\Phi'_1(\zeta)$, $z<0$,
with a unique zero $\mathcal{z}_{0}(\zeta) := -  \bar\Phi_{1}(\zeta)\slash  \bar\Phi_{1}'(\zeta) < 0$. We claim that
\begin{equation}\label{B.20}
  \lim_{\zeta\searrow 0} \mathcal{z}_{0}(\zeta) = -\infty,\quad  \lim_{\zeta\to \infty} \mathcal{z}_{0}(\zeta) =0.
\end{equation}
From \eqref{ST}, the first limit is obvious. To see the second limit, consider an arbitrary $c>0$ and the function
\begin{align*}
  & \bar \Phi^c_1(\zeta) :=   \bar\Phi_{1}(\zeta) - c  \bar\Phi'_{1}(\zeta)= \E\bigg[\frac{1+ q\big(1-\frac{c\zeta}{ \sigma^2}\big) e^{\bar \eta_1(\zeta)}}{ \big(1+ q e^{\bar \eta_1(\zeta)}\big)^2}\bigg].
\end{align*}
Recall that $W\overset{\rm d.}{=}\text{Normal}(0, \sigma^2)$. By (\ref{ST}), $\bar \eta_1(\zeta)={\zeta(\zeta-2W)/(2\sigma^2)} \overset{\rm d.}{=} \text{Normal}({\zeta^2/(2\sigma^2)}, {\zeta^2/\sigma^2})$. After some rearrangement, we obtain
\begin{align*}
  \bar \Phi^c_1(\zeta) &=\int_{\mathbb{R}} \frac{1+ q\big(1-\frac{c\zeta}{\sigma^2}\big) e^x}{ \big(1+ q e^{x}\big)^2} \frac{\sigma}{ \sqrt{2\pi} \zeta}\exp\bigg(-\frac{\sigma^{2}\big(x-\frac{\zeta^2}{2\sigma^2}\big)^2}{ 2 \zeta^2}\bigg)\dd x\\
  & = \frac{\sigma e^{-\frac{\zeta^2}{8\sigma^2}}}{ \sqrt{2\pi} \zeta}\Big(\bar A_1(\zeta) + q\big(1-\tfrac{c\zeta}{\sigma^2}\big) \bar A_2(\zeta)\Big),\\
  \mbox{where}\quad & \bar A_1(\zeta) := \int_{\mathbb{R}} \frac{e^{-\frac{\sigma^2}{ 2\zeta^2} x^2 + \frac{1}{ 2} x }}{ \big(1+ q e^{x}\big)^2} \dd x,\quad\bar A_2(\zeta) := \int_{\mathbb{R}} \frac{e^{-\frac{\sigma^2}{ 2\zeta^2} x^2 + \frac{3}{ 2} x }}{ \big(1+ q e^{x}\big)^2} \dd x.
\end{align*}
Then, note that
\begin{align*}
  \lim_{\zeta\to\infty} \bar A_1(\zeta) = \int_{\mathbb{R}} \frac{e^{ \frac{1}{2} x }}{ \big(1+ q e^{x}\big)^2} \dd x\in (0, \infty),\quad \lim_{\zeta\to\infty} \bar A_2(\zeta) = \int_{\mathbb{R}} \frac{e^{ \frac{3}{2} x }}{ \big(1+ q e^{x}\big)^2} \dd x\in (0, \infty).
\end{align*}
Thus, it is clear that $ \lim_{\zeta\to\infty}  \big(\bar A_1(\zeta) + q(1-c\zeta\slash \sigma^2) \bar A_2(\zeta)\big) = -\infty$, which implies that $ \bar \Phi^c_1(\zeta) <0$ for $\zeta >0$ sufficiently large. Thus, $ \bar\Phi_{1}(\zeta) - c  \bar\Phi'_{1}(\zeta)<0$ and $\mathcal{z}_{0}(\zeta) = -  \bar\Phi_{1}(\zeta)\slash  \bar\Phi_{1}'(\zeta) \ge -c$ for all sufficiently large $\zeta$. Since $c>0$ is arbitrary, we must have $\lim_{\zeta\to\infty}\mathcal{z}_{0}(\zeta) =0$.

Next we analyze $G^1_1(\zeta)$. In particular, \eqref{A.34} becomes
\begin{equation}\label{G11}
  G^{1}_1(\zeta) =  -\big( \hat\Phi_{1}(\zeta) -1\big) - \mathcal{z}_{1}(\zeta)  \hat\Phi'_{1}(\zeta),\quad \mathcal{z}_{1}(\zeta)=\zeta+ \mathcal{z}_{0}(\zeta).
\end{equation}
By \eqref{B.20}, we have that $ \lim_{\zeta\searrow 0} \mathcal{z}_{1}(\zeta) = -\infty$ and  $\lim_{\zeta\to \infty} \mathcal{z}_{1}(\zeta) =\infty$.
Thus, there exists $\breve{\zeta}>0$ such that $\mathcal{z}_{1}(\breve{\zeta})=0$ and $\mathcal{z}_{1}(\zeta)>0$ for all $\zeta>\breve\zeta$. From \eqref{G11}, clearly $G^{1}_1(\breve\zeta) >0$. Also, recalling \eqref{hatPhiN} and noting that $\hat\eta_1(\zeta) \overset{\rm d.}{=} \text{Normal}(-{\zeta^2/(2\sigma^2)}, {\zeta^2/ \sigma^2})$, in a similar fashion as above,
\begin{align*}
  & G^{1}_1(\zeta) =  \E\bigg[\frac{q e^{\hat \eta_1(\zeta)}}{1+q e^{\hat \eta_1(\zeta)}} - \frac{ q e^{\hat \eta_1(\zeta)}\zeta \mathcal{z}_{1}(\zeta) }{\sigma^2 \big(1+ q e^{\hat \eta_1(\zeta)}\big)^2}\bigg]  = \frac{\sigma e^{-\frac{\zeta^2}{8\sigma^2}}}{ \sqrt{2\pi} \zeta}\Big(\hat A_1(\zeta) - \tfrac{1}{q}\big(\tfrac{\zeta \mathcal{z}_{1}(\zeta)}{ \sigma^2}-1\big)\hat A_2(\zeta)\Big), \\
  &\mbox{where}\quad  \hat A_1(\zeta) := \int_{\mathbb{R}} \frac{e^{-\frac{\sigma^2}{ 2\zeta^2} x^2 + \frac{1}{ 2} x }}{\big(1+ \frac{e^{x}}{q}\big)^{2}} \dd x,\quad\hat A_2(\zeta) := \int_{\mathbb{R}} \frac{e^{-\frac{\sigma^2}{ 2\zeta^2} x^2 + \frac{3}{2} x }}{ \big(1+ \frac{e^{x}}{q}\big)^2} \dd x.
\end{align*}
Noting that $\lim_{\zeta\to\infty} \hat A_1(\zeta),\lim_{\zeta\to\infty} \hat A_2(\zeta)\in(0,\infty)$, and $\lim_{\zeta\to \infty} \mathcal{z}_{1}(\zeta) =\infty$, we have $ \lim_{\zeta\to\infty} \big(\hat A_1(\zeta) - {(\zeta \mathcal{z}_{1}(\zeta) / \sigma^2)} \hat A_2(\zeta)\big)=-\infty$, and so $G^{1}_1(\zeta) <0$ for $\zeta$ sufficiently large. Since $G^{1}_1(\breve\zeta) >0$, then there exists $\zeta^*>\breve \zeta$ such that $G^{1}_1(\zeta^*)=0$. Thus, following the same arguments as in Step 2 of the proof of assertion (i), $(\mathcal{Z}^{\zeta^*}, P_1(\mathcal{Z}^{\zeta^*};\cdot))$ is a $\gamma$-limiting equilibrium in the sense of Definition \ref{def:2}.

\medskip

\noindent\underline{Step 2}\quad We now prove the ``only if'' direction. Assume that there exists a $\gamma$-limiting equilibrium.

\smallskip

\noindent\underline{Case 3}\quad $\gamma<1/2$. Then, from (\ref{3.1}), $\tilde P^*_\gamma = p$, and so by \eqref{B.15} and \eqref{limit1},
\begin{equation*}
  \tilde J^\gamma_{\infty}(\tilde P^*_\gamma;\tilde{z},v) =
  \begin{cases}(v-p) \tilde z,\qquad \qquad\quad&~\text{if }\gamma <\beta,\\ (v-p) \tilde z(\chi e^{-\lambda(z)}-\chi_0),&~\text{if }\gamma = \beta,\\  -\chi_0 (v-p) \tilde z, &~\text{if }\gamma >\beta, \end{cases} \quad \tilde z \in \mathbb{R}_{v}.
\end{equation*}
Clearly, in the cases  $\gamma <\beta$ and $\gamma > \beta$ with $\chi_{0}>0$, $\tilde J^\gamma_{\infty}(\tilde P^*_\gamma;\cdot,v)$ has no maximum point in $\mathbb{R}_{v}$. In the case $\gamma > \beta$ with $\chi_{0}=0$, $\tilde J^\gamma_{\infty}(\tilde P^*_\gamma;\tilde{z},v)=0$ for all $\tilde z \in \mathbb{R}_{v}$, but since in Definition \ref{def:2} condition (i) we require $\tilde J^\gamma_{\infty}(\tilde P^*_\gamma; \tilde Z^*_\gamma(v),v)\neq 0$, there is no such $\tilde Z^*_\gamma(v)$. Thus, we must have $\gamma =\beta$, and since $\gamma <1/2$, we obtain $\gamma =\min\{\beta, 1/2\}$.

\smallskip

\noindent\underline{Case 4}\quad $\gamma= 1/2$. If $\beta<1/2$, then by \eqref{B.18} and \eqref{limit1}, $\tilde J^\gamma_{\infty}(\tilde P^*_\gamma;\tilde{z},v)  =  - \chi_0Q_1(\tilde Z^*_\gamma; \tilde z, v)$, $ \tilde z \in \mathbb{R}_{v}$.
By Lemma \ref{lem:1}, $Q_1(\tilde Z^*_\gamma; \tilde z, v)$ is strictly log-concave in $z\in \mathbb{R}_{v}$; hence, if $\chi_0>0$, $\tilde J^\gamma_{\infty}(\tilde P^*_\gamma;\cdot,v)$ has no maximum point in $\mathbb{R}_{v}$, while if $\chi_{0}=0$, then $\tilde J^\gamma_{\infty}(\tilde P^*_\gamma;\cdot,v)=0$, also a violation, and it must be that $\beta \ge 1/2$; thus, $\gamma =\min\{\beta, 1/2\}$.

\smallskip

\noindent\underline{Case 5}\quad $\gamma> 1/2$. Then, \eqref{B.15} and \eqref{limit1} imply that
\begin{equation*}
 \tilde J^\gamma_{\infty}(\tilde P^*_\gamma;\tilde{z},v) =\tilde z \big(v- \1_{\{\tilde z > z^*_\gamma\}} -\tfrac{1}{2} \1_{\{\tilde z = z^*_\gamma\}}\big) \lim_{N\to\infty} \big(\chi e^{-\lambda(N^{\gamma-\beta}\tilde z)}  -\chi_0\big),\quad  \tilde z \in \mathbb{R}_{v},
\end{equation*}
with the limit being finite. When $z^*_\gamma \ge 0$, we have $ \tilde J^\gamma_{\infty}(\tilde P^*_\gamma;\tilde{z},0)= 0$ for all $\tilde z<0$, which again violates the requirement $\tilde J^\gamma_{\infty}(\tilde P^*_\gamma; \tilde Z^*_\gamma(0),0)\neq 0$. Similarly, when $z^*_\gamma < 0$, we have $ \tilde J^\gamma_{\infty}(\tilde P^*_\gamma;\tilde{z},1)= 0$ for all $\tilde z>0$. Therefore, there is no $\gamma$-limiting equilibrium.
\qed

\medskip

\noindent \textbf{Proof of Theorem \ref{thm:1} assertion (iii).}\quad
Let Assumption \ref{as:1} still hold. We now prove the convergence of a finite-$N$ equilibrium $(Z^*_N, P^*_N)$ (Definition \ref{def:1}) towards the unique $\gamma$-limiting equilibrium $(\tilde Z^*_\gamma, \tilde P^*_\gamma =p)$, in the case $\gamma=\beta <1/2$. We proceed in three steps.


\medskip

\noindent \underline{Step 1}\quad In this step, we show that $\tilde z^N_v:= N^{-\gamma} Z^*_N(v)$ is bounded in $N$. Consider $v=0$ first. From Step 2.1 in the proof of Assertion (i), we see that $z^{\diamond,N}_0< Z^*_N(0) <0$, where $z^{\diamond,N}_0= z^\diamond_0$ is the unique zero of $g_{1,N}$ defined in \eqref{A.33}. Noting that
\begin{equation}\label{B.24}
  g_{1,N}(z) = \chi z N^{-\beta} \lambda'(N^{-\beta}z)-\chi+\chi_{0}e^{\lambda(N^{-\beta}z)} = g_{1,1}(N^{-\beta}z),\quad z<0,
\end{equation}
then $z^{\diamond,N}_0 = N^{\beta} z^{\diamond,1}_0$, where $z^{\diamond,1}_0$ is the unique zero of $g_{1,1}$, depending only on the model parameters. Since $\gamma = \beta$, we have  $z^{\diamond,1}_0 \leq \tilde z^N_0 <0$, and so $\tilde z^N_0$ is bounded. We shall put $|\tilde z^N_0|\le K_0$,  for some constant $K_0>0$.

In the same fashion we can prove that $|\tilde z^N_1|\le K_0$ for possibly larger $K_0$. Note that one may not use the function $G^N_1$ from \eqref{A.34} for this purpose -- instead, the following function that is the counterpart of $F^N_0$ in \eqref{A.33} for $v=1$ and satisfies  $G^N_1(\zeta) = F^N_{1}(\zeta; \mathcal{z}_{1}(\zeta))$ should be used:
\begin{equation}\label{B.19}
  F^N_{1}(\zeta;z):=\breve J_{N}((z-\zeta,z);z,1) = g_{1,N}(z) (\hat\Phi_{N}(\zeta)-1) + g_{2,N}(z) \hat\Phi'_{N}(\zeta),\quad z>0.
\end{equation}

%
%
\medskip

\noindent \underline{Step 2}\quad
In this step, we prove the desired uniform convergence of  $F^{N}_{v}(N^\gamma \tilde\zeta;N^\gamma \tilde z)$. Recall $\breve J^\gamma_\infty(\tilde z, v)$ in \eqref{breveJ}. For $v=0$, $\tilde \zeta>0$, and $\tilde z<0$, using \eqref{A.33} further we have
%
%
%
%
%
\begin{align*}
  \big|F^{N}_{0}(N^{\gamma}\tilde{\zeta};N^{\gamma}\tilde{z})-\breve J^\gamma_\infty(\tilde z, 0)\big| &= \Big|g_{1,1}(\tilde z) \bar \Phi_N(N^\gamma \tilde \zeta) + g_{2,1}(\tilde z) N^\gamma\bar \Phi'_N(N^\gamma \tilde \zeta) - p g_{1,1}(\tilde z)\Big|  \\
  &\leq |g_{1,1}(\tilde z)| \big|\bar \Phi_N(N^\gamma \tilde \zeta)  - p\big| + |g_{2,1}(\tilde z)| N^\gamma\bar \Phi'_N(N^\gamma \tilde \zeta).
\end{align*}
From Step 1, we know $|\tilde z^N_v|\le K_0$ and $\tilde \zeta^N:= \tilde z^N_1- \tilde z^N_0 \le 2K_0$, so from now on we shall restrict to $-K_0\le \tilde z<0$ and $0<\tilde \zeta\le 2K_0$. In particular, this implies that $|g_{1,1}(\tilde z)|, |g_{2,1}(\tilde z)|\le K$; here, we use $K>0$ to denote a generic constant that depends only on the model parameters, and its value may vary from line to line.

Since ${x/(1+x)^2}\le {1/ 4}$, by \eqref{ST} we have for $\tilde \zeta\in (0, 2K_0]$,
\begin{align*}
   N^\gamma \bar \Phi'_N(N^\gamma \tilde \zeta) = \frac{N^{2\gamma} \tilde\zeta}{ N\sigma^2}\E\bigg[\frac{qe^{\bar\eta_N(\zeta)}}{ \big(1+ q e^{\bar\eta_N(\zeta)}\big)^2}\bigg]\le   \frac{N^{2\gamma} \tilde\zeta}{ 4 N\sigma^2} \le  K N^{2\gamma -1}.
\end{align*}
Moreover, one can easily verify that $\bar \Phi_N(N^\gamma \tilde \zeta) =\bar \Phi_1\big( N^{\gamma-1/2}\tilde \zeta\big)$, $\bar \Phi_1(0) = p$, as well as $\bar \Phi'_1(0) = -\E\big[ {e^{\bar\eta_1(\zeta)} q(\zeta - W)\slash \sigma^2 \big/  (1+ q e^{\bar\eta_1(\zeta)})^2}\big]\big|_{\zeta=0} = 0$. Then, again, for $\tilde \zeta\in (0, 2K_0]$, and hence $N^{\gamma-1/2}\tilde \zeta\in (0, 2K_0]$, we have
\begin{equation*}
  \big|\bar \Phi_N(N^\gamma \tilde \zeta)  - p\big| = \big|\bar \Phi_1\big( N^{\gamma-\frac{1}{2}}\tilde \zeta\big) - \bar \Phi_1( 0)\big|\le K \big|N^{\gamma-\frac{1}{2}}\tilde \zeta\big|^2\le  K N^{2\gamma -1}.
\end{equation*}
Putting things together, we have shown that
\begin{equation}\label{B.21}
  \sup_{\tilde \zeta \in (0, 2K_0], ~\tilde z \in [-K_0, 0)} \big|F^{N}_{0}(N^{\gamma}\tilde{\zeta};N^{\gamma}\tilde{z})-\breve J^\gamma_\infty(\tilde z, 0)\big|  \le K N^{2\gamma-1}.
\end{equation}
It is similar to establish that $\sup_{\tilde \zeta \in (0, 2K_0], ~\tilde z \in (0, K_0]} \big|F^{N}_{1}(N^{\gamma}\tilde{\zeta};N^{\gamma}\tilde{z})-\breve J^\gamma_\infty(\tilde z, 1)\big| \le K N^{2\gamma-1}$.

\medskip
%
%

\noindent \underline{Step 3}\quad We now prove \eqref{3.2.3}.
Since $F^{N}_{0}(N^{\gamma}\tilde{\zeta}^N;N^{\gamma}\tilde{z}^N_0)=\breve J^\gamma_\infty(\tilde{Z}^{\ast}_{\gamma}(0), 0)=0$, \eqref{B.21} implies
\begin{equation}\label{B.23}
  \big|\breve J^\gamma_\infty(\tilde z^N_0, 0)-\breve J^\gamma_\infty(\tilde{Z}^{\ast}_{\gamma}(0),0)\big|= \big|\breve J^\gamma_\infty(\tilde z^N_0, 0)-F^{N}_{0}(N^{\gamma}\tilde{\zeta}^N;N^{\gamma}\tilde{z}^N_0)\big| \le K N^{2\gamma-1}.
\end{equation}
Note that by \eqref{breveJ},
\begin{equation*}
  \pd_{\tilde z}\breve J^{\gamma}_{\infty}(\tilde{z},0) = p\Big(\chi \lambda'(\tilde z)+ \chi \tilde z \lambda''(\tilde z) + \chi_0 e^{\lambda(\tilde z)}\lambda'(\tilde z)\Big) \le  p\chi \lambda'(\tilde z)<0, \quad \tilde z<0.
\end{equation*}
As $\lambda'$ is increasing (by Assumption \ref{as:1}), 
$\pd_{\tilde z}\breve J^{\gamma}_{\infty}(\tilde{z},0) \le  p\chi \lambda'(\tilde Z^\ast_\gamma(0)/2)<0$ for all $\tilde z \le \tilde{Z}^{\ast}_{\gamma}(0)/2$. Then, denoting $c_0:=  p\chi |\lambda'(\tilde Z^\ast_\gamma(0)/2)|$, by the monotonicity of $\breve J^{\gamma}_{\infty}(\tilde{z},0)$,
\begin{equation*}
  \big|\breve J^\gamma_\infty(\tilde z, 0)-\breve J^\gamma_\infty(\tilde{Z}^{\ast}_{\gamma}(0),0)\big| \ge \tfrac{1}{ 2}c_0 |\tilde{Z}^{\ast}_{\gamma}(0)|>0,\quad \tilde z\in \big[\tfrac{1}{ 2}\tilde{Z}^{\ast}_{\gamma}(0),0\big).
\end{equation*}
By \eqref{B.23}, this further implies that for $N$ large such that $ K N^{2\gamma-1} < c_0 |\tilde{Z}^{\ast}_{\gamma}(0)|\slash 2$, we must have $\tilde z^N_0\le \tilde{Z}^{\ast}_{\gamma}(0)/2$. Thus, with $c_{0}\leq|\pd_{\tilde z}\breve J^{\gamma}_{\infty}(\tilde{z},0)|$ for all $\tilde{z}\leq\tilde{Z}^{\ast}_{\gamma}(0)/2$,
\begin{equation*}
  KN^{2\gamma-1} \ge  \big|\breve J^\gamma_\infty(\tilde z^N_0, 0)-\breve J^\gamma_\infty(\tilde{Z}^{\ast}_{\gamma}(0),0)\big| \ge  c_0 |\tilde z^N_0-\tilde{Z}^{\ast}_{\gamma}(0)|,
\end{equation*}
which immediately implies that
\begin{equation*}
  |\tilde z^N_0-\tilde{Z}^{\ast}_{\gamma}(0)|\le \tfrac{K}{c_0}N^{2\gamma-1}\le KN^{2\gamma-1}.
\end{equation*}
It is similar to show that $|\tilde z^N_1-\tilde{Z}^{\ast}_{\gamma}(1)|\le  KN^{2\gamma-1}$, justifying  \eqref{3.2.3}.


For \eqref{3.2.4}, note that
\begin{equation*}
  P^*_N(y) = P_N(Z^*_N; y) = \tilde P^\gamma_N(N^{-\gamma} Z^*_N; N^{-\max\{\gamma, \frac{1}{ 2}\}} y) = \frac{1}{ 1+q e^{\tilde \kappa_N(\tilde y)}},
\end{equation*}
where $\tilde \kappa_N$ is as in \eqref{A.1} corresponding to $\tilde Z = N^{-\gamma} Z^*_N$ and $\tilde y = N^{-1/2} y$. By Step 1, $|\tilde Z|\le K_0$. Since $\gamma<1/2$, then it follows from \eqref{kNorder} that
\begin{equation*}
  |\tilde \kappa_N(\tilde y)| \le K\big(|\tilde y| N^{\gamma-\frac{1}{ 2}} + N^{2\gamma-1}\big) = K\big(|y| N^{\gamma-1} + N^{2\gamma-1}\big).
\end{equation*}
Therefore, recalling $q=(1-p)/p$, we have
\begin{equation*}
  |P^*_N(y) -p| = \bigg| \frac{1}{ 1+q e^{\tilde \kappa_N(\tilde y)}}-p\bigg| = (1-p)\frac{|1- e^{\tilde \kappa_N(\tilde y)}|}{ 1+q e^{\tilde \kappa_N(\tilde y)}} \le K\big(|y| N^{\gamma-1} + N^{2\gamma-1}\big),
\end{equation*}
proving the first estimate in \eqref{3.2.4}. Finally, the last bound, along with the boundedness of $N^{-\gamma} Z^*_N(v)$ established in Step 1, immediately leads to the second estimate in \eqref{3.2.4}, completing the proof.
\qed

%

\medskip

\renewcommand{\theequation}{C.\arabic{equation}}

\section{Proof of Theorem \ref{thm:2}}\label{C}

We have $\gamma = \beta <1/2$. First, \eqref{3.2.1} and \eqref{B.16} together imply that, for $v\in\{0,1\}$ and $z\in\mathbb{R}_v$,
\begin{equation*}
  N^{-\gamma} J_N(p; z, v) = (v-p) N^{-\gamma} z \big(\chi e^{-\lambda(N^{-\gamma} z)} - \chi_0\big) =\tilde J^\gamma_{\infty}(p;N^{-\gamma} z,v).
\end{equation*}
Then, based on Definition \ref{def:2}, since
\begin{equation*}
  N^{-\gamma} J_N(p; N^\gamma \tilde Z^*_\gamma(v), v) =\tilde J^\gamma_{\infty}(p;  \tilde Z^*_\gamma(v),v) = \sup_{z\in \mathbb{R}_{v}} \tilde J^\gamma_{\infty}(p; N^{-\gamma} z,v) = \sup_{z\in \mathbb{R}_{v}} N^{-\gamma} J_N(p; z, v),
\end{equation*}
we have that \eqref{3.1.7} in Definition \ref{def:3} holds with $\epsilon = 0$.

Next, since $\max\{\gamma, 1/2\}=1/2$, by \eqref{A.1} we have $P_N(N^\gamma \tilde Z^*_\gamma; N^{1/2} \tilde y) = 1/(1+q e^{\tilde\kappa_N(\tilde y)})$, where $\tilde \kappa_N$ corresponds to $\tilde Z =  \tilde Z^*_\gamma$. As $ \tilde Z^*_\gamma$ is independent of $N$, it follows from the exact same arguments for \eqref{3.2.3} that
\begin{equation*}
  \big|P_N(N^\gamma \tilde Z^*_\gamma; N^\frac{1}{ 2} \tilde y) -p\big| \le K \big(|N^\frac{1}{ 2} \tilde y| N^{\gamma-1} + N^{2\gamma-1}\big) \le KN^{\gamma-\frac{1}{ 2}}(|\tilde y|+1),
\end{equation*}
which verifies \eqref{3.1.8} with $\epsilon\equiv\epsilon_N = KN^{\gamma-{1/2}}$, hence (\ref{3.2.5}). The proof is complete.
\qed

\end{appendices}

\bigskip 

\begin{appendices}

\setcounter{section}{0}

\newtheorem{atheorem}{Theorem}[section]
\newtheorem{alemma}{Lemma}[section]
\newtheorem{aassumption}{Assumption}[section]

\renewcommand{\appendixname}{Supplemental Appendix}
\renewcommand{\theequation}{SA.\arabic{equation}}

\renewcommand{\thetheorem}{SA.\arabic{theorem}}
\renewcommand{\thelemma}{SA.\arabic{lemma}}
\renewcommand{\theassumption}{SA.\arabic{assumption}}

\section{Proofs in further discussions}\label{SA}

This supplemental appendix presents formal statements and their detailed proofs in the further discussions on the general scenario involving criminal penalties in Section \ref{sec:5}. The proofs are structurally similar to that of Theorem \ref{thm:1} and Theorem \ref{thm:2} as presented in Appendix \ref{B} and Appendix \ref{C}, but with notable technical differences.


%
%
%

\begin{theorem}\label{thm:3}
Consider the setting of (\ref{3.3.1}), and let Assumption \ref{as:1e} hold. We have the following four assertions. \medskip\\
(i) For every $N\geq1$, there exists an equilibrium $(Z^{\ast}_{N},P^{\ast}_{N})$ in the sense of Definition \ref{def:1}. \\
(ii) There exists a $\gamma$-limiting equilibrium $(\tilde{Z}^{\ast}_{\gamma},\tilde{P}^{\ast}_{\gamma})$ in the sense of Definition \ref{def:2} if and only if (\ref{3.1.2}) holds. Moreover, when $\gamma<1/2$, the $\gamma$-limiting equilibrium $(\tilde{Z}^{\ast}_{\gamma},\tilde{P}^{\ast}_{\gamma})$ is unique with $\tilde{P}^{\ast}_{\gamma}=p$. \\
(iii) Let $\gamma<1/2$ and $(\tilde{Z}^{\ast}_{\gamma},\tilde{P}^{\ast}_{\gamma}\equiv p)$ be the unique $\gamma$-limiting equilibrium from assertion (ii). Then, there exists a constant $K>0$, depending only on the model parameters but not on $N$, such that, for any equilibrium $(Z^{\ast}_{N},P^{\ast}_{N})$ from assertion (i),
\begin{equation}\label{3.1.3}
  |N^{-\gamma}Z^{\ast}_{N}(v)-\tilde{Z}^{\ast}_{\gamma}(v)|\le
  \begin{cases}
    K N^{2\gamma-1},&\!\!\! \text{if }\beta=0 ~\mbox{or}~ \alpha =1, \\
    KN^{\max\{2\gamma-1, -\gamma(\alpha-1)\theta', -\gamma\alpha'\}}, &\!\!\!  \text{if }\beta>0~\mbox{and}~\alpha>1,
  \end{cases}
  ~ v\in\{0,1\},
\end{equation}
and, with $\tilde{P}^{\ast}_{\gamma}=p$, the estimates in \eqref{3.2.4} stand.\\
(iv) As in assertion (iii), let $\gamma<1/2$ and let $(\tilde{Z}^{\ast}_{\gamma}, p)$ be the unique $\gamma$-limiting equilibrium. For every $N\geq1$, $(N^{\gamma}\tilde{Z}^{\ast}_{\gamma},p)$ is an $\epsilon_{N}$-equilibrium in the sense of Definition \ref{def:3}, where, for a constant $K$ depending only on the model parameters,
\begin{equation}\label{3.1.9}
  \epsilon_{N}=
  \begin{cases}
    K N^{\gamma-\frac{1}{2}},&\;\text{if }\beta=0 ~\mbox{or}~ \alpha=1, \\
    K N^{\max\{\gamma-\frac{1}{2}, -\gamma(\alpha-1), -\gamma(\alpha-1)\theta', -\gamma\alpha'\}},
    &\;\text{if }\beta>0~\mbox{and}~ \alpha>1.
  \end{cases}
\end{equation}
\end{theorem}

\medskip

Before proving Theorem \ref{thm:3}, we make a few additional comments on the convergence implications. Compared with Theorem \ref{thm:1}, assertion (iii) in Theorem \ref{thm:3} indicates the joint dependence of the convergence rate for the optimal strategies on both the intensity of investigation ($\beta$) and the severity of criminal penalties ($\alpha$), signaling a tradeoff between small and large values of $\gamma$. Within the maximum brackets in (\ref{3.1.3}), the first component ($2\gamma-1$) stems from the convergence of the price function, whereas the second and third components ($-\gamma(\alpha-1)\theta'$ or $-\gamma\alpha'$) is a result of the large argument behavior for $C_{0}$ (as stated in Assumption \ref{as:1e}). Intuitively, when investigations are sensitive to the population size ($\beta>0$), the imposition of criminal penalties upon prosecution can substantially alter the convergence rate depending on their specific form. In this case, the jump in the convergence rate as $\alpha\searrow1$ in (\ref{3.1.3}) results from the at-most-linear growth of the insider's expected profit with his trade size (see (\ref{3.1.1})), which can be coalesced into the penalty component if and only if $\alpha=1$. Further details can be found in the proof of Theorem \ref{thm:3}.

\medskip

To prepare for the proof of Theorem \ref{thm:3}, we establish a technical lemma.

\begin{lemma}\label{lem:2}
Consider the setting of Lemma \ref{lem:1}, and define
\begin{equation}\label{R}
  R_N(Z;z, v) := e^{\lambda_N(z)} \pd_z \big( (\chi e^{-\lambda_{N}(z)} -\chi_{0} )Q_{N}(Z;z,v)\big),\quad z\in\mathbb{R}_{v},\;v\in\{0,1\}.
\end{equation}
Then, the following two assertions hold. \medskip\\
(i) $R_N(Z; z, 0)$, $z<0$, admits a unique zero $\breve{z}_{0}<0$. Also, $R_N(Z; z, 0)>0$ for $z<\breve z_0$, and $R_N(Z; z, 0)$ is negative and strictly decreasing for $z\in(\breve{z}_{0},0)$. \\
(ii) $R_{N}(Z;z,1)$, $z>0$, admits a unique zero $\breve{z}_{1}>0$. Also, $R_N(Z; z, 1)<0$ for $z>\breve z_1$, and $R_N(Z; z, 1)$ is positive and strictly decreasing for $z\in(0,\breve{z}_{1})$.
\end{lemma}

\noindent\textbf{Proof.}\quad
We only consider the case $v=0$, with $z<0$; the case $v=1$ (with $z>0$) can be proved in the same way. As before, without loss of generality, we assume in the proof that $\lambda(z)$ is twice-differentiable for $z\neq0$, and for notational simplicity we suppress the symbol $Z$ in all the functions.

By Step 1 in the proof of Theorem \ref{thm:1} assertion (i), the function $(\chi e^{-\lambda_{N}(z)} -\chi_{0} )Q_{N}(z,0)$, $z<0$, admits a unique maximum point $\breve{z}_{0}<0$, and it satisfies $R_N(\breve{z}_{0}, 0)=0$,   $R_{N}(z,0) >0$ for $z<\breve z_0$, and $R_{N}(z,0) <0$ for $z\in(\breve{z}_{0},0)$. Moreover, since $Q_N>0$, it is clear that $\chi e^{-\lambda_{N}(\breve z_0)} -\chi_{0}>0$, and then by the monotonicity of $\lambda_N$, we see that $\chi e^{-\lambda_{N}( z)} -\chi_{0}>0$ for all $z\in (\breve z_0, 0)$.

It remains to show that $R_{N}(z,0)$ is strictly decreasing in $(\breve{z}_{0},0)$. Let us write
\begin{equation}\label{H}
  H_N(z) := (\chi  -\chi_{0} e^{\lambda_{N}( z)}) \Phi_N(z)>0,\quad z\in(\breve{z}_{0},0).
\end{equation}
Then, $(\chi e^{-\lambda_{N}(z)} -\chi_{0} )Q_{N}(z,0) = (-z) e^{-\lambda_{N}(z)} H_N(z)$, and thus, for $z\in (\breve z_0, 0)$,
\begin{align}\label{A.e2}
  R_{N}(z,0)&= (z \lambda_N'(z)-1) H_N(z) -zH'_{N}(z) <0; \\
\label{SA.6}
  {\pd_z} R_{N}(z,0)&=(\lambda'_{N}(z)+z\lambda''_{N}(z))H_{N}(z) + (z\lambda'_{N}(z)-2)H'_{N}(z)-zH''_{N}(z).
\end{align}
Recall that $\chi  -\chi_{0} e^{\lambda_{N}( z)}>0$ for $z\in (\breve z_0, 0)$, and similar to \eqref{lambda"}, we have
\begin{equation*}
  \frac{\dd^2}{ \dd z^2} \log \big(\chi  -\chi_{0} e^{\lambda_{N}( z)}\big) = -\frac{\chi_0e^{\lambda_{N}(z)}( (\lambda'_N(z))^2+\lambda_{N}''(z)) }{ \chi  -\chi_{0} e^{\lambda_{N}( z)}}- \frac{(\chi_{0} \lambda_{N}'(z) e^{\lambda_{N}(z)})^2}{ (\chi  -\chi_{0} e^{\lambda_{N}( z)})^2}  \leq0,
\end{equation*}
implying that $\chi  -\chi_{0} e^{\lambda_{N}( z)}$ is log-concave. From the proof of Lemma \ref{lem:1}, we know that $\Phi_N$ is also log-concave, and so $H_N$ is log-concave in $(\breve z_0, 0)$. This implies that $H_{N}(z)H''_{N}(z)-(H'_{N}(z))^{2}\leq 0$.
Hence, from \eqref{SA.6} we obtain
\begin{align*}
  {\pd_z} R_{N}(z,0)&\le (\lambda'_{N}(z)+z\lambda''_{N}(z))H_{N}(z) + (z\lambda'_{N}(z)-2)H'_{N}(z)-z \frac{(H'_{N}(z))^{2} }{ H_N(z)}\\
  &=  (\lambda'_{N}(z)+z\lambda''_{N}(z))H_{N}(z) +\frac{H'_{N}(z) }{ H_N(z)} \big((z\lambda'_{N}(z)-2)H_{N}(z)-z H_N'(z)\big) \\
  &=  (\lambda'_{N}(z)+z\lambda''_{N}(z))H_{N}(z) +\frac{H'_{N}(z) }{ H_N(z)} \big(R_N(z,0)-H_{N}(z)\big),
\end{align*}
where the last line is due to \eqref{A.e2}. Note further that for $z\in (\breve z_0, 0)$,
\begin{equation*}
  H'_N(z) :=  -\chi_{0} e^{\lambda_{N}( z)}\lambda'_N(z)  \Phi_N(z) +  (\chi  -\chi_{0} e^{\lambda_{N}( z)}) \Phi_N'(z)>0,
\end{equation*}
and thus, by \eqref{A.e2} again and Assumption \ref{as:1e}, ${\pd_z} R_{N}(z,0)<0$ for $z\in (\breve z_0, 0)$.
\qed

\medskip

\noindent\textbf{Proof of Theorem \ref{thm:3} assertion (i).}\quad
Consider the equilibrium in Definition \ref{def:1} for fixed $N\geq1$. Similar to the proof of Theorem \ref{thm:1} assertion (i), we proceed in two steps. 

\medskip

\noindent \underline{Step 1}\quad In this step we show that, given $Z=(Z(0),Z(1))\in\mathbb{R}_{0}\times\mathbb{R}_{1}$, the insider's objective function $\acute J_N(Z;z,v)$ from (\ref{2.3.4}) has a unique maximum point $z^*_v\in\mathbb{R}_{v}$.  Without loss of generality, assume as before that $\lambda$ and $C_0$ are twice-differentiable for $z\neq0$, and again we shall only consider the case $v=0$.

First, by (\ref{3.3.1}) and (\ref{A.30}) we have that, for $z<0$,
\begin{equation*}
  \acute J_{N}(Z;z,0)= (\chi e^{-\lambda_{N}(z)} -\chi_{0} )Q_{N}(Z;z,v)-  (1-e^{-\lambda_N(z)})C_{0}(z).
 \end{equation*}
Then, recalling \eqref{R}, we have
\begin{align}\label{breveJN0}
  &\breve J_N(Z; z, 0) := e^{\lambda_N(z)} \partial_z \acute J_N(Z; z, 0)  = R_N(Z; z, 0)+ A_N(z),\\
  \label{A.4}
  &\mbox{where}\quad A_{N}(z):=-\lambda'_{N}(z)C_{0}(z)-(e^{\lambda_{N}(z)}-1)C'_{0}(z).
\end{align}
By Assumption \ref{as:1e}, it is clear that $A_N(z)\geq0$ for $z<0$, and
\begin{equation}\label{AN'}
  A_N'(z) =-\lambda''_{N}(z)C_{0}(z) -(e^{\lambda_{N}(z)}+1)\lambda'_{N}(z)C'_{0}(z)-(e^{\lambda_{N}(z)}-1)C''_{0}(z)\leq0.
\end{equation}
Let $\breve z_0$ be as in Lemma \ref{lem:2}. Then, it follows from \eqref{breveJN0} that
\begin{equation}\label{SA.12}
  \breve J_N(Z; z, 0) >0, ~ z\le \breve z_0 \quad\text{and}\quad \pd_z \breve J_N(Z; z, 0) <0,~z\in (\breve z_0, 0).
\end{equation} Note further that, by \eqref{A.e2} and \eqref{H},
\begin{equation}\label{SA.13}
  \lim_{z\nearrow 0} \breve J_N(Z; z, 0)  = \lim_{z\nearrow 0} R_N(Z; z, 0) = -\Phi_N(Z; 0) <0,
\end{equation}
which means that $\breve J_N(Z; \cdot, 0) $ has a unique zero $z^*_0 \in (\breve z_0, 0)$, and by \eqref{SA.12}, $\breve J_N(Z; z, 0) > 0$ for $z<z^*_0$ and $\breve J_N(Z; z, 0) <0$ for $z\in (z^*_0,0)$. By continuity, this implies that $z^*_0$ is the unique maximum point of $\acute J_N(Z; \cdot, 0)$ on $\mathbb{R}_{0}=(-\infty,0)$.

%
\medskip

\noindent \underline{Step 2}\quad
This step is concerned with the existence of a finite-$N$ equilibrium (Definition \ref{def:1}). As established in Step 1, $Z^{\ast}_{N}=(Z^{\ast}_{N}(0),Z^{\ast}_{N}(1))\in\mathbb{R}_{0}\times\mathbb{R}_{1}$ is an equilibrium strategy, i.e., $(Z^*_N, P_N(Z^*_N;\cdot))$ is an equilibrium, if and only if it satisfies the equilibrium conditions
\begin{equation}\label{1storder}
  \breve J_N(Z^*_N; Z^*_N(v), v)= R_N(Z^*_N; Z^*_N(v), v)+  A_N(Z^*_N(v))=0,\quad v\in\{0,1\},
\end{equation}
again with $Z^{\ast}_{N}(0),Z^{\ast}_{N}(1)\neq0$.

Consider (\ref{1storder}) for $v=0$. With $\zeta^*_N= Z^*_N(1)-Z^*_N(0)\ge 0$, fixing arbitrary $\zeta>0$ as a parameter, we introduce the following function on $(-\infty,0)$:
\begin{equation}\label{F0}
  F^N_0(\zeta; z) := \breve J_N((z, \zeta+ z); z, 0) =  R_N((z, \zeta+ z); z, 0)+  A_N(z),\quad z< 0.
\end{equation}
Recalling \eqref{ST} and \eqref{A.33}, we have
\begin{equation}\label{F02}
  F^N_0(\zeta; z) = g_{1,N}(z) \bar \Phi_N(\zeta) + g_{2,N}(z) \bar \Phi'_N(\zeta)+  A_N(z),\quad z<0.
\end{equation}
By Step 2.1 in the proof of Theorem \ref{thm:1} (i), $g_{1,N}$ has a unique zero $z^\diamond_0<0$ such that $g_{1,N}'(z)<0$ and $g_{2,N}'(z)<g_{1,N}(z)<0<g_{2,N}(z)$ for $z\in (z^\diamond_0, 0)$. Since $\bar \Phi_N(\zeta)>0$, $\bar \Phi_N'(\zeta)>0$, and $A_N(z)\geq0$, $A_N'(z)\le0$, then $F^N_0(\zeta; z^\diamond_0)>0$ and $\pd_z F^N_0(\zeta; z) <0$ for $z\in (z^\diamond_0, 0)$.  Similarly as in \eqref{SA.13}, we have $\lim_{z\nearrow 0} F^N_0(\zeta; z)  = -\bar \Phi_N(\zeta) <0$, and so $F^N_0(\zeta; \cdot)$ has a unique zero in $(z^\diamond_0, 0)$, denoted as $\mathcal{z}_{0}(\zeta)$, which is a continuous function of $\zeta$. Moreover, since $\bar \Phi_N(0) = 1$ and $\bar \Phi'_N(0) = 0$, we have
\begin{equation*}
  0 = \lim_{\zeta\searrow 0}  F^N_0(\zeta; \mathcal{z}_{0}(\zeta)) = \lim_{\zeta\searrow 0} \big( g_{1,N}(\mathcal{z}_{0}(\zeta)) + A_N(\mathcal{z}_{0}(\zeta))\big).
\end{equation*}
Again, noting that $g_{1,N}'<0$ and $A_N'\le0$, then by \eqref{A.33} and \eqref{A.4}, $\lim_{z\nearrow 0} ( g_{1,N}(z) + A_N(z)) = -1 < 0$, and so we have $\lim_{\zeta\searrow 0} \mathcal{z}_{0}(\zeta) < 0$.

Next, consider (\ref{1storder}) for $v=1$. As in the proof of Theorem \ref{thm:1} assertion (i), we denote $\mathcal{z}_{1}(\zeta):=\zeta+\mathcal{z}_{0}(\zeta)$ and $\mathcal{Z}(\zeta):=(\mathcal{z}_{0}(\zeta),\mathcal{z}_{1}(\zeta))$. Recalling \eqref{hatPhiN}, by \eqref{A.34},
\begin{equation*}
  G^{N}_1(\zeta):=  \breve J_{N}(\mathcal{Z}^\zeta;\mathcal{z}_{1}(\zeta),1) =  g_{1,N}(\mathcal{z}_{1}(\zeta))\big( \hat\Phi_{N}(\zeta) -1\big) + g_{2,N}(\mathcal{z}_{1}(\zeta)) \hat\Phi'_{N}(\zeta) + A_N(\mathcal{z}_{1}(\zeta)),
\end{equation*}
where $\hat\Phi_N$ and $\hat\Phi'_N$ are given in \eqref{hatPhiN}. Since $\lim_{\zeta\searrow 0} \mathcal{z}_{0}(\zeta) < 0$ and $\mathcal{z}_{0}(\zeta) \in (z^\diamond_0, 0)$, $\lim_{\zeta\searrow 0} \mathcal{z}_{1}(\zeta) < 0$ and $\lim_{\zeta\to \infty} \mathcal{z}_{1}(\zeta) =\infty$. Hence, there exists $\breve \zeta >0$ such that $\mathcal{z}_{1}(\breve \zeta) =0$. Let $\breve \zeta$ be the largest one having this property, so that $\mathcal{z}_{1}(\zeta) >0$ for all $\zeta > \breve \zeta$. Note that $\lim_{z\nearrow 0}  g_{1,N}(z)=-1$, $\lim_{z\nearrow 0}  g_{2,N}(z)=0$, and $ \lim_{z\nearrow 0} A_N(z) = 0$, and so $G^N_1(\breve\zeta) = 1- \hat\Phi_{N}(\breve\zeta)>0$. Moreover, similarly to Step 2.2 in the proof of Theorem \ref{thm:1} assertion (i), there exists $\hat\zeta > \breve \zeta$ such that $g_{1,N}(\mathcal{z}_{1}(\hat\zeta))=0$ and $g_{2,N}(\mathcal{z}_{1}(\hat\zeta))<0$. Note that $\hat \Phi'_N(\hat\zeta)>0$ and, by \eqref{A.4}, $A_N(\mathcal{z}_{1}(\hat\zeta))\le0$ (as $\mathcal{z}_{1}(\hat\zeta)>0$), and then $G^N_1(\hat\zeta)<0$. Thus, there exists $\zeta^*\in (\breve \zeta, \hat\zeta)$ such that $G^N_1(\zeta^*)=0$. Therefore, $(\mathcal{Z}^{\zeta^*}, P_N(\mathcal{Z}^{\zeta^*};\cdot))$, with $\mathcal{Z}^{\zeta^*}=\big(\mathcal{z}_{0}(\zeta^*), \mathcal{z}_{1}(\zeta^*)\big)$, is a finite-$N$ equilibrium in the sense of Definition \ref{def:1}.
\qed

\medskip

\noindent \textbf{Proof of Theorem \ref{thm:3} assertion (ii).}\quad As before, we prove the ``if'' direction and the ``only if'' direction separately. In a similar fashion as in \eqref{B.15}, especially for $\tilde \Phi^\gamma_N(\tilde Z^*_\gamma; \tilde z)$, given any $\gamma$-limiting equilibrium $(\tilde Z^*_\gamma,\tilde P^*_\gamma)$, we have that for $\tilde z \in \mathbb{R}_{v}$, $v\in\{0,1\}$,
\begin{equation}\label{SA.17}
  \tilde J^\gamma_{\infty}(\tilde P^*_\gamma;\tilde{z},v) = \lim_{N\to\infty}\Big( \tilde z~\big(\chi e^{-\lambda(N^{\gamma-\beta}\tilde z)}  -\chi_0\big)\big( v -\tilde \Phi^\gamma_N(\tilde Z^*_\gamma; \tilde z)\big) - (1-e^{-\lambda(N^{\gamma-\beta}\tilde z)} ) N^{-\gamma} C_0(N^\gamma \tilde z)\Big).
\end{equation}
Let us recall \eqref{B.15} and the limits in \eqref{limit1}. Also, under Assumption \ref{as:1e}, the following additional limits are easily justified: For $\tilde z\neq 0$,
\begin{align}\label{limit2}
  &\lim_{N\to\infty}N^{-\gamma} C_0(N^\gamma \tilde z) =\begin{cases}C_0(\tilde z),&\quad\text{if }\gamma =0,\\ K_\alpha|\tilde z|,&\quad\text{if }\gamma >0, \; \alpha=1,\\ \infty,&\quad\text{if }\gamma >0, \; \alpha>1;\end{cases}\\
\label{limit3}
  & \lim_{N\to\infty}(1-e^{-\lambda(N^{\gamma-\beta}\tilde z)} ) N^{-\gamma} C_0(N^\gamma \tilde z)= \begin{cases}0,&\quad\text{if } 0<\gamma < \frac{\beta\theta}{\theta+\alpha-1},\\ K_\theta K_\alpha |\tilde z|^{\theta+\alpha},&\quad\text{if }\gamma = \frac{\beta\theta}{\theta+\alpha-1}<\beta,\\ \infty ,&\quad\text{if } \frac{\beta\theta}{\theta+\alpha-1}<\gamma<\beta.\end{cases}
\end{align}

\medskip

\noindent\underline{Step 1}\quad In this step we prove the ``if'' direction in two cases. Assume \eqref{3.1.2} holds, and note that $ {\beta\theta}/(\theta+\alpha-1) \le \beta$, with equality holding if and only if $\beta =0$ or $\alpha=1$.

\smallskip

\noindent \underline{Case 1}\quad  $\gamma= \beta\theta/(\theta+\alpha-1) < 1/2$.  We must set $\tilde P^*_\gamma = p$, and it is unique.

\noindent\underline{Case 1.1}\quad $\gamma =\beta=0$. In this case, \eqref{SA.17} becomes
\begin{equation}\label{B.16'}
  \tilde J^\gamma_{\infty}(\tilde P^*_\gamma;\tilde{z},v) =(v-p)  \tilde z~\big(\chi e^{-\lambda(\tilde z)}  -\chi_0\big) - (1-e^{-\lambda(\tilde z)} )C_0(\tilde z), \quad \tilde z \in \mathbb{R}_{v},\;v\in\{0,1\}.
\end{equation}
Analogous to \eqref{breveJ}, for $A_N(\tilde z)$ (from (\ref{A.4})) with $N=1$,
\begin{align}\label{SAbreveJ1}
  \breve J^{\gamma}_{\infty}(\tilde{z},v)&:= e^{\lambda(\tilde z)}\pd_{\tilde{z}}\tilde{J}^{\gamma}_{\infty}(\tilde{P}^*_{\gamma};\tilde{z},v) = (v-p)\big(\chi - \chi \tilde z \lambda'(\tilde z) - \chi_0 e^{\lambda(\tilde z)}\big) + A_1(\tilde z); \nonumber\\
  \pd_{\tilde z}\breve J^{\gamma}_{\infty}(\tilde{z},v)&= -(v-p)\big(\chi \lambda'(\tilde z)+ \chi \tilde z \lambda''(\tilde z) + \chi_0 e^{\lambda(\tilde z)}\lambda'(\tilde z)\big) + A_1'(\tilde z),\quad z<0.
\end{align}
For $v=0$, under Assumption \ref{as:1e}, it is clear that all the statements in \eqref{B.17} remain true. Thus, by following the same arguments after \eqref{B.17} we see that $\tilde{J}^{\gamma}_{\infty}(\tilde{P}^*_{\gamma};\cdot,0)$ has a unique maximum point $\tilde Z^*_\gamma(0)< 0$. For $v=1$, it is similar that $\tilde{J}^{\gamma}_{\infty}(\tilde{P}^*_{\gamma};\cdot,1)$ has a unique maximum point $\tilde Z^*_\gamma(1)>0$. Therefore, $\tilde Z^*_\gamma = (\tilde Z^*_\gamma(0), \tilde Z^*_\gamma(1))$, together with $\tilde P^*_\gamma\equiv p$, forms the unique $\gamma$-limiting equilibrium.

\noindent\underline{Case 1.2}\quad $\beta >0$ and $\alpha =1$. Then, $\gamma = \beta > 0$, and \eqref{SA.17} becomes
\begin{equation}\label{SAbreveJ2}
  \tilde J^\gamma_{\infty}(\tilde P^*_\gamma;\tilde{z},v) =(v-p)  \tilde z~\big(\chi e^{-\lambda(\tilde z)}  -\chi_0\big) - (1-e^{-\lambda(\tilde z)} )K_\alpha|\tilde z|, \quad \tilde z \in \mathbb{R}_{v},
\end{equation}
which is the same as \eqref{B.16'} with $C_0(\tilde z) = K_\alpha|\tilde z|$. Thus, we claim again that there exists a unique $\gamma$-limiting equilibrium.

\noindent\underline{Case 1.3}\quad $\beta >0$ and $\alpha >1$. Then, $\gamma ={\beta\theta}/(\theta+\alpha-1)<\beta$, and \eqref{SA.17} becomes
\begin{align}
\label{SAbreveJ3}
 \tilde J^\gamma_{\infty}(\tilde P^*_\gamma;\tilde{z},v) =(v-p)  \tilde z - K_\theta K_\alpha  |\tilde z|^{\theta + \alpha}, \quad \tilde z \in \mathbb{R}_{v}.
\end{align}
Since $\theta+\alpha >1$, it is clear that $\tilde J^\gamma_{\infty}(\tilde P^*_\gamma;\tilde{z},v)$ above has a unique maximum point $\tilde Z^*_\gamma(v)\in \mathbb{R}_{v}$, and so there exists a unique $\gamma$-limiting equilibrium.

\smallskip

\noindent\underline{Case 2}\quad  $\gamma= 1/2 \le {\beta\theta}/(\theta+\alpha-1)\le \beta$.

\noindent\underline{Case 2.1}\quad $\beta = \gamma = 1/2$. In this case, we have $\alpha = 1$, and based on Assumption \ref{as:1e} conditions (i) and (iii), observe that for some constant $K^v_\alpha\in \mathbb{R}$,
\begin{equation}\label{C02}
  C_0(z) \sim K_\alpha |z| + K^v_\alpha,\quad \text{as }|z|\to\infty.
\end{equation}
Indeed, when $v=1$, $C_0(z)-K_\alpha z$ is still convex and $C_0(z)-K_\alpha z = o(z)$ as $z\to \infty$, and so the only possibility is that $C_0(z)-K_\alpha z$ is constant when $z>0$ is large. The same idea goes for $v=0$.

With this observation, \eqref{SA.17} specializes to
\begin{equation}\label{SA.21}
  \tilde J^\gamma_{\infty}(\tilde P^*_\gamma;\tilde{z},v) =  \big(\chi e^{-\lambda(\tilde z)}  -\chi_0\big)Q_1(\tilde Z^*_\gamma; \tilde z, v) - (1-e^{-\lambda(\tilde z)} ) K_\alpha |\tilde z|, \quad \tilde z \in \mathbb{R}_{v},
\end{equation}
which is a special case of (\ref{3.3.1}) with $N=1$ and $C_0(z) = K_\alpha |z|$. Thus, by assertion (i), a $\gamma$-limiting equilibrium exists.

\noindent\underline{Case 2.2}\quad $\gamma= 1/2 = {\beta\theta}/(\theta+\alpha-1)< \beta$. Then, \eqref{SA.17} becomes
\begin{equation*}
  \tilde J^\gamma_{\infty}(\tilde P^*_\gamma;\tilde{z},v) =  Q_1(\tilde Z^*_\gamma; \tilde z, v) - K_\theta K_\alpha  |\tilde z|^{\theta + \alpha},\quad \tilde z\in\mathbb{R}_v.
\end{equation*}
Adapting the analysis in the proof of assertion (i), one can easily check that in this case, the condition \eqref{A.33} becomes (using $z$ instead of $\tilde z$):
\begin{equation*}
  F^1_{0}(\zeta;z) = - \bar\Phi_{1}(\zeta) -z \bar\Phi'_1(\zeta) + K_\theta K_\alpha(\theta+\alpha)  |z|^{\theta + \alpha-1},\quad z<0.
\end{equation*}
Since $\theta+\alpha>1$, then clearly,
\begin{equation*}
  \lim_{z\nearrow 0}   F^1_{0}(\zeta;z) = - \bar\Phi_{1}(\zeta)<0,\quad    \lim_{z\to -\infty}   F^1_{0}(\zeta;z) =\infty,\quad \pd_z F^1_{0}(\zeta;z)  <0.
\end{equation*}
Thus,  $F^1_{0}(\zeta;z)$ has a unique zero $\mathcal{z}_{0}(\zeta) < 0$. Also, it is clear that $\mathcal{z}_{0}(\zeta) \ge -  \bar\Phi_{1}(\zeta)\slash  \bar\Phi_{1}'(\zeta)$. By \eqref{B.20}, we have $\lim_{\zeta\to \infty} \mathcal{z}_{0}(\zeta) =0$. Moreover, with $ \lim_{\zeta\searrow 0}\bar\Phi_{1}(\zeta)=1$ and $ \lim_{\zeta\searrow 0}\bar\Phi'_{1}(\zeta)=0$, we easily see that $\lim_{\zeta\searrow 0}  \mathcal{z}_{0}(\zeta) = - \big(K_\theta K_\alpha(\theta+\alpha)\big)^{1/(\theta + \alpha-1)}  <0$.
In particular, this implies that, for $\mathcal{z}_{1}(\zeta)=\zeta+ \mathcal{z}_{0}(\zeta)$,
\begin{equation}\label{SA.22}
  \lim_{\zeta\searrow 0}  \mathcal{z}_{1}(\zeta)<0,\quad \lim_{\zeta\to \infty} \mathcal{z}_{1}(\zeta) =\infty.
\end{equation}

On the other hand, the condition \eqref{G11} becomes
\begin{equation}\label{SA.23}
  G^{1}_1(\zeta) =  -\big( \hat\Phi_{1}(\zeta) -1\big) - \mathcal{z}_{1}(\zeta)  \hat\Phi'_{1}(\zeta) - K_\theta K_\alpha(\theta+\alpha)  |\mathcal{z}_{1}(\zeta)|^{\theta + \alpha-1}.
\end{equation}
By \eqref{SA.22}, there exists  $\breve{\zeta}>0$ such that $\mathcal{z}_{1}(\breve{\zeta})=0$, and $\mathcal{z}_{1}(\zeta)>0$ for all $\zeta>\breve\zeta$. By \eqref{SA.23} and \eqref{SA.22}, it is also clear that $G^{1}_1(\breve\zeta) >0$ and $\lim_{\zeta\to\infty} G^{1}_1(\zeta) = -\infty$, and so there exists $\zeta^*>\breve \zeta$ such that $G^{1}_1(\zeta^*)=0$. Therefore, it follows from the same arguments as in the proof of Theorem \ref{thm:1} assertion (i) that, upon setting $\tilde Z^*_\gamma:= (\mathcal{z}_{0}(\zeta^*), \mathcal{z}_{1}(\zeta^*))$, $(\tilde Z^*_\gamma,P_{1}(\tilde Z^*_\gamma;\cdot)\big)$ is a $\gamma$-limiting equilibrium.

\noindent\underline{Case 2.3}\quad $\gamma= 1/2 < {\beta\theta}/(\theta+\alpha-1)\le \beta$. Then, \eqref{SA.17} reduces to $\tilde J^\gamma_{\infty}(\tilde P^*_\gamma;\tilde{z},v) =  Q_1(\tilde Z^*_\gamma; \tilde z, v)$ for $\tilde{z}\in\mathbb{R}_{v}$, which has been analyzed in Case 2 in the proof of Theorem \ref{thm:1} assertion (ii), so there exists a $\gamma$-limiting equilibrium.

\medskip

\noindent\underline{Step 2}\quad We now prove the ``only if'' direction. Assuming that there is a $\gamma$-limiting equilibrium $(\tilde Z^*_\gamma,\tilde P^*_\gamma)$, our goal is to prove that $\gamma$ satisfies \eqref{3.1.2}. Notably, the limit \eqref{limit2} is used when $\gamma \ge \beta$, while \eqref{limit3} is used when $\gamma < \beta$, and so we may exclude the cases $(\gamma\ge \beta, \gamma>0, \alpha>1)$ and ${\beta\theta}/(\theta+\alpha-1)<\gamma<\beta$. Now, if ${\beta\theta}/(\theta+\alpha-1)<\gamma$, then by the second excluded case, we have $\gamma \ge \beta$, which by the first excluded case further implies $\alpha =1$. Hence, we can focus on the following two cases:
\begin{equation}\label{limit4}
  \gamma \le \frac{\beta\theta}{\theta+\alpha-1}\le \beta\quad \mbox{or}\quad \alpha=1,\; \gamma > \beta = \frac{\beta\theta}{\theta+\alpha-1}.
\end{equation}

%

\smallskip

\noindent\underline{Case 3}\quad $\gamma<1/2$. In the second case of \eqref{limit4}, with $\tilde P^*_\gamma=p$, \eqref{SA.17} gives
\begin{equation*}
  \tilde J^\gamma_{\infty}(\tilde P^*_\gamma;\tilde{z},1) = -\chi_0\tilde z~(1 - p) -  K_\alpha |\tilde z| = -\big(\chi_0(1-p)+K_\alpha\big) \tilde z,\quad \tilde z>0,
\end{equation*}
which has no maximum point in $(0,\infty)$, and so we must have the first case of \eqref{limit4}, but if $\gamma < {\beta\theta}/(\theta+\alpha-1)\le \beta$, then \eqref{SA.17} becomes $\tilde J^\gamma_{\infty}(\tilde P^*_\gamma;\tilde{z},v) =  \tilde z~(v -p)$, with no maximum point in $\mathbb{R}_{v}$ either. Thus, it can only be that $\gamma = {\beta\theta}/(\theta+\alpha-1)$, which verifies \eqref{3.1.2} because $\gamma <1/2$ here.

\smallskip

\noindent\underline{Case 4}\quad $\gamma= 1/2$. Suppose ${\beta\theta}/(\theta+\alpha-1)<1/2 = \gamma$. Then, we have the second case of \eqref{limit4}, and \eqref{SA.17} becomes
\begin{equation*}
  \tilde J^\gamma_{\infty}(\tilde P^*_\gamma;\tilde{z},v) =    -\chi_0Q_1(\tilde Z^*_\gamma; \tilde z, v) -  K_\alpha |\tilde z|,\quad \tilde{z}\in\mathbb{R}_{v}.
\end{equation*}
Note that $Q_1(\tilde Z^*_\gamma; \tilde z, v)>0$ for $\tilde z\ne0$. Then, $\tilde J^\gamma_{\infty}(\tilde P^*_\gamma;\tilde{z},v) <0$ for $\tilde z\neq0$, while $\lim_{\tilde z\to 0} \tilde J^\gamma_{\infty}(\tilde P^*_\gamma;\tilde{z},v)=0$ clearly, which means that $\tilde J^\gamma_{\infty}(\tilde P^*_\gamma;\cdot,v)$ has no maximum point in $ \mathbb{R}\setminus\{0\}$. Therefore, we must have ${\beta\theta}/(\theta+\alpha-1)\ge 1/2 = \gamma$, verifying \eqref{3.1.2}.

\smallskip

\noindent\underline{Case 5}\quad $\gamma> 1/2$. We consider two subcases. Again, we write $z^*_\gamma:=(\tilde Z^*_\gamma(0) + \tilde Z^*_\gamma(1))/2$.

\noindent\underline{Case 5.1}\quad $z^*_\gamma \ge 0$. Note that in this case, for $v=0$, $-(\1_{\{\tilde z > z^*_\gamma\}}+\1_{\{\tilde z =z^*_\gamma\}}/2)=0$ for all $\tilde z<0$. Then, based on \eqref{limit4}, by \eqref{SA.17} we have that for $\tilde z<0$,
\begin{align}
 \label{limit5}
  \tilde J^\gamma_{\infty}(\tilde P^*_\gamma;\tilde{z},0) = \begin{cases} 0,&\quad\text{if } \gamma < \frac{\beta\theta}{\theta+\alpha-1}\le \beta,\\ - K_\theta K_\alpha |\tilde z|^{\theta+\alpha},&\quad\text{if } \gamma = \frac{\beta\theta}{\theta+\alpha-1}< \beta,\\
 - (1-e^{-\lambda(\tilde z)} ) K_\alpha |\tilde z|,&\quad\text{if } \gamma = \frac{\beta\theta}{\theta+\alpha-1}< \beta,\\  - K_\alpha |\tilde z|,&\quad\text{if } \alpha=1, \gamma > \beta = \frac{\beta\theta}{\theta+\alpha-1},\end{cases}
\end{align}
and in all these cases, $\tilde J^\gamma_{\infty}(\tilde P^*_\gamma; \tilde Z^*_\gamma(0),0) = \sup_{\tilde z<0}\tilde J^\gamma_{\infty}(\tilde P^*_\gamma;\tilde{z},0) =0$. This violates the requirement $\tilde J^\gamma_{\infty}(\tilde P^*_\gamma; \tilde Z^*_\gamma(0),0)\neq 0$ in Definition \ref{def:2}.

\noindent\underline{Case 5.2}\quad $z^*_\gamma < 0$. In this case, for $v=1$, $1-(\1_{\{\tilde z > z^*_\gamma\}}+\1_{\{\tilde z =z^*_\gamma\}}/2) = \1_{\{\tilde z < z^*_\gamma\}}+\1_{\{\tilde z =z^*_\gamma\}}/2  = 0$ for all $\tilde z>0$. Then, $\tilde J^\gamma_{\infty}(\tilde P^*_\gamma;\tilde{z},1)$ has the same expression as in \eqref{limit5} except with $\tilde z>0$, which results in the same violation. Thus, we conclude that no $\gamma$-limiting equilibrium exists in the case $\gamma>1/2$.
\qed

\medskip

\noindent \textbf{Proof of Theorem \ref{thm:3} assertion (iii).}\quad
Let Assumption \ref{as:1e} hold; by (\ref{3.1.2}), let $\gamma={\beta\theta/(\theta + \alpha-1)} <1/2$. We proceed as in the proof of Theorem \ref{thm:1} assertion (iii), particularly focusing on the case $v=0$.

\medskip

\noindent \underline{Step 1}\quad Let us note that in \eqref{F02}, the zero $Z^*_N(0)$ of the function $F^N_0(Z^*_N(1)-Z^*_N(0);\cdot)$ satisfies $z^\diamond<Z^*_N(0)<0$, where $z^\diamond\equiv z^\diamond_0$ is the unique zero of $g_{1,N}$ defined in \eqref{A.33}. Write $\tilde z^N_0:= N^{-\gamma} Z^*_N(0)$ and $\hat z^N_0:= N^{-\beta} Z^*_N(0)$. By \eqref{B.24} and the subsequent arguments, we have that $|\hat z^N_0|\le K_1$ for some $K_1>0$; in particular, when $\gamma = \beta$, $\tilde z^N_0=\hat z^N_0$ is also bounded in $N$ by $K_1$.

Next, we justify the boundedness of  $\tilde z^N_0$ in the case $\gamma <\beta$, which is recalled to imply $\alpha>1$ and $\gamma>0$. Denote as before $\zeta^*_N:= Z^*_N(1)-Z^*_N(0)$, and note that $F^N_0(\zeta^*_N, Z^*_N(0))=0$, where $F^N_0$ is defined in  \eqref{F02}. From the arguments after \eqref{B.9}, we see that $g_{2,N}$ is decreasing in $(z^\diamond,0)$ with $\lim_{z\nearrow 0} g_{2,N}(z)=0$, so $g_{2,N}(z)>0$ for $z\in (z^\diamond,0)$, and in particular $g_{2,N}(Z^*_N(0))>0$. Recall \eqref{A.4} and note that $-\lambda'_{N}(Z^*_N(0))C_{0}(Z^*_N(0))>0$, $-(e^{\lambda_{N}(Z^*_N(0))}-1)C'_{0}(Z^*_N(0))>0$.
Also note that $\lambda_{N}(Z^*_N(0)) = \lambda(\hat z^N_0)$ and $\lambda_{N}'(Z^*_N(0)) = N^{-\beta}\lambda'(\hat z^N_0)$. Since for $\zeta>0$, $0<\bar\Phi_N(\zeta)<1$ and $\bar\Phi_N'(\zeta)>0$, by \eqref{F02} and \eqref{A.33},
\begin{align}\label{SA.28}
  &\quad N^{-\beta}\big|\lambda'(\hat z^N_0)C_{0}(Z^*_N(0))\big| = \big|\lambda_N'(Z^*_N(0))C_{0}(Z^*_N(0))\big|\le |g_{1,N}(Z^*_N(0)) \bar\Phi_N(\zeta^*_N)|\nonumber\\
  & \le \big|\chi Z^*_N(0)\lambda'_{N}(Z^*_N(0))-\chi+\chi_{0}e^{\lambda_{N}(Z^*_N(0))}\big|=\big|\chi \hat z^N_0\lambda'(\hat z^N_0)-\chi+\chi_{0}e^{\lambda(\hat z^N_0)}\big|\le K_2,
\end{align}
for some constant $K_2>0$, where we have used the crucial fact that $|\hat z^N_0|\le K_1$. Moreover, by Assumption \ref{as:1e} condition (ii), we have $K_3:= \sup_{z\in [-K_1, 0)} {|z|^{\theta-1}/|\lambda'(z)|} <\infty$, and so $|\lambda'(\hat z^N_0)|\ge |\hat z^N_0|^{\theta-1}/K_3$. Further note that by assuming without loss of generality that $|\tilde z^N_0|\ge 1$ (as otherwise we have established the desired boundedness of $\tilde z^N_0$), then $|Z^*_N(0)|\ge N^\gamma$, and thus by Assumption \ref{as:1e} (iii) we have $C_0(Z^*_N(0)) \ge K_\alpha |Z^*_N(0)|^\alpha/2$ for $N$ large. Therefore, we obtain
\begin{align*}
  &\quad \tfrac{2K_2K_3}{K_\alpha} \ge \tfrac{2K_2}{ K_\alpha} N^{-\beta}\big|\lambda'(\hat z^N_0)C_{0}(Z^*_N(0))\big| \ge N^{-\beta} |\hat z^N_0|^{\theta-1} |Z^*_N(0)|^\alpha \\
  &= N^{-\beta} |N^{\gamma-\beta}\tilde z^N_0|^{\theta-1} |N^\gamma \tilde z^N_0|^\alpha = N^{-\beta + (\gamma-\beta)(\theta-1) + \gamma \alpha} |\tilde z^N_0|^{\theta+\alpha-1} =  |\tilde z^N_0|^{\theta+\alpha-1},
\end{align*}
where the last equality is due to $\gamma = {\beta\theta/(\theta+\alpha-1)}$. This immediately implies that $|\tilde z^N_0|\le K_0$ for some $K_0>0$.

Furthermore, in this case \eqref{B.19} generalizes into
\begin{align}
\label{SA.27}
  F^N_{1}(\zeta;z):= g_{1,N}(z) (\hat\Phi_{N}(\zeta)-1) + g_{2,N}(z) \hat\Phi'_{N}(\zeta) + A_N(z),\quad z>0,
\end{align}
and in the same manner we can show that $\tilde z^N_1:= N^{-\gamma} Z^*_N(1)$ is also bounded by $K_0$.

\medskip

\noindent \underline{Step 2}\quad In this step we prove the uniform convergence of  $F^{N}_{v}(N^\gamma \tilde\zeta;N^\gamma \tilde z)$, for $0< \tilde \zeta \le 2K_0$ and $\tilde z\in  \mathbb{R}_{v}$ with $|\tilde z|\le K_0$, and we shall only do it for $v=0$. The case $v=1$ can be proved analogously. We consider three cases, corresponding to, respectively, Case 1.1, Case 1.2, and Case 1.3 in the proof of assertion (ii).

\smallskip

\noindent\underline{Case 1}\quad $\gamma = \beta=0$. Recalling \eqref{F02} and \eqref{SAbreveJ1},  we have
\begin{align*}
  &\big|F^{N}_{0}(N^\gamma \tilde\zeta;N^\gamma \tilde z) - \breve J^\gamma_\infty(\tilde z, 0)\big|\le I_N + \big|A_N(N^\gamma \tilde z) - A_1(\tilde z)\big|, \\
  \mbox{where}&\quad I_N := \big|g_{1,N}(N^\gamma) \bar \Phi_N(N^\gamma \tilde\zeta) + g_{2,N}(N^\gamma \tilde z) \bar \Phi'_N(N^\gamma \tilde \zeta) - p g_{1,1}(\tilde z)\big|,
\end{align*}
where $I_N$ is exactly the (absolute) difference indicated in \eqref{B.21}, with $|I_N|\le K N^{2\gamma -1}$. Moreover, from \eqref{A.4} we see that $A_N(N^\gamma \tilde z) = A_1(\tilde z)$ when $\gamma = \beta =0$ or $\gamma = \beta>0$ and $C_0(z)=K_\alpha |z|$. Thus,
\begin{equation}\label{SA.29}
  \big|F^{N}_{0}(N^\gamma \tilde\zeta;N^\gamma \tilde z) - \breve J^\gamma_\infty(\tilde z, 0)\big|\le  K N^{2\gamma -1}.
\end{equation}

\smallskip

\noindent\underline{Case 2}\quad $\gamma = \beta >0$ and $\alpha =1$. This is the same as Case 1, with the specialization $C_0(z) = K_\alpha|z|$, and thus \eqref{SA.29} remains true.

\smallskip

\noindent\underline{Case 3.}\quad $\beta>0$ and $\alpha >1$, so that $\gamma = {\beta \theta / (\theta+\alpha -1)} < \beta$. In this case, recalling \eqref{SAbreveJ3},
\begin{equation}\label{SA.34}
  \breve J^\gamma_\infty(\tilde z, 0) := \pd_{\tilde{z}}\tilde{J}^{\gamma}_{\infty}(\tilde{P}^*_{\gamma};\tilde{z},v) =  -p  + K_\theta K_\alpha (\theta+\alpha) |\tilde z|^{\theta+\alpha-1}.
\end{equation}
Then, by \eqref{F02}, \eqref{A.33}, and \eqref{A.4}, we have
\begin{align*}
  &\quad\big|F^{N}_{0}(N^\gamma \tilde\zeta;N^\gamma \tilde z) - \breve J^\gamma_\infty(\tilde z, 0)\big|\\
  &\le \big|g_{1,N}(N^\gamma \tilde z) \bar\Phi_N(N^\gamma \tilde\zeta) +p\Big| + \Big|g_{2,N}(N^\gamma \tilde z) \bar\Phi'_N(N^\gamma \tilde\zeta)\big| \\
  &\quad + \big|-\lambda'_{N}(N^\gamma \tilde z)C_{0}(N^\gamma \tilde z)-(e^{\lambda_{N}(N^\gamma \tilde z)}-1)C'_{0}(N^\gamma \tilde z) - K_\theta K_\alpha (\theta+\alpha) |\tilde z|^{\theta+\alpha-1}\big|.
\end{align*}
Recall that $N^\gamma \bar\Phi_N'(N^\gamma \tilde\zeta) + \big|\bar\Phi_N(N^\gamma \tilde\zeta) -p\big|\le K N^{2\gamma -1}$ for $0< \tilde\zeta \le 2K_0$, and in the following $K$ again denotes a generic positive constant that can vary. Moreover, by Assumption \ref{as:1e} conditions (ii) and (iii), defining $\tilde \theta':= \theta(1+\theta')$, $\tilde \alpha' := \alpha-\alpha'$, for $\tilde z\in [-K_0, 0)$,
\begin{align*}
  &  g_{1,N} (N^\gamma \tilde z) = \chi N^\gamma \tilde z N^{-\beta}\lambda'(N^{\gamma-\beta} \tilde z)-\chi+\chi_{0}e^{\lambda(N^{\gamma-\beta} \tilde z)}\\
  &\quad =\chi \tilde z N^{\gamma-\beta} \big(-K_\theta \theta |N^{\gamma-\beta} \tilde z|^{\theta-1} + O(|N^{\gamma-\beta}|^{\tilde\theta'-1})\big) - \chi + \chi_0\big(1+K_\theta |N^{\gamma-\beta} \tilde z|^\theta + O(|N^{\gamma-\beta}|^{\tilde\theta'})\big)\\
  &\quad = -1 + O(N^{(\gamma-\beta)\theta(1+\theta')});\\
  & g_{2,N}(N^\gamma \tilde z)=(\chi_{0}e^{\lambda(N^{\gamma-\beta} \tilde z)}-\chi)N^\gamma \tilde z = N^\gamma\tilde z \big(-1+o(1)\big);\\
  &\lambda'_{N}(N^\gamma \tilde z)C_{0}(N^\gamma \tilde z)=N^{-\beta} \lambda'(N^{\gamma-\beta} \tilde z)C_{0}(N^\gamma \tilde z) \\
  &\quad =N^{-\beta}\big( -K_\theta \theta|N^{\gamma-\beta} \tilde z|^{\theta-1} + O(|N^{\gamma-\beta}|^{\tilde\theta'-1} )\big)\big(K_\alpha |N^\gamma \tilde z|^\alpha + O(|N^\gamma|^{\tilde\alpha'})\big)\\
  &\quad = - K_\theta K_\alpha \theta |\tilde z|^{\theta+\alpha-1} N^{-\beta + (\gamma-\beta)(\theta-1) + \gamma \alpha}+O\big(N^{-\beta +\max\{(\gamma-\beta)(\tilde\theta'-1)+\gamma \alpha, (\gamma-\beta)(\theta-1)+\gamma \tilde\alpha'\} }\big)\\
  &\quad =  - K_\theta K_\alpha \theta |\tilde z|^{\theta+\alpha-1} +O\big(N^{\max\{(\gamma-\beta)\theta\theta', -\gamma\alpha'\}}\big);\\
  &(e^{\lambda_{N}(N^\gamma \tilde z)}-1)C'_{0}(N^\gamma \tilde z) = (e^{\lambda(N^{\gamma -\beta}\tilde z)}-1)C'_{0}(N^\gamma \tilde z) \\
  &\quad = \big(-K_\theta |N^{\gamma -\beta}\tilde z|^\theta + O(N^{(\gamma -\beta)\tilde\theta'})\big) \big(K_\alpha \alpha |N^\gamma \tilde z|^{\alpha-1}+ O(N^{\gamma(\tilde\alpha'-1)})\big)\\
  &\quad =  -K_\theta K_\alpha \alpha |\tilde z|^{\theta+\alpha-1} N^{(\gamma-\beta)\theta + \gamma (\alpha-1)} + O\big(N^{\max\{(\gamma-\beta)\theta + \gamma (\tilde\alpha'-1), (\gamma-\beta)\tilde\theta' + \gamma (\alpha-1)\}}\big)\\
  &\quad =  - K_\theta K_\alpha \theta |\tilde z|^{\theta+\alpha-1} + O\big(N^{\max\{(\gamma-\beta)\theta\theta', -\gamma\alpha'\}}\big),
\end{align*}
where the notations $O$ and $o$ are understood as being uniform in $\tilde z$. Then, since by $\gamma ={\beta\theta/(\theta + \alpha-1)}$, $\beta = \gamma (1+{(\alpha-1)/\theta})$,
\begin{align}\label{SA.31}
  &\quad\big|F^{N}_{0}(N^\gamma \tilde\zeta;N^\gamma \tilde z) - \breve J^\gamma_\infty(\tilde z, 0)\big|\nonumber\\
  &\le \big|\big(-1 + O(N^{(\gamma-\beta)\theta(1+\theta')})\big)\big(p+O(N^{2\gamma-1})\big) +p\big| + \big|O(N^{2\gamma-1})\big|+ \big|O\big(N^{\max\{(\gamma-\beta)\theta\theta', -\gamma\alpha'\}}\big)\big|\nonumber\\
  &\le K\big(N^{2\gamma-1} + N^{\max\{(\gamma-\beta)\theta\theta', -\gamma\alpha'\}}\big)\le  K  N^{\max\{2\gamma-1, -\gamma(\alpha-1)\theta', -\gamma\alpha'\}}.
\end{align}

\medskip

\noindent \underline{Step 3}\quad We now prove \eqref{3.1.3}. First, in Case 1 above, by \eqref{AN'} we see $A_1'(\tilde z) \le 0$, $\tilde z<0$. Then, based on Step 3 in the proof of Theorem \ref{thm:1} assertion (iii) as well as \eqref{SAbreveJ1}, we have
$\pd_{\tilde z}\breve J^{\gamma}_{\infty}(\tilde{z},0)\le p\chi \lambda'(\tilde z) \le -p \chi |\lambda'(\tilde{Z}^{\ast}_{\gamma}(0)/2)|$, for all $\tilde z \le \tilde{Z}^{\ast}_{\gamma}(0)/2$. Thus, by \eqref{SA.29} and exactly the same arguments as in Step 3 in the proof of Theorem \ref{thm:1} assertion (iii), we obtain $|\tilde z^N_0-\tilde{Z}^{\ast}_{\gamma}(0)|\le  KN^{2\gamma-1}$. Similarly, the same estimate holds for Case 2 above.

For Case 3 above,
by \eqref{SA.34},
\begin{align*}
  \pd_{\tilde z}\breve J^{\gamma}_{\infty}(\tilde{z},0) &= \pd_{\tilde z}\Big(-p  + K_\theta K_\alpha (\theta+\alpha) |\tilde z|^{\theta+\alpha-1}\Big) =  - K_\theta K_\alpha (\theta+\alpha)(\theta+\alpha-1) |\tilde z|^{\theta+\alpha-2}\\
  &\le  - K_\theta K_\alpha (\theta+\alpha)(\theta+\alpha-1) \big|\tfrac{1}{ 2}\tilde{Z}^{\ast}_{\gamma}(0)\big|^{\theta+\alpha-2},\quad \tilde z \le \tfrac{1}{ 2}\tilde{Z}^{\ast}_{\gamma}(0),
\end{align*}
from where similar arguments and \eqref{SA.31} yield the last estimate in \eqref{3.1.3}.

Finally, the case $v=1$ can be proved similarly. Additionally, the same estimate in \eqref{3.2.4} in the setting of Theorem \ref{thm:1} stands -- by the exact same arguments in its proof.
\qed

\medskip

\noindent \textbf{Proof of Theorem \ref{thm:3} assertion (iv).}\quad We follow the proof of Theorem \ref{thm:2} presented in Appendix \ref{C}. In Definition \ref{def:3}, first, \eqref{3.1.8} is justified based on Theorem \ref{thm:2} with $\epsilon=\epsilon_N = KN^{\gamma-1/2}$, so it suffices to verify \eqref{3.1.7} with some appropriate $\epsilon$. To this end, note that by \eqref{3.3.1},
\begin{align}\label{SA.32}
  N^{-\gamma} J_N(p; N^\gamma \tilde z, v) &=N^{-\gamma} e^{-\lambda(N^{\gamma -\beta}\tilde z)}(v-p) N^\gamma \tilde z- N^{-\gamma} (1-e^{-\lambda(N^{\gamma-\beta}\tilde z)})C_{0}(N^{\gamma}\tilde z)\nonumber\\
  &= (v-p)  \tilde z (\chi e^{-\lambda(N^{\gamma -\beta}\tilde z)} - \chi_0)-  (1-e^{-\lambda(N^{\gamma-\beta}\tilde z)}) N^{-\gamma}C_{0}(N^{\gamma}\tilde z),\quad z\in\mathbb{R}_{v}.
\end{align}
Since, again, $\gamma = {\beta \theta/(\theta + \alpha -1)}<1/2$ and $\tilde P^*_\gamma =p$, we turn to the two cases in Step 2 in the above proof of assertion (iii).

\smallskip

In Case 1, namely $\gamma=\beta=0$, \eqref{B.16'} gives that
\begin{equation*}
  N^{-\gamma} J_N(p; N^\gamma \tilde z, v) = (v-p)\tilde z (\chi e^{-\lambda(\tilde z)} - \chi_0)- (1-e^{-\lambda(\tilde z)})C_{0}(\tilde z) =\tilde J^\gamma_{\infty}(p; z,v).
\end{equation*}
Then, by Definition \ref{def:2},
\begin{equation*}
  N^{-\gamma} J_N(p; N^\gamma \tilde Z^*_\gamma(v), v) =\tilde J^\gamma_{\infty}(p;  \tilde Z^*_\gamma(v),v) = \sup_{\tilde z\in \mathbb{R}_{v}} \tilde J^\gamma_{\infty}(p; \tilde z,v) = \sup_{z\in \mathbb{R}_{v}} N^{-\gamma} J_N(p; z, v),
\end{equation*}
which implies \eqref{3.1.7} with $\epsilon = 0$. Therefore, combined with \eqref{3.1.8}, $\epsilon=\epsilon_N=KN^{\gamma-1/2}$. The same holds in Case 2: $\gamma=\beta >0$ and $\alpha=1$.

It remains to consider Case 3, namely $\beta >0$ and $\alpha>1$. We will verify \eqref{3.1.7}  only for $v=0$; the case $v=1$ can be considered similarly. Recall \eqref{SA.32} and denote
\begin{align}\label{SA.33}
  &\breve J_N(p; \tilde z, 0) :=  e^{-\lambda(N^{\gamma -\beta}\tilde z)} \pd_{\tilde z}\Big(N^{-\gamma} J_N(p; N^\gamma \tilde z, 0)\Big) = - p\big(\chi-\chi_0 e^{\lambda(N^{\gamma -\beta}\tilde z)}\big) \nonumber \\
  &\quad + p\chi N^{\gamma -\beta} \tilde z \lambda'(N^{\gamma -\beta} \tilde z) - N^{-\beta} \lambda'(N^{\gamma -\beta} \tilde z) C_0(N^\gamma \tilde z) - (e^{\lambda(N^{\gamma -\beta} \tilde z)}-1)C_0'(N^\gamma \tilde z),\quad \tilde z<0,
\end{align}
with
\begin{equation*}
  \lim_{\tilde z\searrow 0} J_N(p; N^\gamma \tilde z, 0)= 0,\quad  \lim_{\tilde z\to -\infty} J_N(p; N^\gamma \tilde z, 0) =-\infty,\quad \lim_{\tilde z\searrow 0} \breve J_N(p; \tilde z, 0) = - p <0.
\end{equation*}
By the first and third limits, it is clear that $\sup_{\tilde z<0} J_N(p; N^\gamma \tilde z, 0)>0$. Then, by the first two limits above,
we see that  $J_N(p; \cdot, 0)$ has a maximum point $\tilde z^*_N <0$, which satisfies $\breve J_N(p; \tilde z^*_N, 0)=0$.

We claim that $|\tilde z^*_N|\le K_0$ for some $K_0>0$. Indeed, note that all three terms in the second line of \eqref{SA.33} are positive. Then, denoting $\hat z^*_N:= N^{\gamma -\beta} \tilde z^*_N$, we must have
\begin{equation*}
  |p\chi \hat z^*_N \lambda'(\hat z^*_N)|+\big|N^{-\beta} \lambda'(\hat z^*_N) C_0(N^\gamma \tilde z^*_N)\big|  \le p\big(\chi-\chi_0 e^{\lambda(\hat z^*_N)}\big)\le p\chi.
\end{equation*}
In particular, $|\hat z^*_N \lambda'(\hat z^*_N)|\le 1$. Then, clearly $|\hat z^*_N|\le K_1$ for some $K_1>0$, and
\begin{equation*}
  \big|N^{-\beta} \lambda'(\hat z^*_N) C_0(N^\gamma \tilde z^*_N)\big|  \le  p\chi,
\end{equation*}
which is \eqref{SA.28} with $K_2=p\chi$. Then, employing the arguments following \eqref{SA.28}, we obtain the desired boundedness of $ \tilde z^*_N$.

Now, by \eqref{SAbreveJ3}, we have
\begin{align*}
 \tilde J^\gamma_{\infty}(\tilde P^*_\gamma;\tilde{z},0) =-p  \tilde z - K_\theta K_\alpha  |\tilde z|^{\theta + \alpha}, \quad \tilde z <0,
\end{align*}
from which using that $|\tilde z|\le K_0$, by \eqref{SA.32} we obtain that as $N\to\infty$,
\begin{align*}
  &\quad\big|N^{-\gamma} J_N(p; N^\gamma \tilde z, 0) -  \tilde J^\gamma_{\infty}(\tilde P^*_\gamma;\tilde{z},0)\big|\\
  &=\big| -p  \tilde z (\chi e^{-\lambda(N^{\gamma -\beta}\tilde z)} - \chi_0) -  (1-e^{-\lambda(N^{\gamma-\beta}\tilde z)}) N^{-\gamma}C_{0}(N^{\gamma}\tilde z)+p  \tilde z + K_\theta K_\alpha  |\tilde z|^{\theta + \alpha}\big|\\
  &\le K\big|e^{-\lambda(N^{\gamma -\beta}\tilde z)}-1\big| + \big|  (1-e^{-\lambda(N^{\gamma-\beta}\tilde z)}) N^{-\gamma}C_{0}(N^{\gamma}\tilde z)- K_\theta K_\alpha  |\tilde z|^{\theta + \alpha}\big| \\
  &\le K\big|\lambda(N^{\gamma -\beta}\tilde z)\big| + \big|  \big(\lambda(N^{\gamma-\beta}\tilde z) + O(\lambda^2(N^{\gamma-\beta}\tilde z))\big) N^{-\gamma}C_{0}(N^{\gamma}\tilde z)- K_\theta K_\alpha  |\tilde z|^{\theta + \alpha}\big|\\
  &\le K|N^{\gamma -\beta}\tilde z|^{\theta} +  \big|  \big(K_\theta |N^{\gamma-\beta}\tilde z|^{\theta} + O(  |N^{\gamma-\beta}|^{\theta(1+\theta')}) + O(|N^{\gamma-\beta}|^{2\theta})\big) \\
  &\qquad \times N^{-\gamma} \big(K_\alpha|N^{\gamma}\tilde z|^\alpha + O(|N^{\gamma}|^{\alpha-\alpha'})\big) - K_\theta K_\alpha  |\tilde z|^{\theta + \alpha}\big| \\
  &\le K\big( N^{(\gamma-\beta)\theta} + N^{(\gamma-\beta)\theta -\gamma + \gamma (\alpha-\alpha')} +  N^{(\gamma-\beta)\theta(1+\theta')-\gamma + \gamma\alpha } + N^{(\gamma-\beta)2\theta - \gamma + \gamma\alpha}\big)\\
  &\le K\big( N^{-\gamma(\alpha-1)} + N^{-\gamma\alpha'} +  N^{-\gamma(\alpha-1)\theta'} \big)\le K N^{\max\{-\gamma(\alpha-1),  -\gamma\alpha', -\gamma(\alpha-1)\theta'\}} =: \tfrac{1}{2} \epsilon'_N.
\end{align*}
Therefore,
\begin{align*}
  &\quad\sup_{z<0} N^{-\gamma} J_N(p; z, 0) = N^{-\gamma} J_N(p; N^\gamma \tilde z^*_N, 0)\le \tilde J^\gamma_{\infty}(\tilde P^*_\gamma;\tilde z^*_N,0) + \tfrac{1}{2}\epsilon'_N\\
  &\le \tilde J^\gamma_{\infty}(\tilde P^*_\gamma;\tilde Z^*_\gamma(0),0) + \tfrac{1}{ 2}\epsilon'_N \le N^{-\gamma} J_N(p; N^\gamma \tilde Z^*_\gamma(0) , 0) +\epsilon'_N,
\end{align*}
which verifies \eqref{3.1.7} with $\epsilon=\epsilon'_N$ and whose combination with \eqref{3.1.8} yields that in this case, $\epsilon=\max\{\epsilon_{N},\epsilon'_{N}\}=KN^{\max\{\gamma-1/2, -\gamma(\alpha-1),  -\gamma\alpha', -\gamma(\alpha-1)\theta'\}}$. This completes the proof.
\qed

\renewcommand{\theequation}{SB.\arabic{equation}}

\section{Auxiliary details}\label{SB}

This supplemental appendix provides auxiliary details for minor illustrative or computational arguments made throughout the paper, including all those not covered in the preceding appendices.

\medskip

\noindent \textbf{Details for hazard rate (\ref{2.1.3})} \smallskip\\
Let $N=\sigma=1$ without loss of generality. Then, for (\ref{2.1.2}), it is easy to see that
\begin{equation*}
  \mathcal{p}_{1}(z)=D(z)\PP\{|W+z|\geq\bar{y}\}=\frac{D(z)}{2}\bigg(\erfc\frac{\bar{y}+z}{\sqrt{2}}+\erfc\frac{\bar{y}-z}{\sqrt{2}}\bigg),\quad z\in\mathbb{R},
\end{equation*}
where $\bar{y}>0$. Matching this to (\ref{2.1.1}) and using that $\lambda_{1}$ is positive-valued gives (\ref{2.1.3}).

Further, for the complementary error function, it is familiar that (see Abramowitz and Stegun (\citeyear{AS72}) \text{Chap.} 7)
\begin{equation*}
  \erfc x=1-\tfrac{2x}{\sqrt{\pi}}+O(x^{2}),\quad\text{as }x\to0.
\end{equation*}
Thus,
\begin{align}\label{B.1}
  \lambda_{1}(z) \equiv\lambda(z)&=-\log\bigg(1-\frac{D(z)}{2}\bigg(\erfc\frac{\bar{y}+z}{\sqrt{2}}+\erfc\frac{\bar{y}-z}{\sqrt{2}}\bigg)\bigg) \nonumber\\
  &=-\log(1-D(z))+O(z^{2}) =O(|z|^{\min\{\theta_{D},2\}})\quad \text{as }z\to0,
\end{align}
assuming that $D(z)\sim K_{D}|z|^{\theta_{D}}$ as $|z|\searrow0$, with $K_{D}>0$ and $\theta_{D}\geq1$. Note that since $|\erfc x|<1$, $\forall x\in\mathbb{R}$, the function $D$ is required to be bounded by 1, or else (\ref{B.1}) is not real-valued. Moreover, if $D$ is twice continuously differentiable, then so is $\lambda$ in (\ref{B.1}). Using that $\erfc'x=-2e^{x^{2}}/\sqrt{\pi}$, for $z>0$ we obtain
\begin{equation}\label{B.2}
  \lambda'(z)=\frac{\frac{D(z)}{\sqrt{2\pi}}\big(e^{-(\bar{y}+z)^{2}/2}+e^{-(\bar{y}-z)^{2}/2}\big) +\frac{D'(z)}{2}\big(\erfc\frac{\bar{y}+z}{\sqrt{2}}+\erfc\frac{\bar{y}-z}{\sqrt{2}}\big)} {1-\frac{D(z)}{2}\big(\erfc\frac{\bar{y}+z}{\sqrt{2}}+\erfc\frac{\bar{y}-z}{\sqrt{2}}\big)}.
\end{equation}
Since the function $\erfc((\bar{y}+z)/\sqrt{2})+\erfc((\bar{y}-z)/\sqrt{2})$, $z>0$, is strictly increasing, (\ref{B.2}) is clearly increasing provided that $D'$ is also increasing; in particular,
\begin{equation*}
  \lambda'(z)\geq \tfrac{1}{\sqrt{2\pi}}e^{-(\bar{y}+z)^{2}/2}(e^{2\bar{y}z}-1)D(z)\geq \sqrt{\tfrac{2}{\pi}}\bar{y}e^{-\bar{y}^{2}/2}D(z),\quad z>0.
\end{equation*}
It follows from the oddity of $\lambda'$ that $|\lambda'(z)|\geq bD(z)$, $z\neq0$, with $b=2\bar{y}e^{-\bar{y}^{2}/2}/\sqrt{2\pi}$. Therefore, for $D(z)=K_{D}|z|^{\theta_{D}}$,  we have $|\lambda'(z)|\geq bD(z)=bK_{D}|z|^{\theta_{D}}$, $z\neq 0$. Additionally, with $D'(0)$ and $D''(0)$ both finite, a further expansion (\ref{B.2}) around $z=0$ directly shows that $\lambda'(z)=\lambda'(0)+O(|z|)$ as $z\to0$.
\qed

\medskip

\noindent\textbf{Details for Example \ref{ex:1}.}\quad
Using Theorem \ref{thm:1}, since $\gamma=\beta<1/2$, the equilibrium price function $\tilde{P}^{\ast}_{\gamma}=p\in(0,1)$, and, as in \eqref{1storder2}, the limiting equilibrium conditions are the decoupled equations $\breve J^{\gamma}_{\infty}(\tilde{z},v)=0$, $v\in\{0,1\}$, for $\breve J^{\gamma}_{\infty}(\tilde{z},v)$ defined in \eqref{breveJ}.  With $\lambda(\tilde{z})=\tilde{z}^{2}$ and $\chi_{0}=\chi-1\geq0$, they specialize to
\begin{equation}\label{SB.3}
  \breve J^{\gamma}_{\infty}(\tilde{z},v)=(v-p)\big(\chi - 2\chi\tilde{z}^{2}-\chi_{0}e^{\tilde{z}^{2}}\big)=0,
\end{equation}
to be solved for $\tilde{z}\in\mathbb{R}_{v}$. Note that $\breve J^{\gamma}_{\infty}(\tilde{z},1)=-(1-p)/p \breve J^{\gamma}_{\infty}(-\tilde{z},0)$ for $\tilde{z}>0$, which explains the symmetry (about 0) of their solutions. By letting $x = 1/2- \tilde{z}^{2}$, \eqref{SB.3} is equivalent to
\begin{equation*}
  2\chi x - \chi_0 e^{\frac{1}{2}- x}=0 \quad\Longleftrightarrow\quad xe^x = \tfrac{\sqrt{e}\chi_0}{2\chi}.
\end{equation*}
whose unique nonnegative solution can be expressed in terms of the Lambert $\mathrm{W}$ function as $x=\mathrm{W}_0(\sqrt{e}\chi_{0}/(2\chi))$, and $\tilde{z}=\pm\sqrt{1/2-x}$. Thus, setting $\chi=3$ (and $\chi_{0}=2$) gives us the $\gamma$-limiting equilibrium strategy specified in \eqref{4.5}.
\hfill {\scriptsize$\blacklozenge$}

\medskip

\noindent \textbf{Details for Example \ref{ex:2}.}\quad
From the proof of Theorem \ref{thm:3} assertion (ii), since $\gamma=\beta<1/2$, we clearly have $\tilde{P}^{\ast}_{\gamma}=p=1/3$, while the limiting equilibrium conditions are again the decoupled system $\breve J^{\gamma}_{\infty}(\tilde{z},v)=0$, $v\in\{0,1\}$, for $\breve J^{\gamma}_{\infty}(\tilde{z},v)$ in \eqref{SAbreveJ1}, both for $\beta=0$ and for $\beta >0$, the latter case being due to $C_0(z) = |z|$ as used in \eqref{SAbreveJ2} (with $K_\alpha=1$). With $\lambda(z)=|\tilde{z}|$ further, and recalling $\chi=1$, by \eqref{SAbreveJ1} and \eqref{A.4} we find that
\begin{equation}\label{SB.4}
  \breve J^{\gamma}_{\infty}(\tilde{z},v)=\big(v-\tfrac{1}{3}\big)(1-|\tilde z|)+ (-1)^{v}\big(|\tilde z| + e^{|\tilde z|}-1\big) = 0,
\end{equation}
to be solved for $\tilde{z}\in\mathbb{R}_{v}$, $v\in\{0,1\}$. Let $\tilde z_v\in \mathbb{R}_{v}$ denote the solution, and set $x_v = 1-|\tilde z_v|$. Then, for $v=0$, \eqref{SB.4} is equivalent to
\begin{equation*}
  -\tfrac{1}{ 3} x_0 +\big(- x_0 + e^{1-x_0}\big)=0 \quad\Longleftrightarrow\quad x_0e^{x_0} = \tfrac{3e}{4}.
\end{equation*}
Then, $x_0 = \mathrm{W}_0(3e\slash 4)$, and so $\tilde z_0 = \mathrm{W}_0(3e\slash 4)-1$. Similarly, for $v=1$, \eqref{SB.4} is equivalent to
\begin{equation*}
  \tfrac{2}{ 3} x_1 -\big(- x_1 + e^{1-x_1}\big)=0 \quad\Longleftrightarrow\quad x_1e^{x_1} = \tfrac{3e}{5}.
\end{equation*}
Then, $x_1 = \mathrm{W}_0(3e/ 5)$, and so $\tilde z_1 = 1-\mathrm{W}_0(3e/ 5)$.
\qed

\medskip


\noindent \textbf{Details for Example \ref{ex:3}.}\quad
With $\gamma=\beta=0$, we have $\tilde{P}^{\ast}_{0}=p=1/3$, and the limiting equilibrium strategy amounts to maximizing the limiting objective functions in \eqref{B.16'}, with $\chi=1$ and $\chi_0=0$.
By plugging in the given expressions of $\lambda$ and $C_{0}$ and by straightforward calculations, we obtain
\begin{equation*}
  \tilde J^{0}_{\infty}(p;\tilde{z},0)=
  \begin{cases}
    \frac{1}{4},&\text{if }\tilde{z}\leq-6, \\
    \frac{\tilde{z}(\tilde{z}+48)}{144(\tilde{z}-1)},&\text{if }-6<\tilde{z}<0,
  \end{cases}
  \quad
  \tilde J^{0}_{\infty}(p;\tilde{z},1)=
  \begin{cases}
    \frac{\tilde{z}(96-25\tilde{z})}{144(\tilde{z}+1)},&\text{if }0<\tilde{z}<\frac{6}{5}, \\
    \frac{1}{4},&\text{if }\tilde{z}\ge \frac{6}{5}.
  \end{cases}
\end{equation*}
Since ${\tilde{z}(\tilde{z}+48)}/{(144(\tilde{z}-1))} <1/4$ for $-6<\tilde{z}<0$ and $ {\tilde{z}(96-25\tilde{z})}/{(144(\tilde{z}+1))} < 1/4$ for $0<\tilde{z}<6/5$, the continuum of (limiting) equilibrium strategies in \eqref{SB.5} is justified.
\qed

\bigskip

\noindent \textbf{Details for calibration experiments in Section \ref{sec:4}.} \smallskip\\
Suppose that there exists $\bar{z}>0$ such that $D(z)=1$ for $|z|\geq\bar{z}$, meaning that as the insider increases his trade size indefinitely, the probability of prosecution conditional on investigation can reach 1. Then, using that (see, again, Abramowitz and Stegun (\citeyear{AS72}) \text{Chap.} 7)
\begin{equation*}
  \erfc x=
  \begin{cases}
    2+e^{-x^{2}}\Big(\frac{1}{\sqrt{\pi}x}+O(x^{-2})\Big),&\;\text{as }x\to-\infty,\\
    e^{-x^{2}}\Big(\frac{1}{\sqrt{\pi}x}+O(x^{-2})\Big),&\;\text{as }x\to\infty,\\
  \end{cases}
\end{equation*}
we further see that the hazard rate (\ref{2.1.3}) satisfies
\begin{equation*}
  \lambda_{1}(z)\equiv\lambda(z)\sim \frac{z^{2}}{2\sigma^{2}},\quad\text{as }|z|\to\infty,
\end{equation*}
justifying the choice of the approximation $\lambda(z)=Kz^{2}$ with $K=1/(2\sigma^{2})$. For example, Figure \ref{fig:3} below gives an illustration with $\sigma=1000$, $\beta=0$, $\bar{y}=3$, and $D(z)=\min\{z^{2},1\}$ (with $\bar{z}=1$).

\begin{figure}[H]
  \centering
  \includegraphics[width=3in]{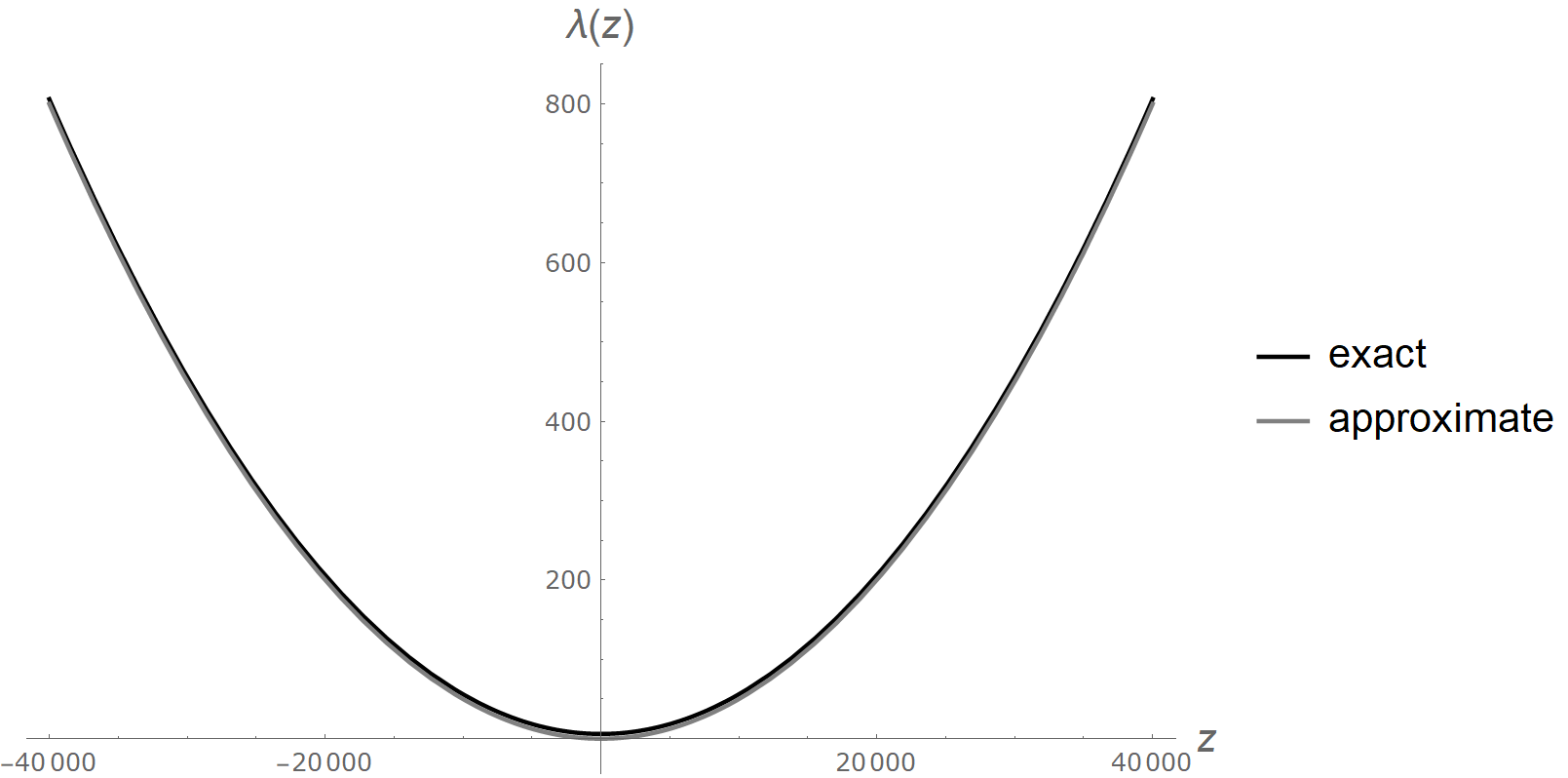}\\
  \caption{Hazard rate approximation}
  \label{fig:3}
\end{figure}

The formula (\ref{4.1.1}) follows straight from the law of total probability:
\begin{equation*}
  \E\big[|Z(V)|\big|B_{N}(Z(V))=1\big]=\frac{\E\big[|Z(V)|\1(B_{N}(Z(V))=1)\big]}{\PP\{B_{N}(Z(V))=1\}} =\frac{\E[|Z(V)|(1-e^{-\lambda_{N}(Z(V))})]}{\E[1-e^{-\lambda_{N}(Z(V))}]},
\end{equation*}
recalling that $V\overset{\rm d.}{=}\text{Bernoulli}(p)$. Similarly, for (\ref{4.1.2}) and (\ref{4.1.3}), note that $X_N\overset{\rm d.}{=}\text{Normal}(N\mu, N(\sigma^{2}-\mu^{2}))$, independent of $V$, and then
\begin{align*}
  \E\big[X_{N}+|Z(V)|\big|B_N(Z(V))=1\big]&=\E[X_N]+\E\big[|Z(V)|\big|B_N(Z(V))=1\big] \\
  &= N\mu+\frac{\sum^{1}_{v=0} p(v)|Z(v)|\big(1-e^{-\lambda_{N}(Z(v))}\big)} {\sum^{1}_{v=0} p(v)\big(1-e^{-\lambda_{N}(Z(v))}\big)}, \\
  \PP\bigg\{\frac{X_N+|Z(V)|}{|Z(V)|}>x\Big|B_N(Z(V))=1\bigg\}&=\PP\bigg\{X_N> (x-1)|Z(V)|\Big|B_N(Z(V))=1\bigg\} \\
  &=\frac{\sum^{1}_{v=0}p(v)\big(1-e^{-\lambda_{N}(Z(v))}\big) \frac{1}{2}\erfc\frac{|Z(v)|(x-1)-N\mu}{\sqrt{2N(\sigma^{2}-\mu^{2})}}}{\sum^{1}_{v=0} p(v)\big(1-e^{-\lambda_{N}(Z(v))}\big)}.
\end{align*}

For the calibration, by writing $Z=N^{\gamma}\tilde{Z}^{\ast}_{\gamma}$ in terms of (\ref{4.1.4}) and using the symmetry $|\tilde{Z}^{\ast}_{\gamma}(0)|=\tilde{Z}^{\ast}_{\gamma}(1)>0$, the equations (\ref{4.1.1}), (\ref{4.1.2}), and (\ref{4.1.3}) can be significantly simplified, permitting explicit solutions. In particular,
\begin{equation*}
  \text{(\ref{4.1.1})}=N^{\gamma}\sigma\times\mathfrak{a},\quad\text{(\ref{4.1.2})}=N\mu+N^{\gamma}\sigma\times\mathfrak{a},\quad \text{(\ref{4.1.3})}=\tfrac{1}{2}\erfc\tfrac{(N^{\gamma}\sigma\times\mathfrak{a})(x-1)-N\mu}{\sqrt{2N(\sigma^{2}-\mu^{2})}}.
\end{equation*}

For conciseness, let us denote the average insider share volume, the average total share volume, and the median insider-to-total volume ratio from data by $\mathfrak{i}$, $\mathfrak{v}$, and $\mathfrak{r}$, respectively, and recall that $\mathfrak{a}=\sqrt{1-2\mathrm{W}_{0}(\sqrt{e}\chi_{0}/(2\chi))}$, with $\chi_{0}=\chi-1$.

The first calibration condition is due to (\ref{4.1.1}) and \eqref{4.1.4},
\begin{equation}\label{B.3}
  N^{\gamma}\sigma\times\mathfrak{a}=\mathfrak{i}.
\end{equation}
Similarly, from (\ref{4.1.2}) we have the second calibration condition,
\begin{equation}\label{B.4}
  N\mu+N^{\gamma}\sigma\times\mathfrak{a}=\mathfrak{v}.
\end{equation}
The third calibration condition is obtained in the same manner by evaluating (\ref{4.1.3}). In particular, since $\erfc-1$ is an odd function, we obtain
\begin{equation}\label{B.5}
  N^{\gamma}\sigma\times\mathfrak{a}\times(\mathfrak{r}^{-1}-1)-N\mu=0.
\end{equation}

Notably, in calibration experiment I, the average total trading volume is associated with the standard error 10,246, respectively, as reported by Meulbroek (\citeyear{M92}). Since there are a total of 588 insider trading episodes included in the data set, a reasonable estimate for the standard deviation of the trading volume, given independence between insider trades and normal trades, is
\begin{equation}\label{B.s}
  \mathfrak{s}=\sqrt{10246^{2}\times 588}\approx248452.
\end{equation}
Thus, the additional condition $N(\sigma^{2}-\mu^{2})=\mathfrak{s}^2$ can be used along with (\ref{B.3}) and (\ref{B.4}) to obtain an estimate for the parameter $\mu>0$, namely
\begin{equation}\label{B.mu}
  \hat{\mu}=\frac{\sqrt{\mathfrak{s}^{4}+4\sigma^{2}\times(\mathfrak{v}-\mathfrak{i})^{2}}-\mathfrak{s}^{2}} {2\times(\mathfrak{v}-\mathfrak{i})},
\end{equation}
which we simply take as $\mu$.

Taking (\ref{B.mu}) as given, the preceding calibration conditions are easily solvable by elementary means. In particular, from (\ref{B.3}) and (\ref{B.4}) we obtain
\begin{equation}\label{SB.e1}
  \hat{N}=\frac{\mathfrak{v}-\mathfrak{i}}{\mu},\quad\hat{\gamma}=\frac{\log\frac{\mathfrak{i}}{\sigma\times\mathfrak{a}}} {\log\hat{N}};
\end{equation}
if (\ref{B.5}) is used instead of (\ref{B.4}),
\begin{equation}\label{SB.e2}
  \hat{N}=\frac{\mathfrak{i}\times(1-\mathfrak{r})}{\mu\times\mathfrak{r}},\quad \hat{\gamma}=\frac{\log\frac{\mathfrak{i}}{\sigma\times\mathfrak{a}}}{\log\hat{N}};
\end{equation}
when only (\ref{B.4}) and (\ref{B.5}) are used, we have
\begin{equation}\label{SB.e3}
  \hat{N}=\frac{\mathfrak{v}\times(1-\mathfrak{r})}{\mu},\quad\hat{\gamma}=\frac{\log\frac{\mathfrak{v}\times\mathfrak{r}} {\sigma\times\mathfrak{a}}}{\log\hat{N}}.
\end{equation}

According to the statistics shown in Table \ref{tab:1}, setting $\mathfrak{i}=9819$, $\mathfrak{v}=113909$, $\mathfrak{r}=0.113$, $\sigma=1000$, and $\chi\in\{1,2,3\}$ and evaluating (\ref{B.mu}) verifies $\mu=1.68625$, and subsequently evaluating (\ref{B.mu}), (\ref{SB.e1}), (\ref{SB.e2}), and (\ref{SB.e3}) gives rise to Table \ref{tab:2} for calibration experiment I.


\smallskip

In the same vein, for calibration experiment II, Table \ref{tab:5} is a result of evaluating (\ref{SB.e2}) with $\mathfrak{i}=4900$, $\mathfrak{r}=0.026$, $\mu=1.68625$, $\sigma=1000$, and $\chi\in\{1,2,3\}$, based on the information in Table \ref{tab:4}.

Moreover, the formula (\ref{4.2.1}) follows directly from the law of total probability, $\E[B(Z(V))]=\E[1-e^{-\lambda_{N}(Z(V))}]$. Then, using that $|\tilde{Z}^{\ast}_{\gamma}(0)|=\tilde{Z}^{\ast}_{\gamma}(1)>0$, the prosecution probability implied by the equilibrium is
\begin{equation}\label{SB.6}
  1-e^{-\lambda_{N}(N^{\gamma}\tilde{Z}^{\ast}_{\gamma}(1))}=1-e^{-\lambda(\tilde{Z}^{\ast}_{\gamma}(1))},
\end{equation}
which upon setting $\gamma=\beta$ becomes independent of $N$ and $\gamma$. Putting the solved finite-population and limiting strategy values from Table \ref{tab:6} into (\ref{4.2.1}) (with $Z=Z^{\ast}_{N}$) and (\ref{B.6}), respectively, gives the last column of Table \ref{tab:6}.
\qed
\end{appendices}

\end{document}